\documentclass{article}
\usepackage[utf8]{inputenc}
\usepackage{amsmath,amsthm,amsfonts} 
\usepackage{bm}
\usepackage{makeidx}
\usepackage{graphicx}
\usepackage{xcolor}
\usepackage[numbers]{natbib}
\usepackage{hyperref}

\usepackage{subfig}
\usepackage{pifont}
\usepackage{appendix}
\usepackage{authblk}
\usepackage{caption}
\usepackage{subfloat}
\usepackage{subfig}
\newcommand{\f}{\frac}

\newcommand{\p}{\partial}

\usepackage{ulem}

\title{A purely mechanical model with asymmetric features for early morphogenesis of rod-shaped bacteria micro-colony }

\author{Marie Doumic\footnote{Sorbonne Universit\'{e}, Inria, Universit\'{e} Paris-Diderot, CNRS,  Laboratoire Jacques-Louis Lions, F-75005 Paris, France. Email adress: marie.doumic@inria.fr}, $\quad$ Sophie Hecht, \footnote{Sorbonne Universit\'{e}, Inria, Universit\'{e} Paris-Diderot, CNRS,  Laboratoire Jacques-Louis Lions, F-75005 Paris, France. Email adress: sophie.hecht@inria.fr}$\;$\footnote{corresponding author} $\quad$ Diane Peurichard\footnote{Sorbonne Universit\'{e}, Inria, Universit\'{e} Paris-Diderot, CNRS,  Laboratoire Jacques-Louis Lions, F-75005 Paris, France. Email adress: diane.peurichard@inria.fr} }
\date{\today}

\begin{document}

\maketitle

\begin{abstract}
To model the morphogenesis of rod-shaped bacterial micro-colony, several individual-based models have been proposed in the biophysical literature. When studying the shape of micro-colonies, most models present interaction forces such as attraction or filial link. In this article, we propose a model where the bacteria interact only through non-overlapping constraints. We
 consider  the asymmetry of the bacteria, and its influence on the friction with the substrate. Besides, we consider asymmetry in the mass distribution of the bacteria along their length. These two new modelling assumptions allow us to retrieve mechanical behaviours of micro-colony growth without the need of interaction such as attraction. We compare our model to various sets of experiments, discuss our results, and propose several quantifiers to compare model to data in a systematic way.
\end{abstract}
          
\section{Introduction}

Bacteria are ubiquitous unicellular organisms, whose biomass exceeds that of all other living organisms, and on which our survival is dependent.
From a single organism, they quickly develop into organised micro-colonies and biofilm structures. The self-organisation of the colony into a dense aggregate is the result of the interplay of various chemical and biological signalling as well as mechanical interactions. These interplays, while increasingly studied in the past decade, are still only partly understood. In particular, the influence of the mechanical or chemical interactions between the particles such as attraction, repulsion or alignment on the global shape of the colony is not clear. 

To model mathematically and simulate
the self-organisation of bacteria, scientists have used microscopic and macroscopic models. On the one hand, microscopic models consider each particle individually and interactions are represented by forces or constraints. These models give a high level of description but also result in computationally costly numerical results. In the context of bacterial growth, microscopic models mainly take the form of individual-based models (IBM) where each rod-shaped bacterium is described by a spherocylinder \cite{Duvernoy2018,Ghosh:2015aa}. In \cite{You2018,Boyer2011,Volfson2008}, the non-overlapping constraint on neighbouring bacteria is achieved via a repulsive force based on Hertzian theory. Some other models represent bacteria as hard-spheres \cite{Jonsson:2005aa} or spring-mass systems \cite{Grant:2014aa,Storck:2014aa} and consider volume exclusion potential \cite{Cho:2007aa}. On the other hand, macroscopic models consider local averages such as densities and describe the evolution of a system through partial differential equations (PDE). This description reduces the computational cost but is less precise than the one of a microscopic model. The macroscopic models found in the literature are based on nematodynamic equations \cite{DellArciprete2018,Duvernoy2018,Doostmohammadi:2016aa}. These have been developed in the context of liquid crystals. The crystal being rod-shaped, similarly to \textit{E. coli} and \textit{pseudomonas} bacteria, the nematic model can be adapted to bacteria development by the addition of growth. However, these macroscopic models are often complex and rely on empirical laws, so that they are  difficult to relate to the reality of a biological system.

In this study, our aim is  to understand how the mechanical interactions between the bacteria drive the growth of micro-colonies, which forces are necessary to take into account and which are not, and to propose quantifiers to estimate the model parameters. Moreover, we focus on the early steps of the micro-colony morphogenesis, which is the phase where a continuous approximation would be the less accurate. We therefore chose to develop an IBM model, which allows us a finer investigation of the influence of each modeling ingredient than a continuous averaging model.

\ 

 Models have been used to study the different steps of a biofilm formation. Studies have shown that free-swimming bacteria migrate on surfaces rich in nutrient and transit to a sessile state before starting the formation of micro-colonies \cite{Ariel:2013aa,Park:2003aa,Jacob:2004aa,Giomi:2013aa}. Once immobile, a given bacterium grows and divides, giving birth to a small cluster of cells called a micro-colony \cite{Stewart2005,Shapiro:1989aa}. The micro-colony first grows in a two-dimensional organisation before developing into a three-dimensional structure \cite{Grant:2014aa,Su:2012aa}. Later on, the micro-colony transforms into a mature biofilm. The morphology (filamentous or mushroom structures \cite{Hall-Stoodley:2004aa}) of these biofilms, as well as their physiology (visco-elastic, viscous \cite{Shaw:2004aa}), have been widely studied. 
 
 The process which is the focus of our study is the formation of a micro-colony from a single individual. In particular we consider the situation of a non-motile 2D growing colony on a controlled non-restrictive-space substrate environment \cite{Stewart2005,Su:2010aa}. A wide range of models has been developed to study this situation. Among them we found models considering the extracellular matrix \cite{Ghosh:2015aa}, nutrient \cite{Farrell:2013aa,Ghosh:2015aa}, substrate adhesion \cite{Duvernoy2018,Storck:2014aa}, bacteria attraction \cite{Duvernoy2018,Storck:2014aa}, fillial link \cite{Storck:2014aa}... These models are, however, up to our knowledge, unable to reproduce some spatial characteristics of micro-colonies. In particular, most models fail to recover the elongated shape of the micro-colony at an early stage \cite{ Farrell:2013aa,Ghosh:2015aa,Grant:2014aa,You2018}.  In~\cite{Duvernoy2018}, the authors show how an asymmetric adhesion can control the shape of the micro-colony, by a comparison of their model with several types of bacteria and more or less adhesive substrates. Building on their study, we aim at considering the simplest possible model able to recover spatial specificity of the micro-colony growth, questioning whether an attractive potential as proposed in~\cite{Duvernoy2018} is mandatory or not, and whether asymmetric adhesion may be taken into account in a simpler way. To compare simulation results to experimental data, we focus mostly on two characteristics: (C1) the arrangement of colonies composed of four bacteria; (C2) the elongated shape of micro-colonies. The first point (C1) is supported by biological experiments \cite{Shapiro:1989aa} which have shown that the first two daughter cells slide side-by-side after the first division, giving rise to a four-cell array organisation. This configuration is however dependent on the substrate adhesion \cite{Su:2010aa}. The second characteristic we want to study (C2) is the elongation of the micro-colony,  which have up to now failed to be reproduced without considering interaction forces other than non overlapping~\cite{Duvernoy2018} or filial link~\cite{Storck:2014aa}. In addition of these two characteristics we consider the organisation of the bacteria inside the colony, the density of bacteria and the angle observed at division.

Our approach relies on the consideration that the bacterium may present asymetric features. The asymmetry can be considered on different levels during the development. In \cite{Stewart2005}, experiments showed that a bacterium divides into two daughter cells with slightly different growth rates. In particular, the growth of the daughter cell keeping the original pole of its mother is slown down with the development of the micro-colony. On the opposite, in  \cite{Wang:2010aa} the growth rate is stable for the different generations of bacteria. These contradictory results have been discussed in \cite{Delyon:2018}. The study concluded that the different observations where likely due to the fact that depending of the study the bacteria were in transient or stationary phase. For the sake of simplicity, in our model, the division is considered symmetrical and the growth rate of the daughters independent of the growth rate of the mother. Nevertheless, the paper \cite{Stewart2005} has brought out the question of the asymmetry of the pole of adhesion of the bacterium. A recent study \cite{Duvernoy2018} has confirmed this possibility by computing the adhesion force of the pole on different substrates. The difference of adhesion of the pole is also suspected to be an explanation for the four-cell array organisation (C1). In this paper, we consider a similar approach, where instead of adhesion, we consider a non-uniform distribution of the mass along the length of the bacteria.  Besides, we also consider the influence of the shape of the bacteria on their movement. This is modelled by the choice of asymmetric friction. It translates the fact that it is easier for a spherocylinder to slide along its longitudinal axis than to slide transversally. If this type of model has already been considered in \cite{Farrell:2017aa}, the study has however not been developed in our case, {\it i.e.} during the early stage of morphogenesis, where nutrient and space are unlimited. 

These new model assumptions allow us to retrieve the spatial configurations (C1) and (C2) and to investigate the influence of each parameter of the system. To validate our approach, we compare our model with two sets of experimental data, respectively published in the two articles \cite{DellArciprete2018,Duvernoy2018}. The model parameters are tuned to fit some characteristics of the experimental micro-colony such as length distribution and growth rate distribution. Then we compare our model with the experimental data for a variety of quantifiers.  The strength of the model we propose relies on its ability to reproduce mechanical properties of colony growth observed. An especially interesting feature is the fact that it did not require the implementation of active attraction or alignment  between the bacteria - interactions which could only be explained by chemical signalling - and solely relies on the asymmetry of the bacteria and mechanical laws.

This paper is divided in the following four sections. In Section~\ref{Section2} we introduce the IBM developed to study the growth of a micro-colony. The influence of the parameters of the model as well as their choice  is studied in Section~\ref{Section3}. Section~\ref{Section4} contains the comparison between the IBM presented in this paper and experimental data. Finally in Section \ref{Section5} a conclusion is presented, together with a discussion of possible improvements and use for further investigation.

\section{The Individual-based model} \label{Section2}

The Individual-based model we propose is based on models found in the literature \cite{Boyer2011,Damme2019,Volfson2008,You2018}. 
We consider a population of nonmotile rod-shaped bacteria growing and dividing on a 2D medium,  and interacting via steric forces with their neighbours. Bacterium motion is supposed to be essentially passive: bacteria repulse each other to avoid overlapping as they grow in length and divide. As the Reynolds number of the bacteria is very small \cite{kumar2010}, we suppose that inertial forces can be neglected and we consider an over-damped regime for bacterial motion. 

\begin{figure}[!ht] 
   \centering
\includegraphics[scale=0.7]{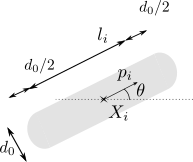}
\caption{Representation of a bacterium $i$.}
\label{fig:1}
\end{figure}

More specifically, each bacterium is modelled by a spherocylinder described by its centre $(X_i)_{i\in[1,N]}$ and orientation vector $p_i=(\cos \theta_i, \sin \theta_i)$. The diameter of a bacterium $i\in[1,N]$ is supposed to be fixed and denoted by $d_0$ while its time-dependent length is denoted by $l_i$. Each bacterium has an associated time-dependent mass $m_i$ (further described). A representation of the bacterium is provided in Fig.~\ref{fig:1}. In the following, we detail each component of the model.

\begin{figure}[!ht] 
   \centering
\includegraphics[scale=0.6]{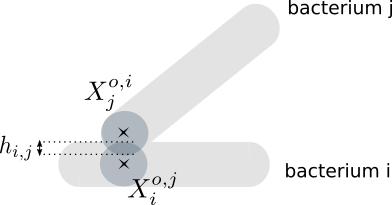}
\caption{Representation of the interaction between two bacteria $i$ and $j$ with an overlapping $h_{i,j}$. The bacteria are represented by grey spherocylinders. The two balls by which we approximate the bacteria for the repulsive force are drawn in blue. }
\label{fig:2}
\end{figure}

\paragraph{ Computation of steric forces.} The force between two spherocylinders $i$ and $j$ is approximated by
the force between two spheres of diameter $d_0$, placed along the major axis of the rods at such positions that their distance is minimal (see Figure \ref{fig:2}). Denoting by $X^{o,j}_{i}$ (resp. $X^{o,i}_{j}$) the point on the spherocylinder $i$ centre line segment (resp. of the spherocylinder $j$) closest to the spherocylinder $j$ centre line segment (resp. $i$), the pairwise interaction force between the spherocylinders $i$ and $j$ is set to, using Hertzian theory \cite{Shafer}:
$$F_{i,j}= Y d_0^{1/2}h_{i,j}^{3/2} n_{i,j},$$
where  $Y$ is the Young's modulus, $h_{i,j} = |X^{o,j}_{i} -X^{o,i}_{j} |-2d_0 $ is the overlap distance between the two spherocylinders, and $n_{i,j}$ is their common unit normal vector given by
$$n_{i,j} = \frac{X^{o,j}_{i} -X^{o,i}_{j}}{|X^{o,j}_{i} -X^{o,i}_{j}|}. $$

 This steric force between the spherocylinders $i$ and $j$ generates a torque acting on the centre of the spherocylinder $i$ of the form: 
$$
T_{i,j} =   \Big ((X^{o,j}_{i}-X_i) \land  F_{i,j} \Big)\cdot z,
$$
where $z$ is the unit vector perpendicular to the plane of the bacterium.

The total force $F^s_i$ and torque $T^s_i$ sensed by bacterium $i$ due to non-overlapping with its contact neighbours are then supposed to be the sum of all these elementary pairwise forces and torques:
\begin{align*}
F^s_i &= \sum_{j} F_{i,j}, \\
T^s_i &= \sum_{j} T_{i,j},
\end{align*}
where the sum runs over all spherocylinders in contact with the spherocylinder $i$.

\paragraph{Computation of asymmetric friction forces.}  In addition to the non-overlapping forces between the bacteria, we consider the friction force on the substrate. This force is usually of the form $F^f_i=-m_i\zeta \frac{d X_i}{dt} $ with $\zeta$ the friction coefficient, i.e the drag per unit length originating from the substrate adhesion. However, in this study, we consider an asymmetric friction, in order to take into account the influence of the shape of the bacteria on the movement \cite{Duvernoy2018,Farrell:2017aa}. For a given bacterium $i$, let $\zeta^{||}_i$ and $\zeta_i^{\perp}$ be the respective coefficients in the directions parallel and perpendicular to the  axis of the cell. In the particular case of rod-shaped bacteria, we consider $\zeta^{||}_i \leq \zeta^{\perp}_i $. Then the friction matrix in the repository of a given bacterium is given by $\begin{pmatrix}
\zeta_i^{||} & 0\\
0 & \zeta_i^{\perp}
\end{pmatrix}$. Therefore with a change of basis, we get that the friction matrix in the general basis is 
$$K_i=\begin{pmatrix}
\zeta_i^{||} \cos(\theta_i)^2+\zeta_i^{\perp} \sin(\theta_i)^2 & (\zeta_i^{||}-\zeta_i^{\perp}) \cos(\theta_i) \sin(\theta_i)\\
(\zeta_i^{||}-\zeta_i^{\perp}) \cos(\theta_i) \sin(\theta_i) & \zeta_i^{\perp} \cos(\theta_i)^2+\zeta_i^{||} \sin(\theta_i)^2
\end{pmatrix}.$$ 
Then the friction force of a bacterium $i$ is given by
$$F^f_i = - m_i K_i \frac{d X_i}{dt}.$$

\paragraph{ Equations of motion.}
 Altogether, using Newton's equations in the over-damped regime, the evolution of the position $X_i$ and the orientation angle $\theta_i$ of the bacterium $i$ is governed by:
\begin{align} 
    \frac{d X_i}{dt} &= K_i^{-1} \frac{1}{m_i} F^s_i, \label{eq:X}\\
    \frac{d \theta_i}{dt} &= \frac{1}{\zeta_{\perp} I_i} T^s_i, \label{eq:Theta}
\end{align}
where $I_i$ denotes the inertia momentum of bacterium $i$ \cite{Farrell:2017aa}. 

The second equation \eqref{eq:Theta} has been obtained by considering the angular momentum $L_{\Delta}(M)$ of a point $M$ of the bacterium $i$ with respect to the axis of rotation $(\Delta)$ passing by $X_i$ orientated by the unit vector $z$ (perpendicular to the plane of the bacterium). We have 
$$ L_{\Delta}(M) = ((M-X_i) \land m_i \frac{d X_i}{dt} ) \cdot z,$$
with $\land$ the two dimensional vector product. It is well know that $L_{\Delta}(M) \approx I_i \frac{d \theta_i}{dt}$ with $I_i$ the inertia momentum of the bacterium from the axis of rotation $(\Delta)$. A small computation gives $$ ((M-X_i) \land F^f_i ) \cdot z = \zeta_{\perp} I_i \frac{d \theta_i}{dt}.$$ 
 Using the law of conservation of angular momentum gives \eqref{eq:Theta}.

\begin{figure}[!ht] 
   \centering
\includegraphics[scale=0.4]{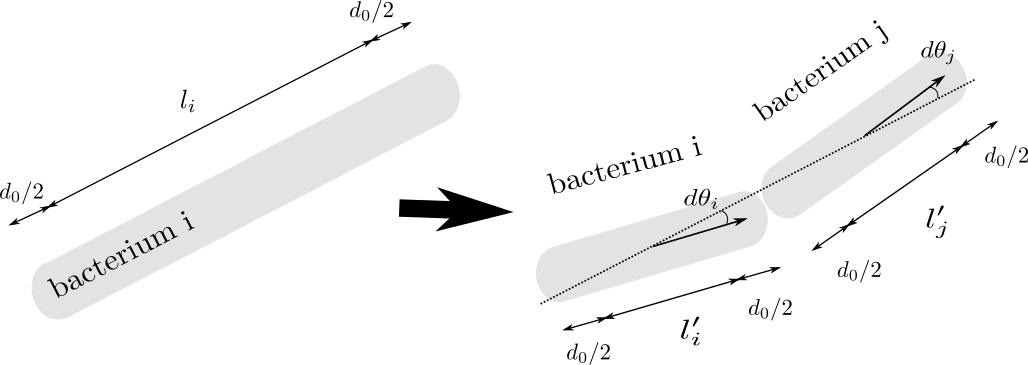}
\caption{Representation of the division of a bacterium $i$ into two daughter bacteria $i$ and $j$. The mother cell is of length $l_i$ and the two daughter cells are of length $l_i'=l_j'=0.5~l^i-d_0$. The angle of the daughters are disrupted by $d\theta_i$ and $d\theta_j$. }
\label{fig:3}
\end{figure}

\paragraph{Exponential growth.}  The cell cycle followed by the bacteria is composed of two steps: (i) first the elongation of the cell with an exponential growth rate, and (ii) the division of the bacterium into two symmetric daughter bacteria. The length growth is supposed to be exponential, as proved in many studies~\cite{robert2014}, which is translated by 
\begin{equation} \label{eq:l}
  \frac{d l_i}{dt} = g_i l_i ,
  \end{equation}
with $g_i$ the growth rate respective to the bacterium $i$. When the increment of length of a bacterium $i$ reaches a given threshold $\epsilon_l^i$, the bacterium divides, giving birth to two daughter cells of length $0.5~l^i-d_0$. At division, we consider a small noise on the orientation $d\theta_i$ in order to break the alignment of the bacterium. A representation of the division is presented in Fig.~\ref{fig:3}.

\paragraph{Distribution of mass} 
In order to model the redistribution of material during cell division, we suppose that the mass $m_i$ of a bacterium $i$ is not necessarily uniform, but rather pole-dependent. To this aim, given a bacterium $i$ daughter of a bacterium $m$, we denote by $X^{p_o}_i$ the old pole, i.e the extremity of the spherocylinder that was already a pole for the bacterium $m$, and by $X^{p_n}_i$ the new pole. We suppose that the masses $m^{p_o}_i$ and $m^{p_n}_i$ associated to these two poles are not necessarily equal but distributed such that $ m^{p_o}_i = \alpha_i m_i $ and $ m^{p_n}_i = (1-\alpha_i) m_i $, with $\alpha_i \in [0,1]$. In order to take into account mass variation for cell division, $\alpha_i$ is chosen to be time-dependent. 
The centre of mass of a bacterium is then given by
$$ X_i = \frac{1}{m_i} (m^{p_o}_i X^{p_o}_i+m^{p_n}_i X^{p_n}_i)=\alpha_i X_i^{p_0} + (1-\alpha_i)X_i^{p_n}.$$

\textit{\textbf{Remark: }} {As the friction force is proportional to the bacterium mass, considering asymmetric mass distribution may be viewed as a way to change the friction or substrate adhesion coefficient along the bacterium, hypothesis which may appear more physically relevant, as shown in~\cite{Duvernoy2018}. }
\medskip

In the numerical simulations the system is always initialised with one bacterium at position $X_1=(0,0)$, with orientation angle $\theta_1=0$, mass $m_1=m^{\mbox{\scriptsize ini}}$ and length $l_1=l^{\mbox{\scriptsize ini}}$. The algorithm implemented to simulate the model is presented in Appendix \ref{app:algo}.

\section{Numerical simulations } \label{Section3}

In this section, we present some numerical simulations of the model introduced in Section~\ref{Section2}. We first explain the choice of the model parameters to fit a set of experiments. We then study the influence of the asymmetric friction and the mass distribution on the growth of the micro-colony.

\subsection{Choice of the model parameters} \label{Section3.1}

The model parameters are listed below:
\begin{itemize}
    \item the parameters related to the initialisation: $l^{\mbox{\scriptsize ini}}$, $d_0$,
    \item the parameters related to the division of the bacteria: the threshold of division $(\epsilon_i)_i$, the noise of the angle at division $(d\theta_i)_i$, the growth rate $(g_i)_i$.
    \item the parameters related to the mass distribution: the (possibly time-dependent) mass ratio $(\alpha_i)_i$,
    \item the parameters related to the non-overlapping force: the Young's modulus $Y$, 
    \item the parameters related to the friction force: $(\zeta_i)_i$, $(\zeta_i^{||})_i$, $(\zeta_i^{\perp})_i$.
    \item the parameters related to the algorithm: the time step $dt$ and the final time $T_{\mbox{\scriptsize max}}$.
\end{itemize}

The length and diameter of the bacterium are initialised depending on the set of experiments we aim to fit. This is also the case for the parameters related to the division of the bacteria. In this paper, we compare our numerical simulations to three sets of experimental data. The first two sets of data originated from \cite{Duvernoy2018}. In the following of the paper, we denote by Dataset 1 the data corresponding to colonies of \textit{E. coli} and Dataset 2 the data corresponding to colonies of \textit{pseudomonas}. The third set of data corresponding to \textit{E. coli} colonies originates from \cite{DellArciprete2018}.  All data have been extracted by image segmentation of pictures taken at fixed time intervals from growing micro-colonies, and have been kindly provided by the authors of~\cite{DellArciprete2018,Duvernoy2018}.

\paragraph{Dataset 1} Data of 7 colonies of \textit{E. coli} bacteria taken  every 3 minutes for final times varying between 138 and 204 minutes. The first data of each colony contains 2 bacteria.

\paragraph{Dataset 2} Data of 10 colonies of \textit{pseudomonas} bacteria taken every 5 minutes for final times varying between 312 and 417 minutes. The first data for each colony contains 2 to 4 bacteria.

\paragraph{Dataset 3} Data of 32 colonies of \textit{E. coli} bacteria taken growing every minute for final times varying between 180 and 341 minutes. The first data for each colony contains 1 to 2 bacteria. Note that the data do not give access to the width of the bacteria. This set of data, while being more consequent than the other two, presents some mistakes in the segmentation and is therefore more delicate to deal with. For the sake of the study we thus did not consider any data corresponding to bacteria observable for less than 15 minutes: we found out that these cases, when observed in more detail, correspond to segmentation errors leading to a bacterium dividing into two and after a short time merging again into one. \\

For each set of experiments, the parameters are defined as follow:
\begin{itemize}
\item the diameter of the bacteria $d_0$ is defined according to the average width of the bacteria available in the dataset. In the case of Dataset 3, the diameter of the bacteria has been estimated from the images of \cite{DellArciprete2018} by linking the length of the bacteria and an image available in the paper \cite{DellArciprete2018}. 
\item the threshold of division $\epsilon_i$: a bacterium divides when its increment of length reaches the threshold $\epsilon_i$. This threshold is randomly chosen according to the law of at-division increments estimated from the experimental data available, see Appendix~\ref{app:division} for more details. In this way, we have data-driven parameters, and the modelling assumptions which are currently the most widely accepted ones for bacterial division~\cite{Amir:2014aa,Taheri-Araghi:2015aa}.
\item the noise in the angle at division $d\theta_i$ is chosen from a uniform distribution $ \mathcal{U}(-\Theta/2,\Theta/2) $. The choice of $\Theta$ is made to fit the angle at division of the experimental data. However given the fact that data are available every 1 or 3 minutes, we do not have access to the real angle at division. The choice of this parameter is further discussed in Section~\ref{Section3.3.3}.
\item the growth rate $g_i$:  After cell division, each daughter cell is assigned a growth rate $g_i$ which is supposed to be constant all along the bacterium lifetime.  This hypothesis is supported by the observation of the evolution of the growth rate in time in the experimental data and previous studies~\cite{robert2014}. The value $g_i$ is chosen according to the growth rate law estimated from the experimental data available. The growth rate  of a bacterium $j$ is computed with the formula 
$$g_j = \frac{1}{t^d_j-t^n_j} \ln(\frac{l^d_j}{l^n_j}),$$
with $t^n_j$, $t^d_j$ the time at birth and death respectively and $l^n_j$, $l^d_j$ the length at  birth and death respectively.
\item the value of the Young's modulus is fixed to $Y=4 MPa$ according to the paper \cite{You2018}.
\item the friction coefficients $\zeta= 200 Pa h$ according to the paper \cite{You2018}. For the sake of simplicity, the longitudinal and normal friction are chosen of the form $\zeta^{||}_i = A_i \zeta_i$ and $\zeta^{\perp}_i = \frac{\zeta_i}{A_i}$. The value of $A_i$ is chosen such that $0<A_i<1$ to represent the fact that it is easier for the bacteria to slide in its direction than perpendicular to it. The choice of the value of $A_i$ is discussed in Section~\ref{Section3.3.1}.
\item the mass ratio verifies $0<\alpha_i<1$ for all bacteria and its choice is discussed in Section~\ref{Section3.3.2}.
\item the time parameters:  the time step is initially chosen to $dt = 10^{-2}$ and then adapted to ensure that the maximal displacement of the bacteria does not exceed a given threshold. The final time $T_{\mbox{\scriptsize max}}$ is chosen to ensure that the simulation-produced colonies reach similar area as the experimental data. 
\end{itemize}

The values of the parameters for the different experiments are summarised in Table~\ref{Table:1}. The values in bold are subject to change along the paper.


\begin{table}[h!]
\centering
\begin{tabular}{ |p{2cm}||p{2cm}|p{2cm}|p{2cm}|  }
 \hline
 \multicolumn{4}{|c|}{Parameter values} \\
 \hline
Parameter & Dataset 1 & Dataset 2 & Dataset 3\\
 \hline
  $l^{\mbox{\scriptsize ini}}$   & 4.45 $\mu m$    & 2.41 $\mu m$  &   3.378 $\mu m$ \\
 $d_0$  &   1.40 $\mu m$   & 0.89 $\mu m$    & 1 $\mu m$ \\
 $\epsilon_i$ & \multicolumn{3}{|c|}{to fit the experimental distribution}\\
 $\Theta$    & $\bm{10^{-5}}$ & $\bm{10^{-5}}$ &  $\bm{10^{-5}}$\\
 $g_i$ & \multicolumn{3}{|c|}{to fit the experimental distribution} \\
  $Y$ & $4 MPa$  & $4 MPa$ & $4 MPa$\\
 $\zeta_i$ & $200 Pa h$  & $200 Pa h$ & $200 Pa h$\\
 $A_i$ & $\bm{1}$  & $\bm{1}$ & $\bm{1}$\\
  $\alpha_i$ & $\bm{0.5}$  & $\bm{0.5}$   & $\bm{0.5}$\\
 $dt$ & \multicolumn{3}{|c|}{such that the movement stays small} \\
 $T_{\mbox{\scriptsize max}}$ & 280 min  & 500 min & 400 min\\
 \hline
\end{tabular}
\caption{\label{Table:1}{Parameter values taken in the absence of specification }}
\end{table}

\begin{figure}[!ht] 
   \centering
\includegraphics[scale=0.28]{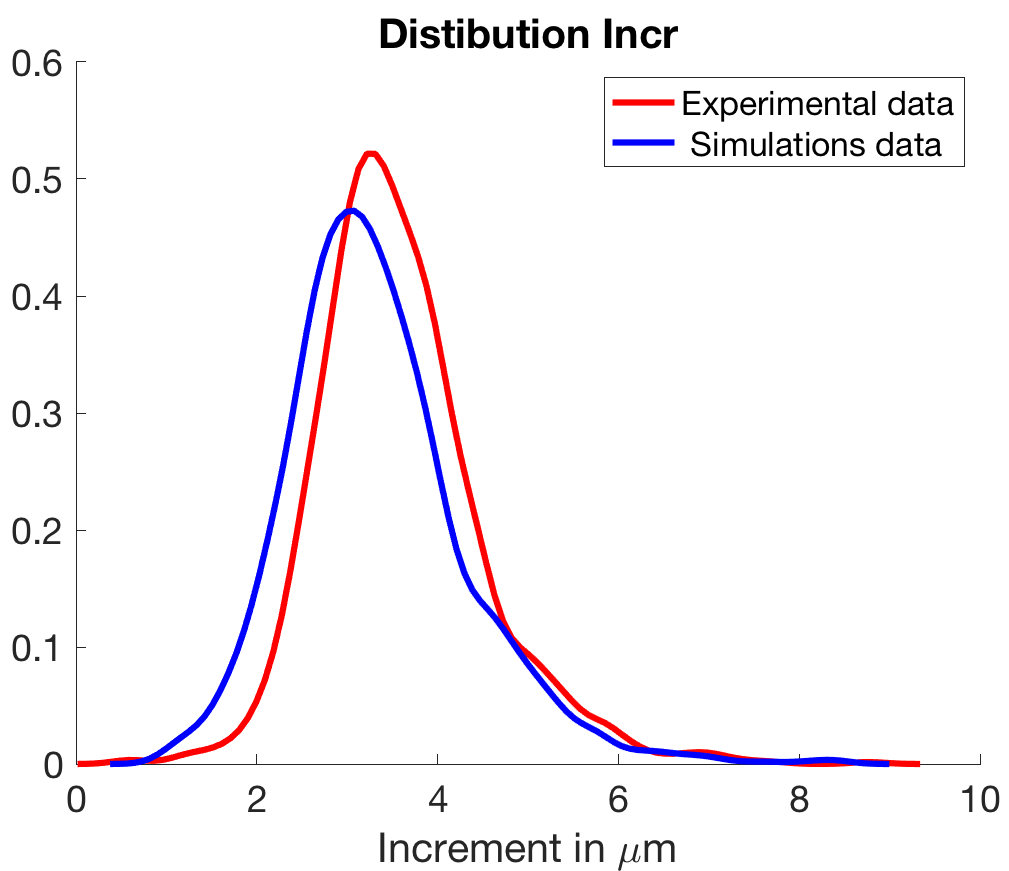}
\includegraphics[scale=0.28]{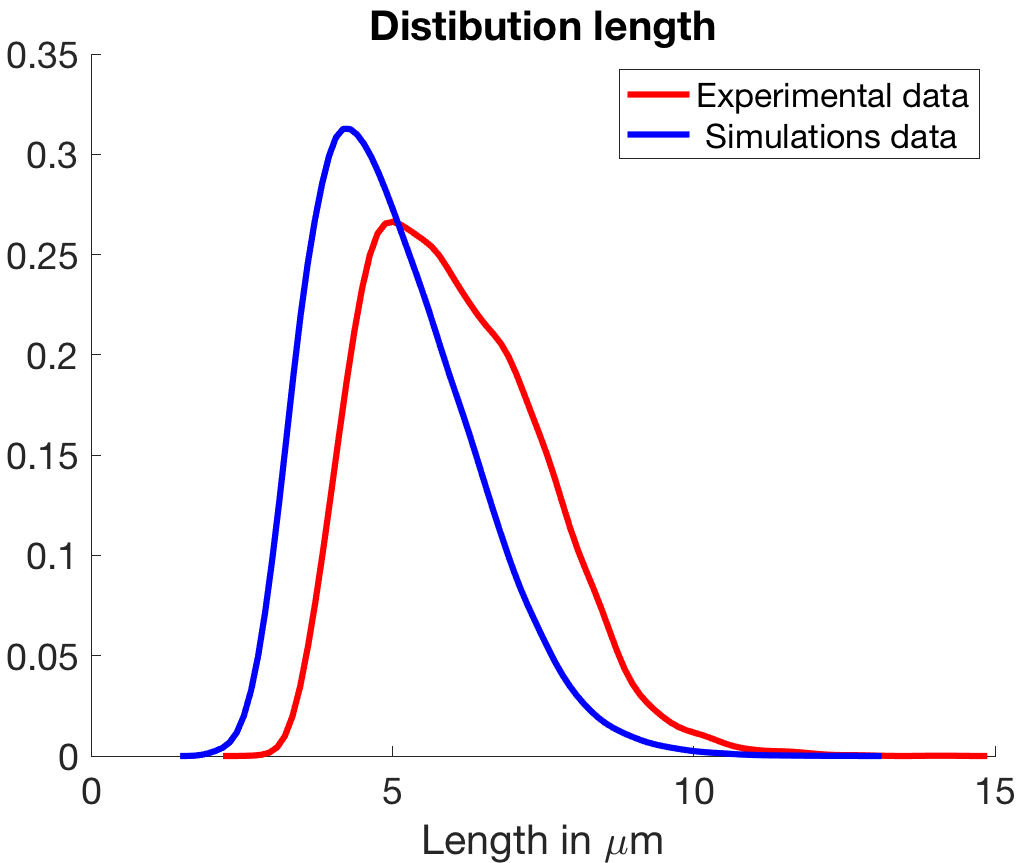}
\includegraphics[scale=0.28]{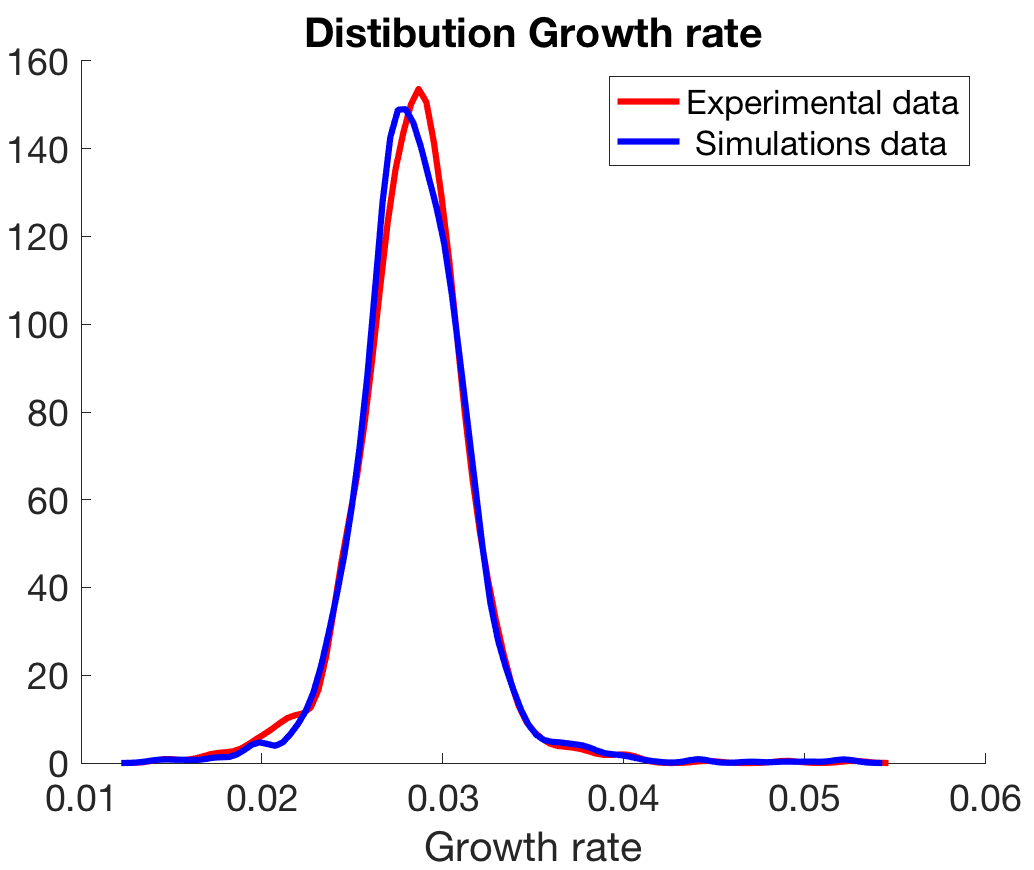}
\caption{Dataset 1: from left to right: distributions of the increment length, lenght, and growth rate for 10 initial configurations. The experimental distribution are plotted in red and the numerical simulation distributions are plotted in blue.}
\label{fig:4}
\end{figure}

\begin{figure}[!ht] 
   \centering
\includegraphics[scale=0.28]{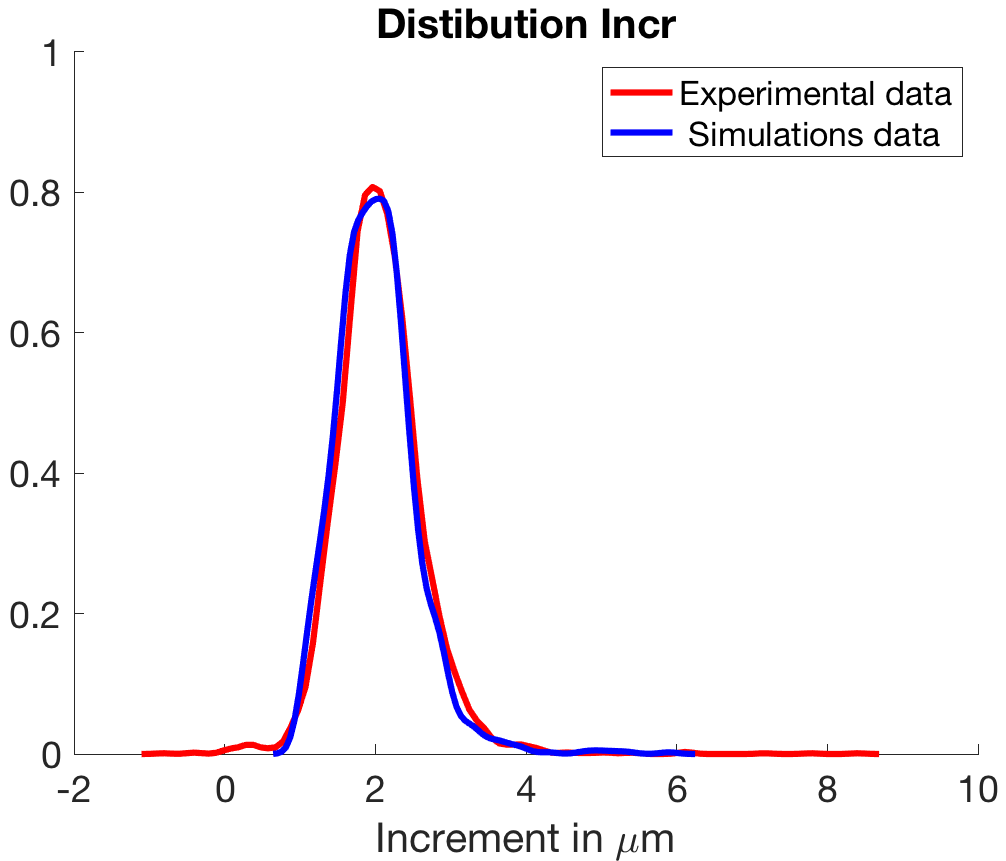}
\includegraphics[scale=0.28]{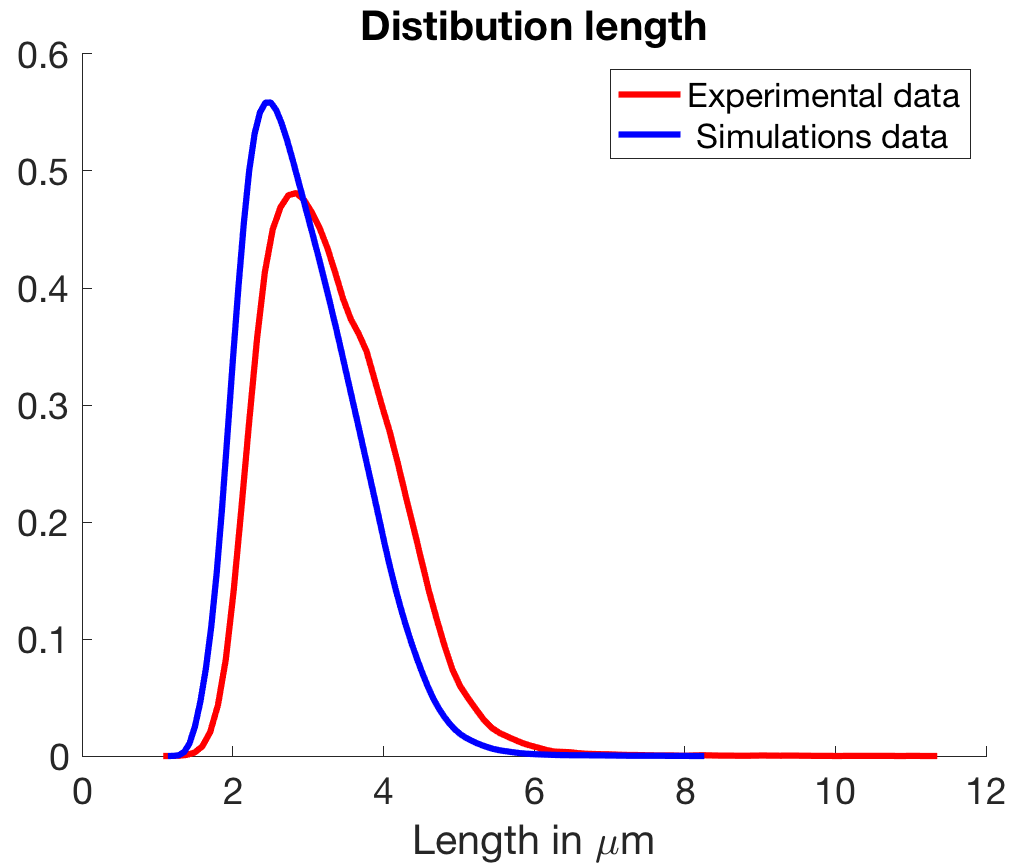}
\includegraphics[scale=0.28]{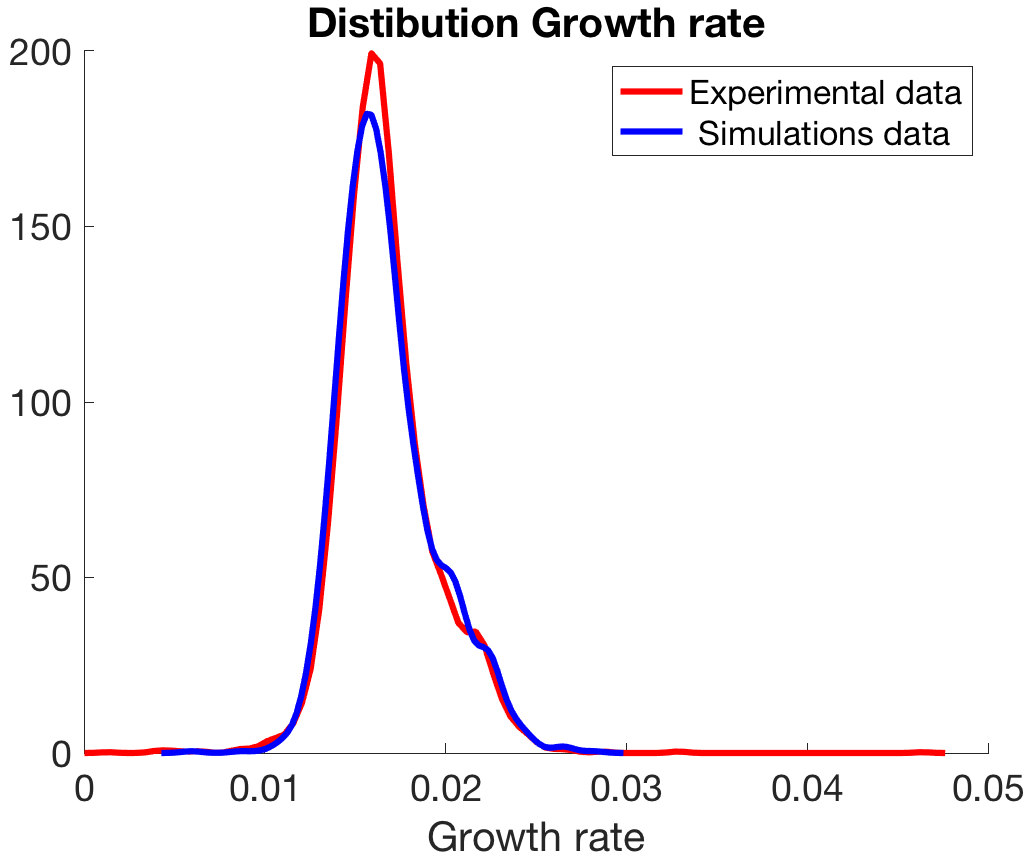}
\caption{Dataset 2: from left to right: distributions of the  increment length, lenght and growth rate for 10 initial configurations. The experimental distribution are plotted in red and the numerical simulation distributions are plotted in blue.}
\label{fig:5}
\end{figure}

\begin{figure}[!ht] 
   \centering
\includegraphics[scale=0.28]{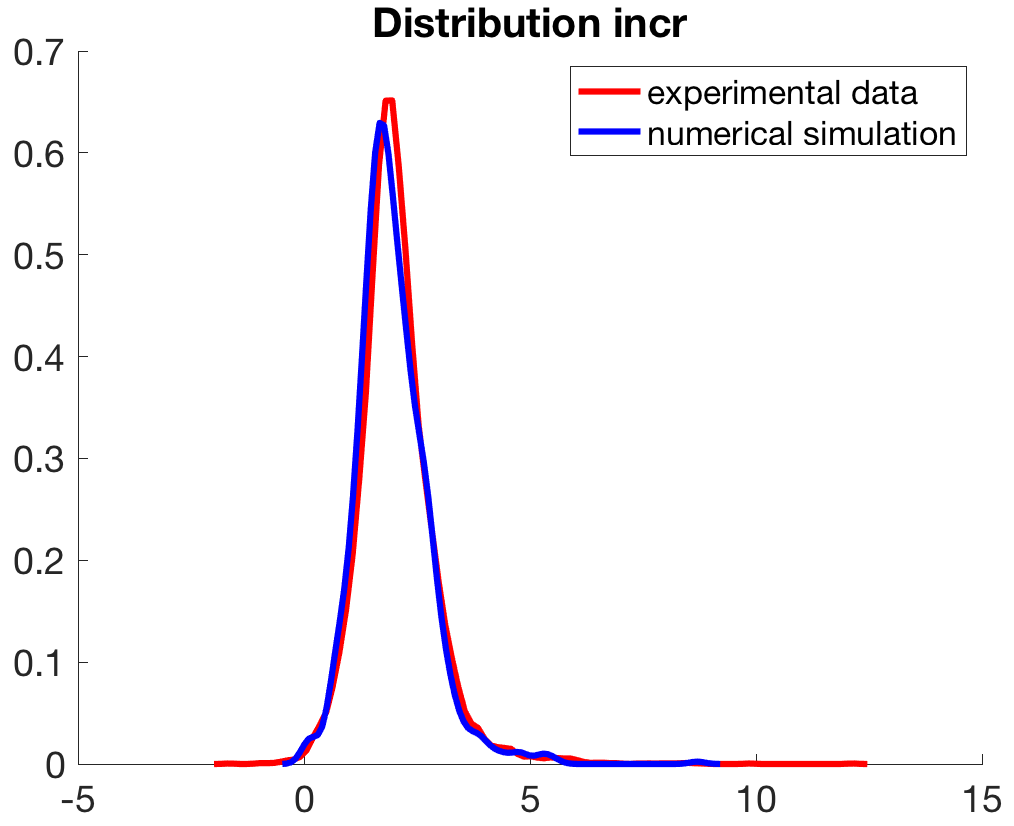}
\includegraphics[scale=0.28]{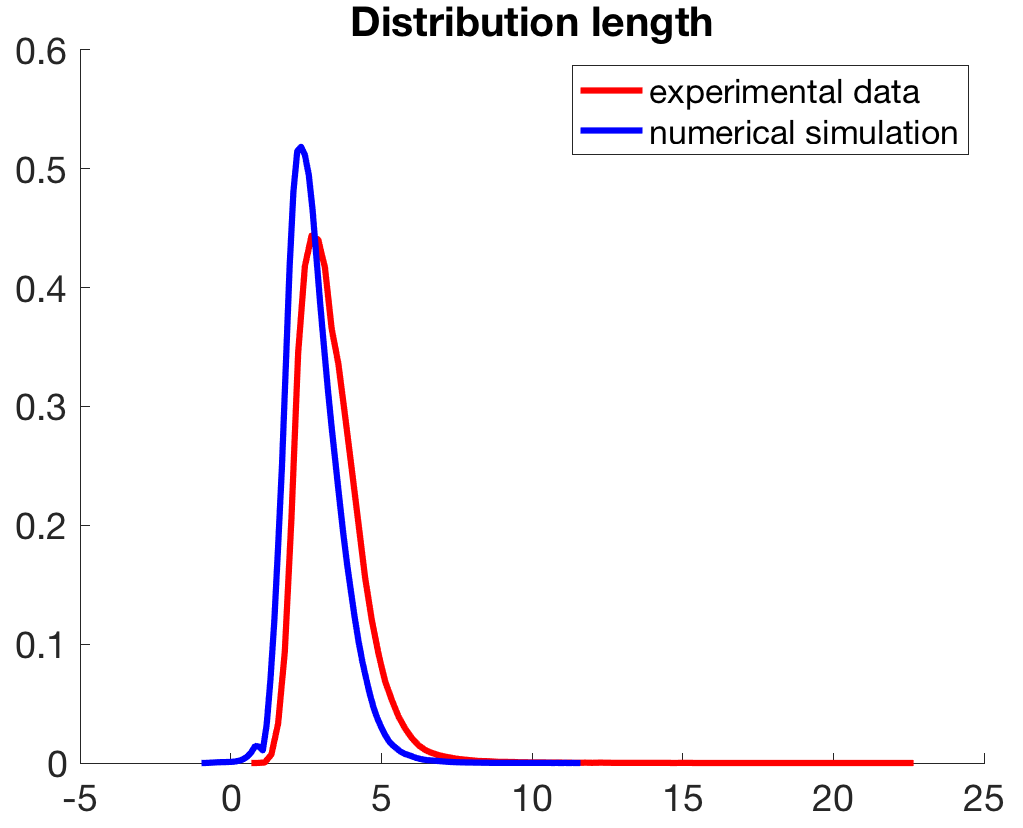}
\includegraphics[scale=0.28]{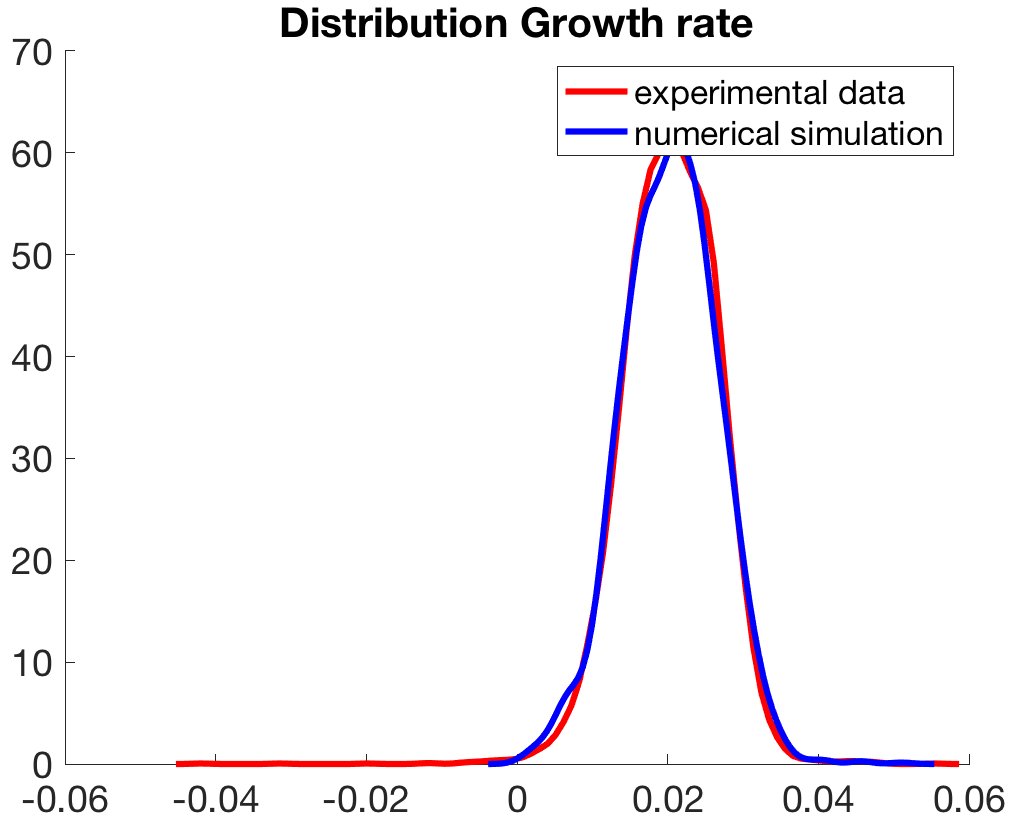}
\caption{Dataset 3: from left to right: distributions of the  increment length, length and growth rate for 10 initial configurations. The experimental distributions are plotted in red and the numerical simulation distributions are plotted in blue.}
\label{fig:6}
\end{figure}

In Figs.~\ref{fig:4}, \ref{fig:5} and \ref{fig:6}, we show  the distributions of the increment length (left figures), length of the bacteria (middle figures),  and growth rate (right figures), computed from the simulations (blue curves) and from the experimental data of Experiments 1, 2 and 3 respectively  (red curves).  

On Figs.~\ref{fig:4}, Figs.~\ref{fig:5} and \ref{fig:6}  we observe a very good agreement between the simulation values and the experimental data for the growth rate. Indeed, the red distribution (experimental data) and the blue distribution (numerical simulations) are almost similar. This is expected as the experimental distribution is given as an input of the code. However, it is noteworthy that we observe a small shift between the two distributions for the increment at division. This is due to the sampling noise:  the more events are sampled, the closest the numerical distribution is going to be to the experimental one. Therefore the difference is due to the fact that the number of event sample is not high enough for a perfect fit. 

Considering the distribution of the lengths for both experiments we observe a shift between the experimental and the numerical distribution. This shift is also observed when looking at the experimental and simulated distributions of the lengths at birth and of the lengths at division (data not shown). These differences are due to the incremental model we use to model the growth and the division of the bacteria. While the literature indicates that the incremental (also called {\it adder})  model \cite{Taheri-Araghi:2015aa,Sauls:2016aa,Amir:2014aa} gives better results to predict the cell distributions than the models based solely on the length \cite{Metz:1986} or on the  age, the incremental model remains based on a simplifying assumption, and we believe that it could still be improved, as shown by our not-so-perfect fits.  However, because solving this issue is not the main objective of this paper, we consider the incremental model to be good enough for our purpose. Nevertheless, we need to be careful about how this affects the results of our numerical simulations. In particular, we noticed that the evolution of the number of bacteria in time in the colony was slower than the ones of the experimental data. This results in smaller colonies. Therefore in the next sections, instead of comparing the experimental and simulated colonies as functions of time, it will be as functions of the colonies area. Because the area of a colony might change from a simulation to another the area are averaged on intervals of size $100 \mu m^2$ for Dataset 1 and $50 \mu m^2$ for Dataset 2 and 3.

\subsection{The quantifiers} \label{Section3.2}

\begin{figure}
     \centering
     \subfloat[Four cell array organisation\label{fig:7:4cell}]{
        \includegraphics[scale=0.22]{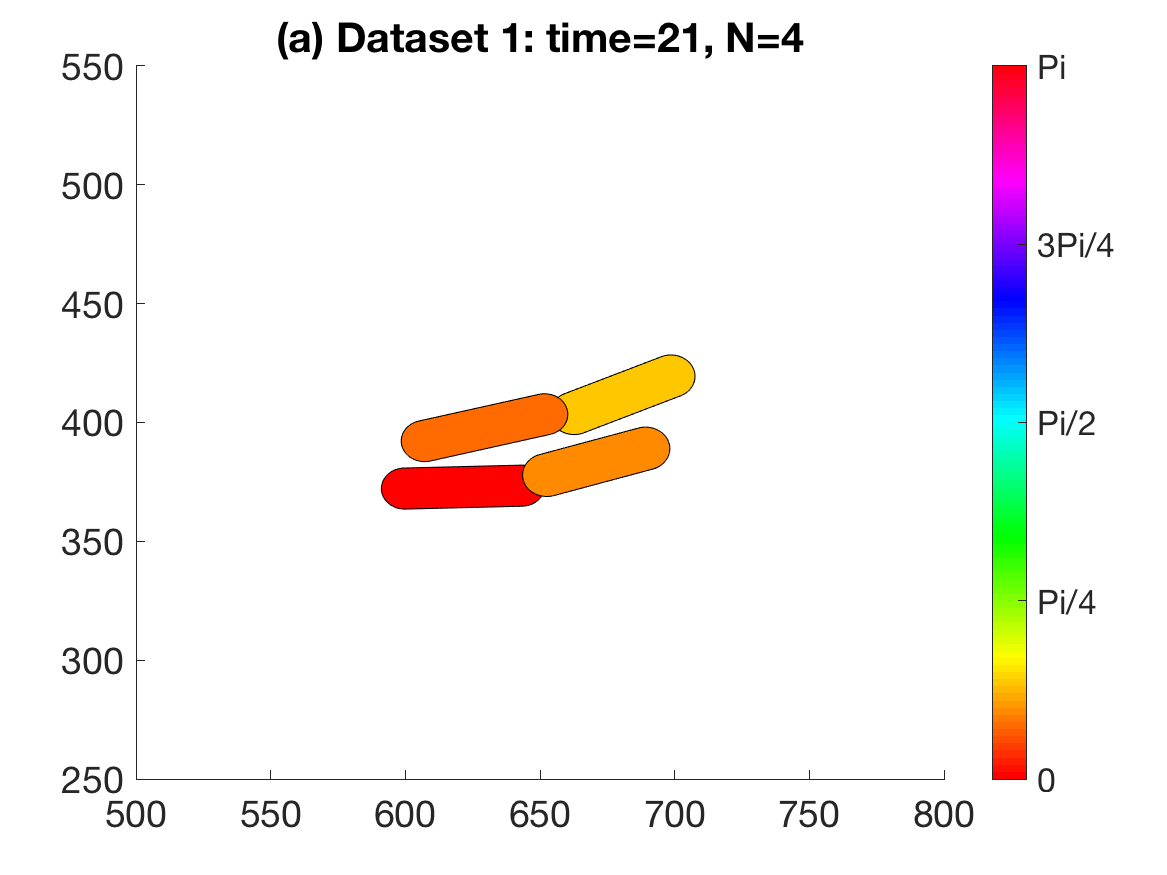}
        \includegraphics[scale=0.22]{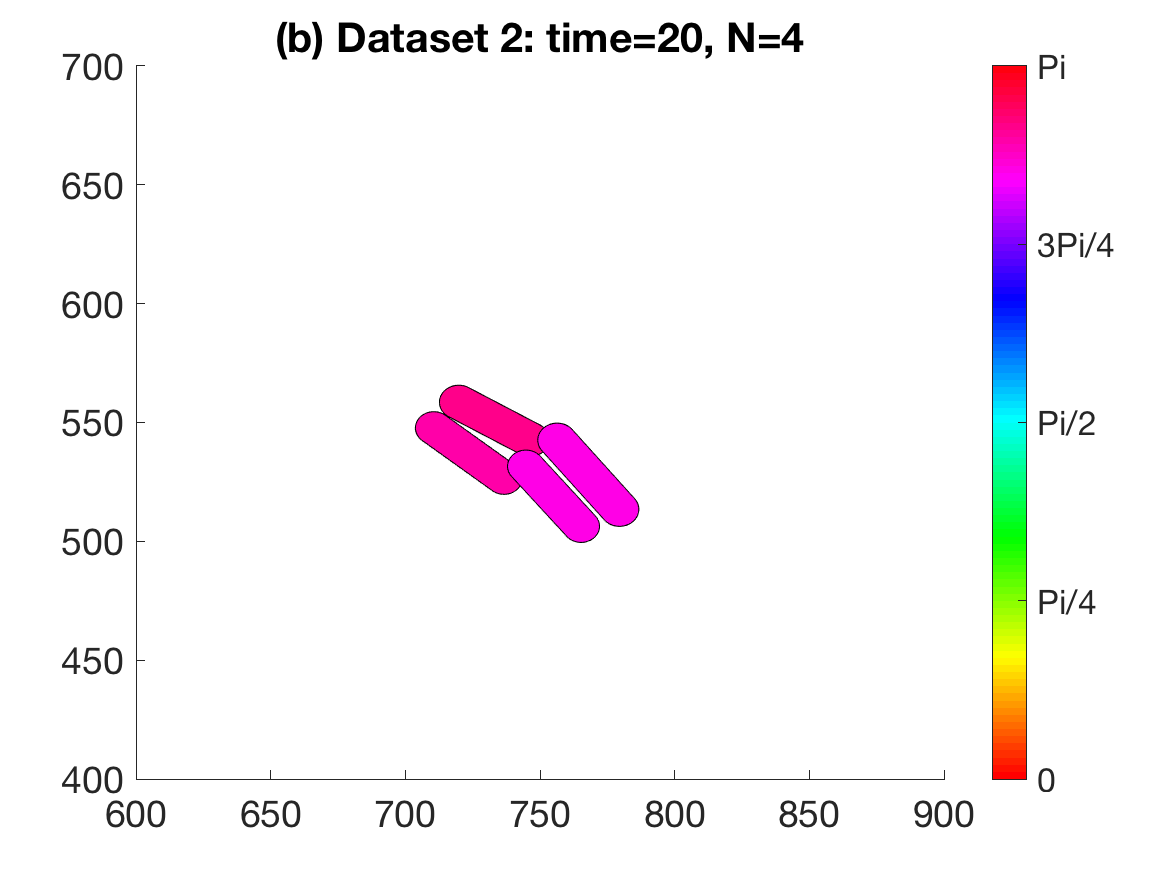}
        \includegraphics[scale=0.22]{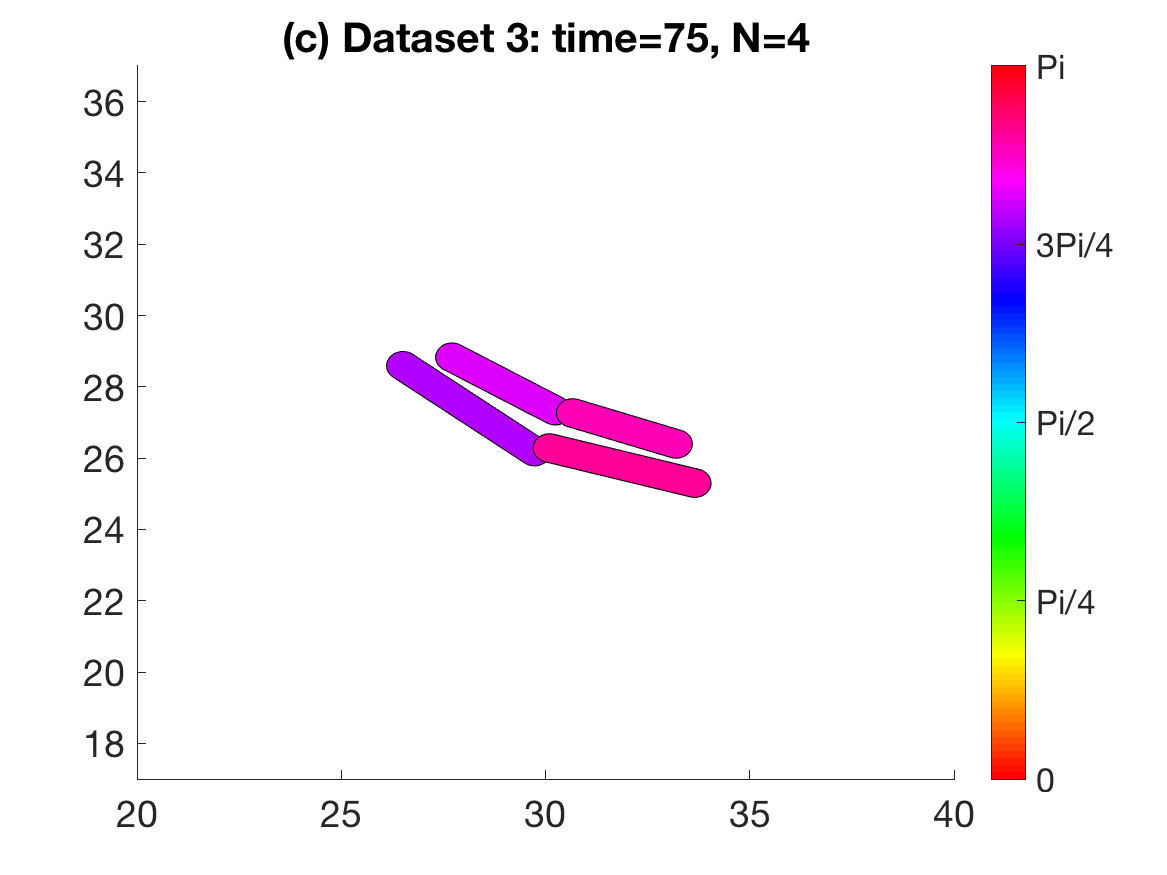}}
     \hfill
     \subfloat[Early stages $N \approx 40$]{
        \includegraphics[scale=0.22]{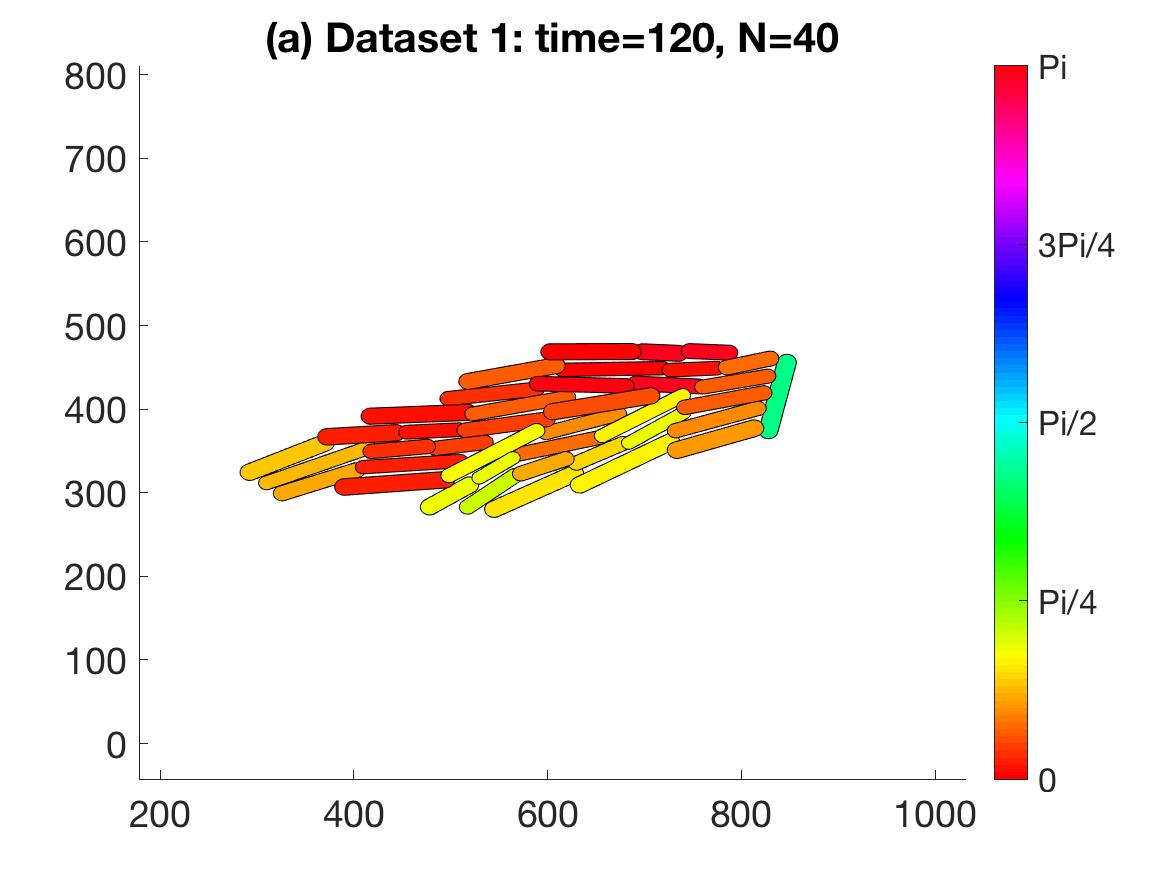}
        \includegraphics[scale=0.22]{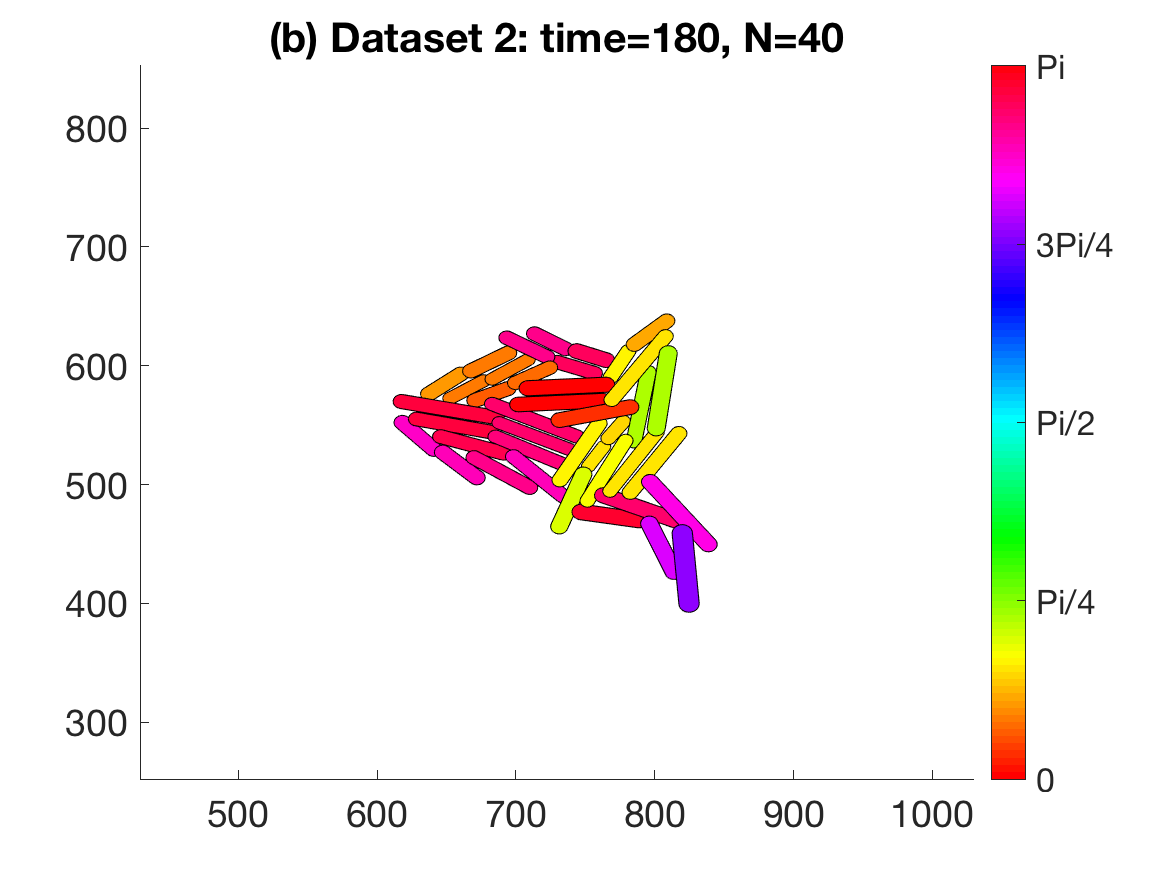}
        \includegraphics[scale=0.22]{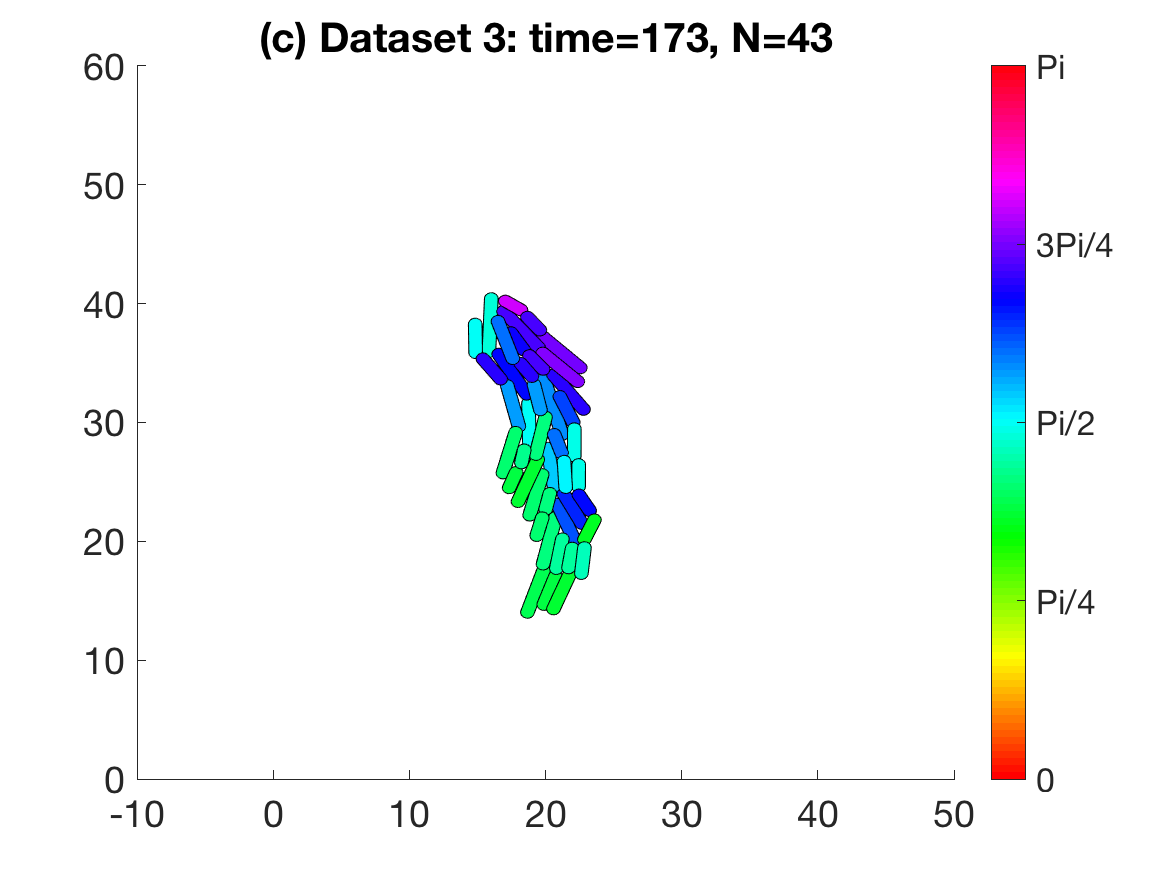}}
     \hfill
     \subfloat[Late stage]{
        \includegraphics[scale=0.22]{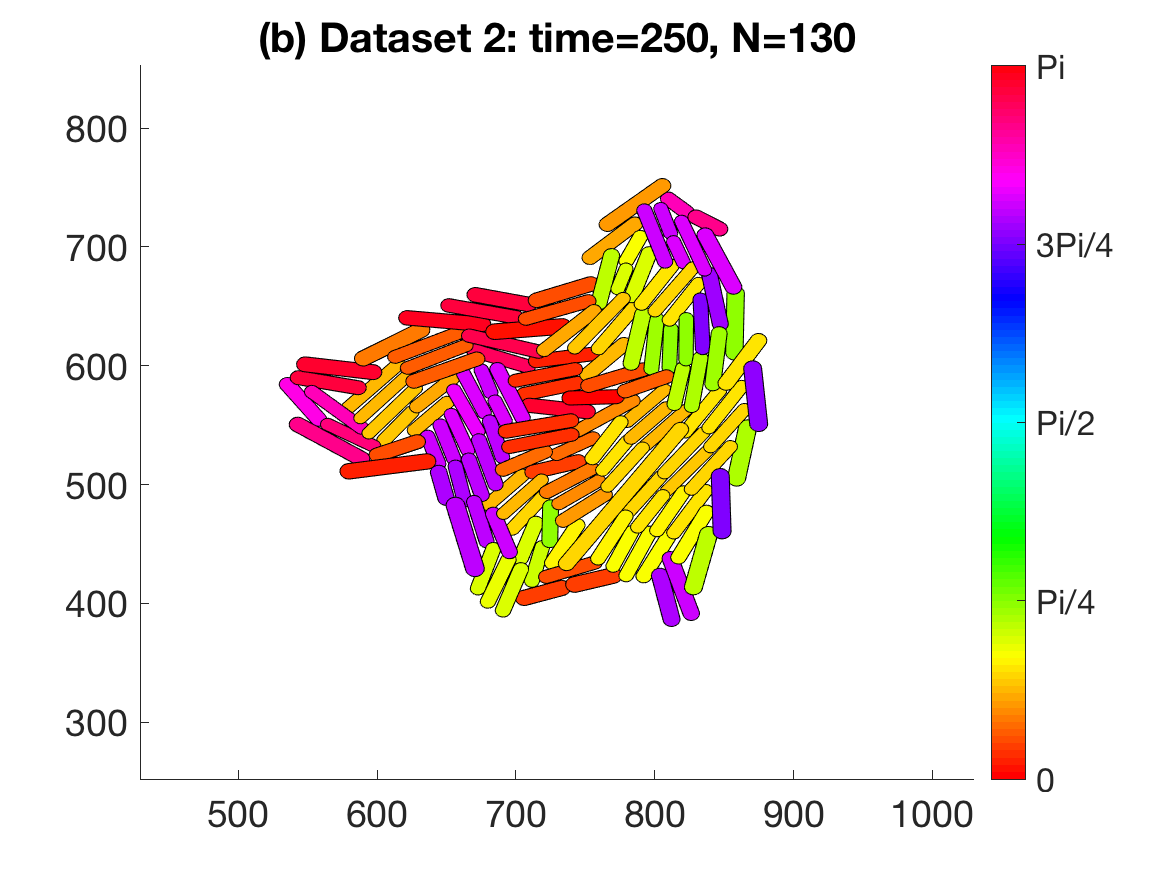}
        \includegraphics[scale=0.22]{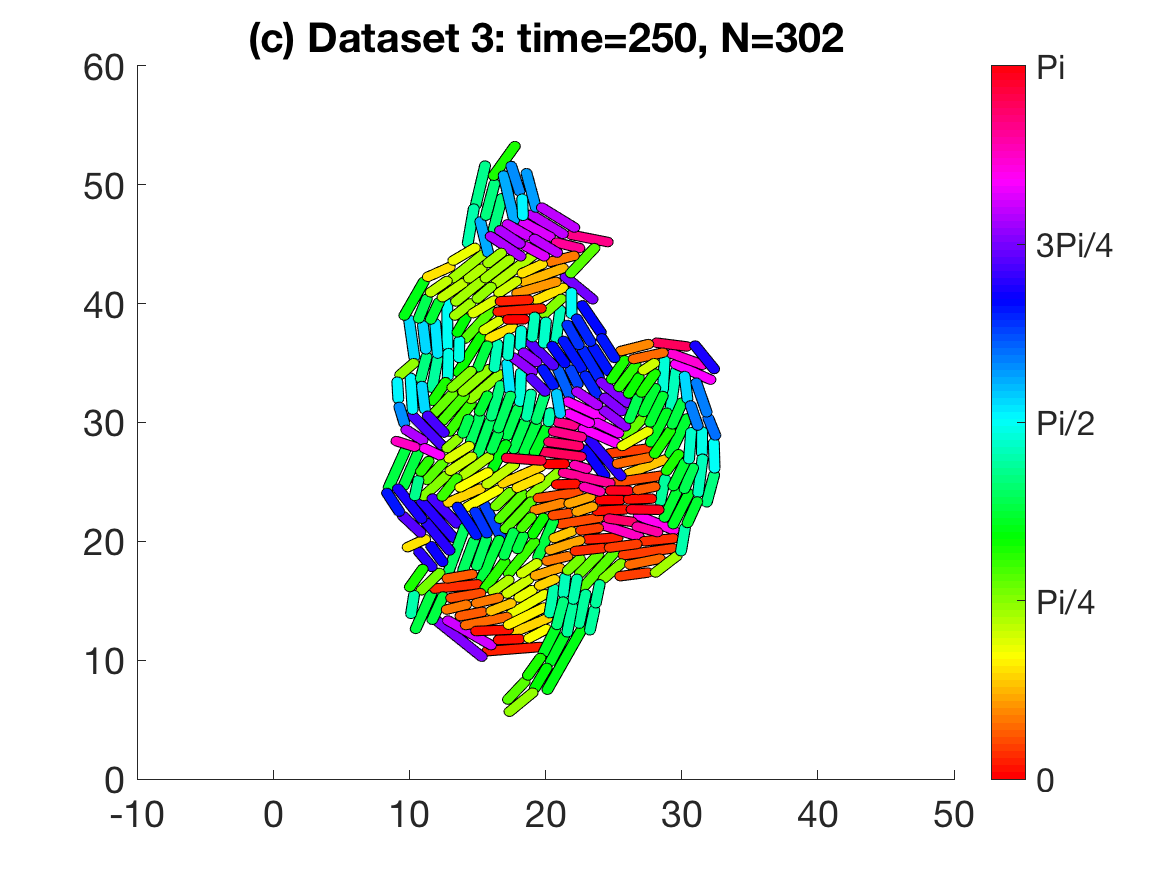}}
        \caption{Plot of a colony from Dataset 1 (a), Dataset 2 (b) and Dataset 3 (c) at time corresponding to: four cell colonies (1), colony composed of $N=40$ cells (2), colony at time $t=250$ min. All the colonies are in the four-cell array organisation. The colours of the bacteria are determined by their orientations. }
        \label{fig:7}
\end{figure}

In this section, we define the quantifiers used to perform the comparison  between the numerical and experimental data and discuss their meaning. Some quantifiers refer to the characteristics (C1) and (C2) of the colonies presented in the introduction; we recall that (C1) is the four-cell array observed in early stages of development, and (C2) is the elongated shape of the colony. In Fig.~\ref{fig:7} we plot the colonies resulting from the segmentation of Datasets~1, 2 and 3. Each segmented bacterium is represented by a  spherocylinder, its colour indicating its directional angle from 0 (red) to $\pi$. Figs.~\ref{fig:7}~(1) present the colonies in the four-cell array organisation corresponding to the characteristic (C1) while Figs.~\ref{fig:7}~(3) show the colonies at a later stage of development. Dataset~1 gives access to data for a smaller duration and therefore is plotted at an earlier time. 

 {\bf \it Remark.} We noticed that using the values of the bacteria diameters extracted from the segmented data of \cite{Duvernoy2018} led to very dense colonies composed of overlapping bacteria. This phenomenon is however not observed in the microscopy images of the colonies before segmentation - we refer, {\it e.g.}, to the figure~1 of~\cite{DellArciprete2018}, to the figure~1, (b) and (c) of~\cite{You2018}, or yet to the supplementary movies~6 and 7 of~\cite{Duvernoy2018}. It can be due  (i) to the flexibility of the real bacteria that is not taken into account in a spherocylinder representation and (ii) to the choice of the parameters for representing the segmented bacteria by spherocylinders (namely their width, length). From real images, we estimated that the actual overlapping amount in the cell colonies from the microscopy images was better fitted by reducing the width of the bacteria of $20\%$ compared to the value provided in the referenced papers. In the remaining of this paper, we therefore use this reduced value for generating the images (for instance Fig. \ref{fig:7}) as well as computing the statistical quantifiers.

 While Panels~(a) and (c) clearly show that \textit{E. coli} colonies tend to organise into elongated structures (characteristic (C2)), this observation is not so clear for the pseudonomas colonies (Panels~(b)). Moreover, we observe the emergence of locally aligned clusters inside the colonies, with high anisotropy in orientation for Dataset~1 (Panels~(a) of Fig.~\ref{fig:7}). In this case, we observe a correlation between the orientation of the bacteria and that of the whole colony, while this correlation becomes less clear for larger colonies (Fig.~\ref{fig:7}~(3), Panels~(b) and (c)).  Therefore, in the following, we will quantify the shape of the overall colony as well as its local anisotropy. Finally, we observe that bacteria seem to be tightly packed. We will therefore take an interest in the density of the micro-colonies.

\ 

Among the quantities we consider we find:
\begin{itemize}
\item the aspect ratio $\alpha_R$ which quantifies the shape of a micro-colony. This measure requires to determine two characteristic lengths for the shape, that we denote $l$ and $L$. There exist various ways to determine two characteristic lengths of a colony. In our study, following the observation made in \cite{DuvernoyPhd}, we define the lengths $l$ and $L$ as the semi-minor and semi-major axis of the ellipse fitted with the same normalised second central moments as the convex envelop of the colony. For a domain $A$ of centroid $(\bar{x},\bar{y})$, the normalised second central moments is defined by the covariance matrix $ \frac{1}{\mu_{00}} \begin{pmatrix}
\mu_{20} & \mu_{11} \\
\mu_{11} & \mu_{02} 
\end{pmatrix}$ with $\mu_{pq} = \int_A (x-\bar{x})^p (y-\bar{y})^q dxdy $. Therefore $\alpha_R=\frac{l}{L}$ with $l$ and $L$ the respective minor an major axis of the ellipse such that the previous covariance matrix is equal for the ellipse and the convex envelop of the colony.
\item the local order parameter $\lambda$ which quantifies the local anisotropy of the bacteria orientations. For each bacterium $i$, we compute the mean of the projection matrices on the orientation vectors of the neighbouring bacteria:
$$O_i = \frac{1}{Card \{j\in[1,N], |X_i-X_j|\leq 3/2R_i\}} \sum_{\substack{ j\in[1,N] \\ |X_i-X_j|\leq 3/2R_i}} \begin{pmatrix}
{\cos \theta_j}^2 & \cos \theta_j \sin \theta_j \\
\cos \theta_j \sin \theta_j & {\sin \theta_j}^2
\end{pmatrix}.$$
 Note that the local anisotropy is computed for bacteria whose centres are located in a ball centred at the bacterium $i$ position and of radius $\frac{3 R_i}{2}$. Then, we define $\lambda_i$ as the largest eigenvalue of the matrix $O_i$, which gives a measure of the local anisotropy in orientations around the bacterium $i$. Notice that when all the bacteria are locally aligned around bacterium $i$, $\lambda_i=1$, while $\lambda_i$ tends to  $\frac{1}{2}$ when the neighbours of the bacterium $i$ are randomly oriented. The local order parameter is defined as the average of all these eigenvalues:
$$\lambda = \frac{1}{N} \sum_{i=1}^N \lambda_i.$$
\item the density parameter $\delta$: this parameter is computed via image analysis tools, by computing the surface area of the envelope of the colony and dividing it by the surface area of the filled colony. 
\item the distance $d_2$ which characterizes the structure of the two-cell colony. This quantifier aims to characterize the four-cell array organisation of a colony (C2). However because it is not trivial to quantify the arrangement of four bacteria, we focus of the   structure of two-cell colonies, right before the division. Indeed, to be into a four-cell structure indicates that, before division, the two bacteria of the colony were side by side longitudinally. Then, in the case of a colony composed of two bacteria, $d_2$ is defined by
$$ d_2 = |\frac{(X_2-X_1)\cdot(X_2-X^{p_o}_2)}{|X_2-X_1||X_2-X^{p_o}_2|}| .$$
The value of $d_2$ is between $0$ and $1$, where $d_2=0$ when the four cells are side-by-side in a four-cell array configuration and $d_2 = 1$ when the two bacteria are aligned.
\item the distribution of the  angle $\Theta$ between the two daughter cells at division. This parameter is observed using the same time intervals as in the experiments (further explained in Section~\ref{Section3.3.3}). 
\end{itemize}

\subsection{Influence of some of the model parameters} \label{Section3.3}

In this section, we discuss the influence of the asymmetric friction and of the distribution of mass which are the two new key components of our model. We also study the influence of the noise of the angle at division $\Theta$. In this section, the parameters used are the ones corresponding to Dataset~1 in Table~\ref{Table:1}. The results are averaged over 10 simulations.

\subsubsection{The asymmetric friction} \label{Section3.3.1}

We first discuss the influence of the asymmetric friction on the colony growth by varying the value of the friction anisotropy $A_i$. For the sake of simplicity, we will consider that this ratio is the same for all bacteria and we denote it by $A$. Therefore, the parallel friction $\zeta^{||}$ and the perpendicular friction $\zeta^{\perp}$  are defined by
$$ \zeta^{||} = A \zeta \quad  \mbox{and} \quad  \zeta^{\perp} = \frac{\zeta}{A}. $$
Note that $A=1$ corresponds to an isotropic friction while $A\neq 1$ supposes a directional dependence of friction. In this paper, we will focus on the case $A \leq 1$ which expresses the fact that it is more difficult for a bacterium to slide in its perpendicular direction  than in its direction.

\begin{figure}[!ht] 
   \centering
\includegraphics[scale=0.35]{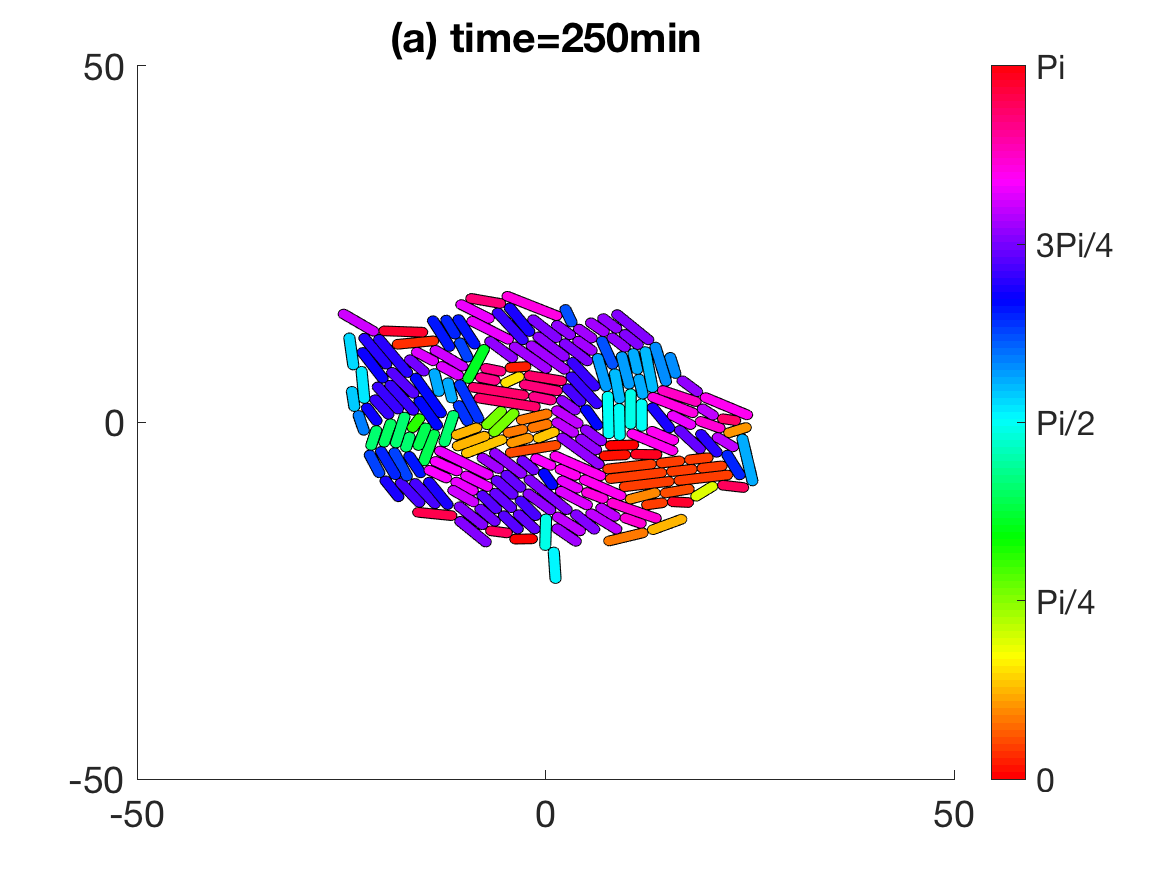}
\includegraphics[scale=0.35]{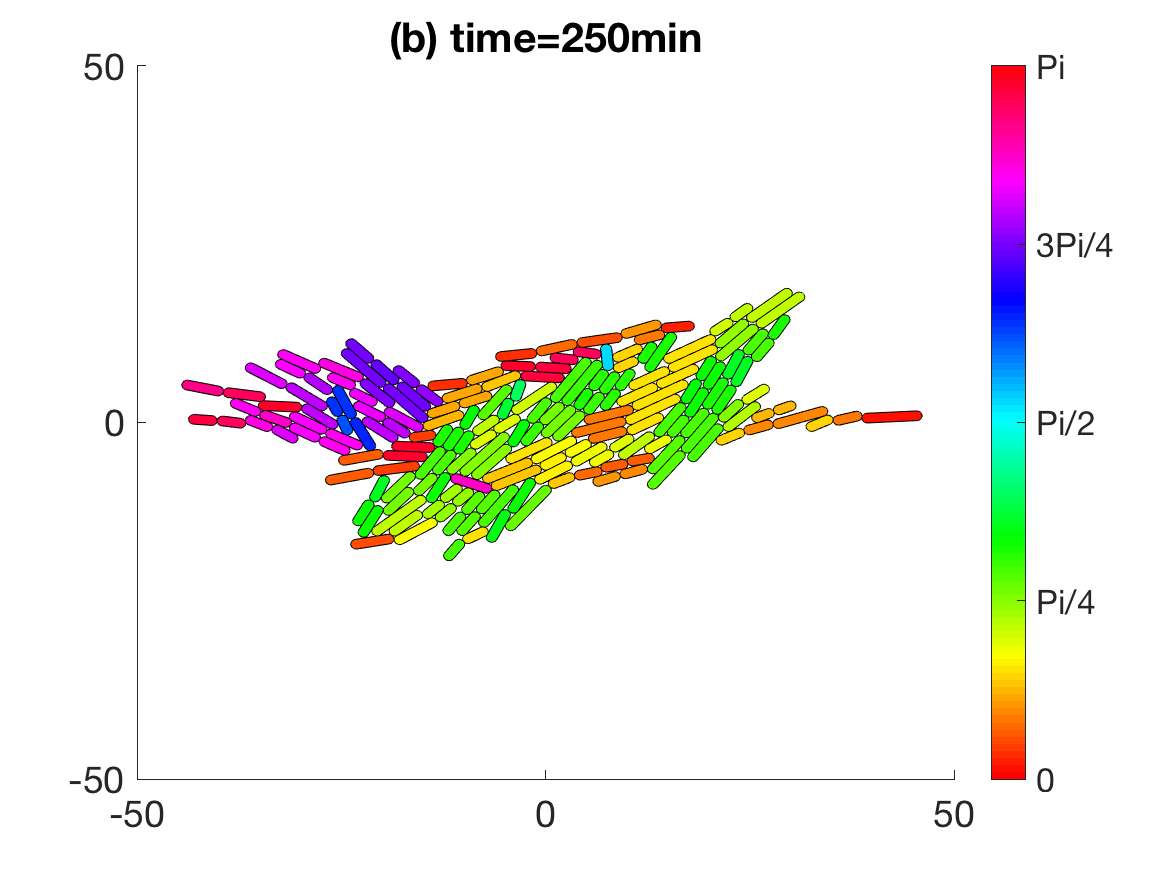}
\caption{Plot of the colony for $A=1$ (a) and $A=0.4$ (b) at $t= 250 \mbox{ min}$. The color of the bacteria are given by their angle from the horizontal axis.}
\label{fig:8}
\end{figure}

In Fig.~\ref{fig:8}  we show the two colonies at time $t= 250 \mbox{ min}$ for $A=1$  (Panel~(a)) and $A=0.4$ (Panel~(b)). From Fig.~\ref{fig:8}, we observe that the friction anisotropy parameter $A$ has a strong influence on the shape of the colony: Anisotropic friction ($A<1$, Panel~(b)) leads to the emergence of elongated bacterial structures coupled with a large anisotropic orientation of the bacteria, while isotropic friction (case $A=1$, Panel~(a)) promotes the formation of round colonies with more variability in the bacteria orientations. 

\begin{figure}[!ht] 
   \centering
\includegraphics[scale=0.35]{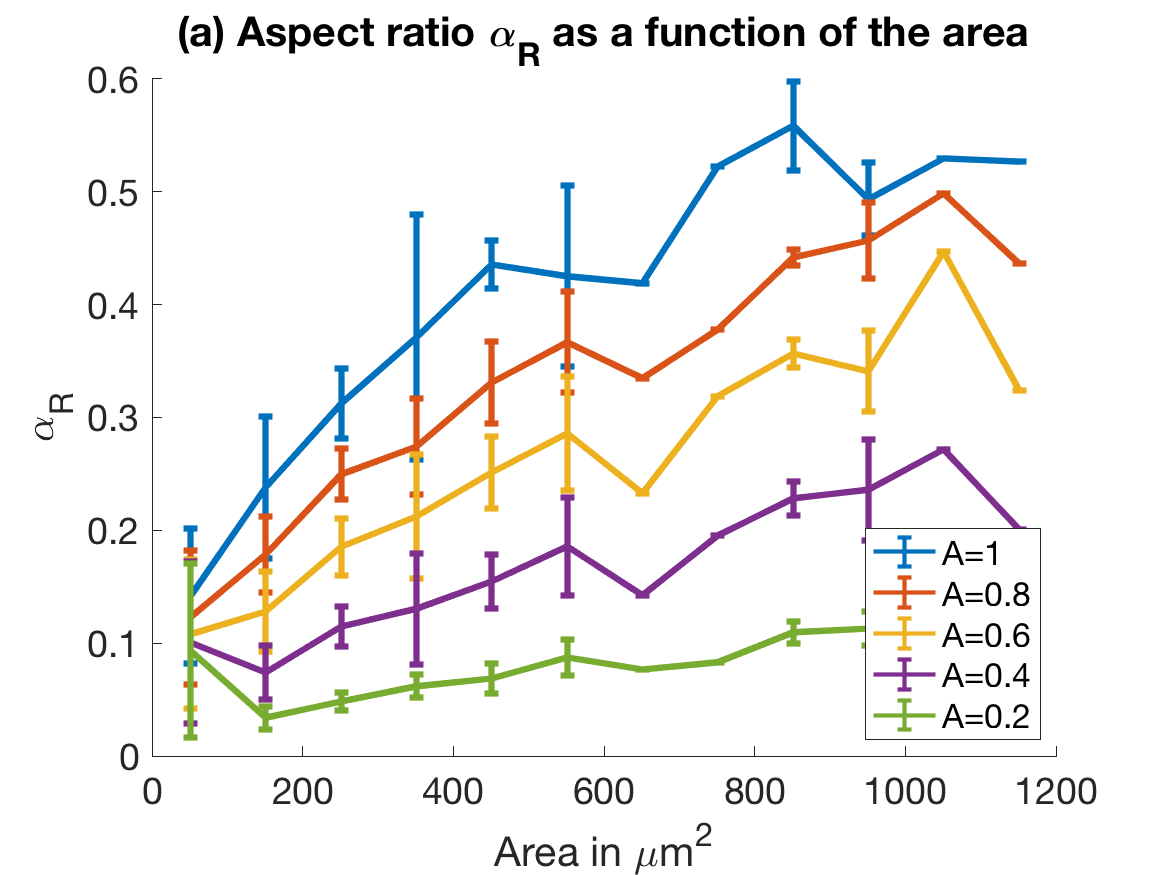}
\includegraphics[scale=0.35]{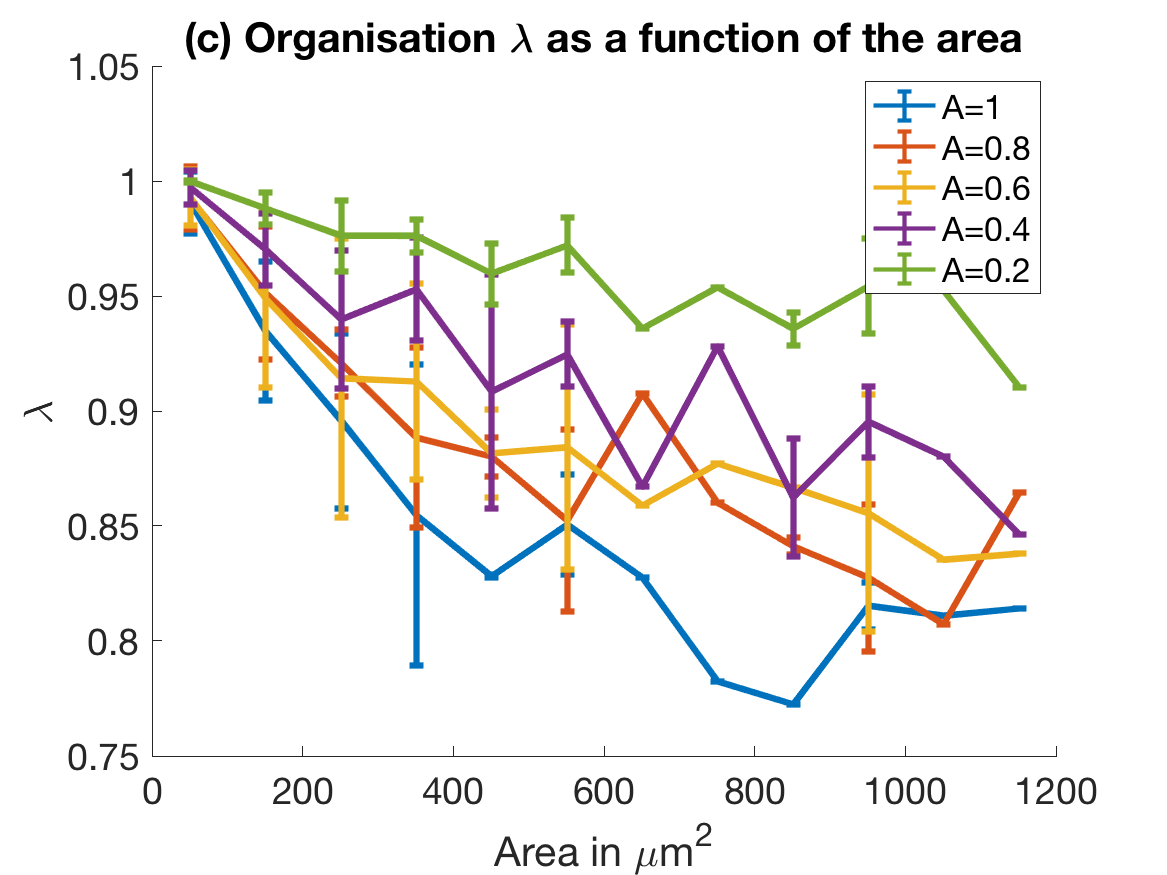}
\includegraphics[scale=0.35]{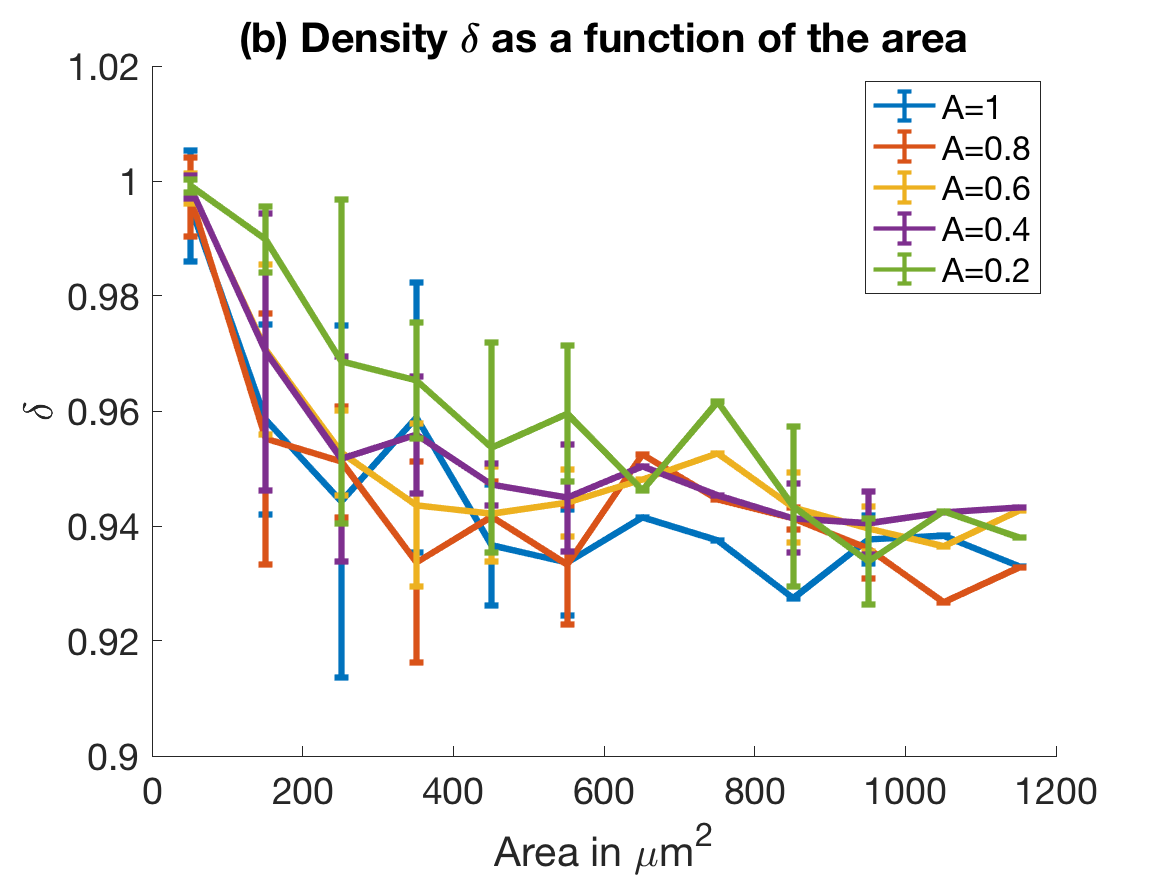}
\includegraphics[scale=0.35]{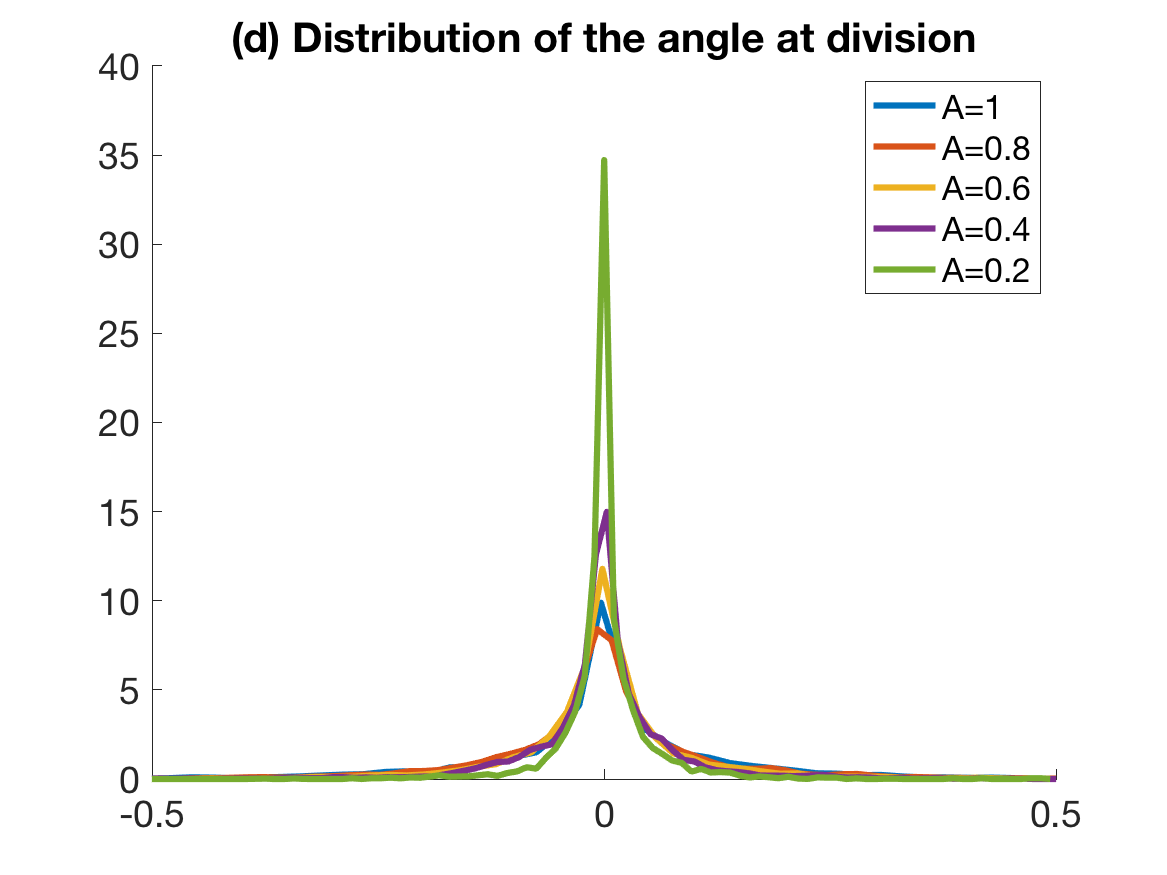}
\caption{ Evolution of the aspect ratio $\alpha_R$ (a), the density (b) and the local order quantifier $\lambda$ (c)  as functions of the area of the colony, and of the distribution of the angle at division (d), for different values of $A$: $A=1$ , $A=0.8$, $A=0.6$, $A=0.4$,  $A=0.2$.}
\label{fig:9}
\end{figure}

To quantify  these observations, we show in Fig.~\ref{fig:9} the evolution of the aspect ratio $\alpha_R$ (Panel (a)), the local order quantifier $\lambda$  (Panel (b)) and the colony density (Panel~(c)), as functions of the area of the colony. Panel~(d) shows the distribution of the angles at division all along the simulation and for all bacteria.  For each figure, we used different values of $A$: $A=1$ (blue curves), $A=0.8$ (red curves), $A=0.6$ (yellow curves), $A=0.4$ (purple curves) and $A=0.2$ (green curves). 

Fig. \ref{fig:9} (a) shows that the aspect ratio $\alpha_R$ of the colony increases as the colony  grows, with rates depending on the anisotropic friction $A$: we observe a fast convergence towards a spherical shape for $A=1$ (isotropic friction, blue curve), while for smaller values of $A$ the colony remains elongated and converges slower towards a spherical shape. For the extreme case $A=0.2$, the colony remains elongated all along the simulation (green curve). Together with these observations, Panel~(c) shows that as the anisotropic friction $A$ decreases, the local organisation of the bacteria $\lambda$ increases (compare blue and green curves of Panel~(c)). We also note however that in all cases the local alignment of the bacteria decreases as the colony grows. These results show that by making it less costly for a bacteria to slide in its longitudinal direction than perpendicular to it, an anisotropic friction favours the alignment of the bacteria and consequently creates anisotropy in their orientation. This results in more elongated overall structures. From Panel~(b), we note that anisotropic friction seems to have little influence on the overall density of the colony, although a slight increase of the density when $A$ decreases is observed at early times of the colony, showing once again that anisotropic friction favours the emergence of more organized and therefore slightly denser colonies. Finally, Panel~(d) shows that anisotropic friction favours slightly more concentrated distributions of angles at division, which shows that the bacterial orientation remains closer to the orientation given at division (since in these simulations we have a very small $\Theta=10^{-5}$): they tend to have more difficulty for rotational movement. In Table~\ref{Table:2}, we show the values of the quantifier $d_2$ which characterizes the type of structure obtained when the system is composed of 4 cells (recall that $d_2 = 1$ when the cells are aligned, $d_2 = 0$ when they are organized in a 4-cell array structure). As one can observe in Table~\ref{Table:2}, anisotropic friction has no influence on the initial organisation of the micro-colony. These are expected results because at very early stage (when only two bacteria are present), there is no mechanical interest for a bacterium to turn. 

\begin{table}[h!]
\centering
\begin{tabular}{|l|l l l|} 
\hline
$A$ & average of $d_2$ & minimum of $d_2$ & maximum of $d_2$ \\ 
\hline
\hline
$A=1$  & 0.999999891665219  & 0.999999459667383 & 0.999999999972967 \\ 
\hline
$A=0.8$  & 0.999999996374091 & 0.999999987537210 & 0.999999999995483
 \\ 
\hline
$A=0.6$  & 0.999999999713114 & 0.999999998770278   & 0.999999999999658 \\ 
\hline
$A=0.4$  & 0.999999999932987 & 0.999999999655075   & 0.999999999999919\\
\hline
$A=0.2$  & 0.999999999962652 & 0.999999999823866   & 0.999999999999236 \\ 
\hline
\end{tabular}
\caption{\label{Table:2}{Influence of asymmetric friction on the four-cell array quantifier $d_2$}
}\end{table}

\subsubsection{The mass distribution} \label{Section3.3.2}

In this section, we consider the distribution of the mass along the length of the bacteria. The most classical approach is to consider that mass is distributed uniformly along the length, however in this paper we explore the possibility of an asymmetric distribution. During the division of the bacteria, the genetic material has to split into two to locate each side. Therefore it is acceptable to consider that the distribution of the weight of the bacteria is not uniform. In particular, we consider that the mass located near the old pole is more important than the one located near the new pole, i.e. $\alpha_i>=0.5$ (recall that $L\alpha_i$ is the distance between the center of mass of bacterium $i$ and its newest pole, and $\alpha_i = 0.5$ when the center of mass corresponds to the geometric center). Note that this distribution of mass could be compared to the existence of asymmetric adhesion force to the substrate which has been studied in \cite{Duvernoy2018}. However, despite the asymmetric friction and some attraction, this study seems not able to recover the four-cell array structure of bacteria micro-colony (see Supplementary Movie~9 of~\cite{Duvernoy2018}). To make our approach more realistic we consider that the parameter $\alpha_i$ may be time-dependent. We will now consider that the value of $\alpha_i$ returns linearly to $0.5$ (uniform mass distribution) in $T_{\alpha_i}$ minutes. We will consider two cases: either ${T_{\alpha_i}}=+\infty$, {\it i.e.} the mass distribution remains constant during all the lifetime of the bacterium (but is however equally shared at division), or $T_{\alpha_i}$ is equal to half the average lifespan of the bacteria of the experimental colonies. For simplicity we will denote this value $T_\alpha$ without the index $i$.

\begin{figure}[!ht] 
   \centering
\includegraphics[scale=0.35]{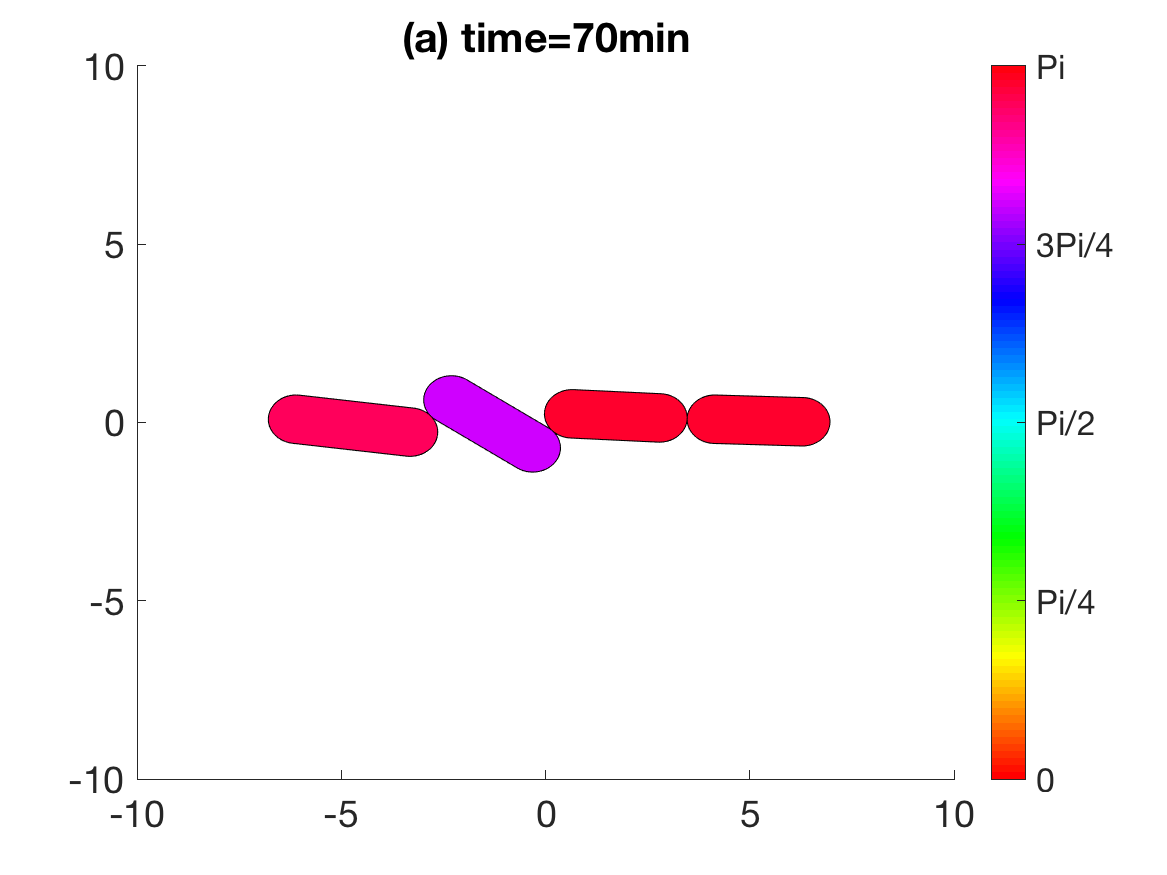}
\includegraphics[scale=0.35]{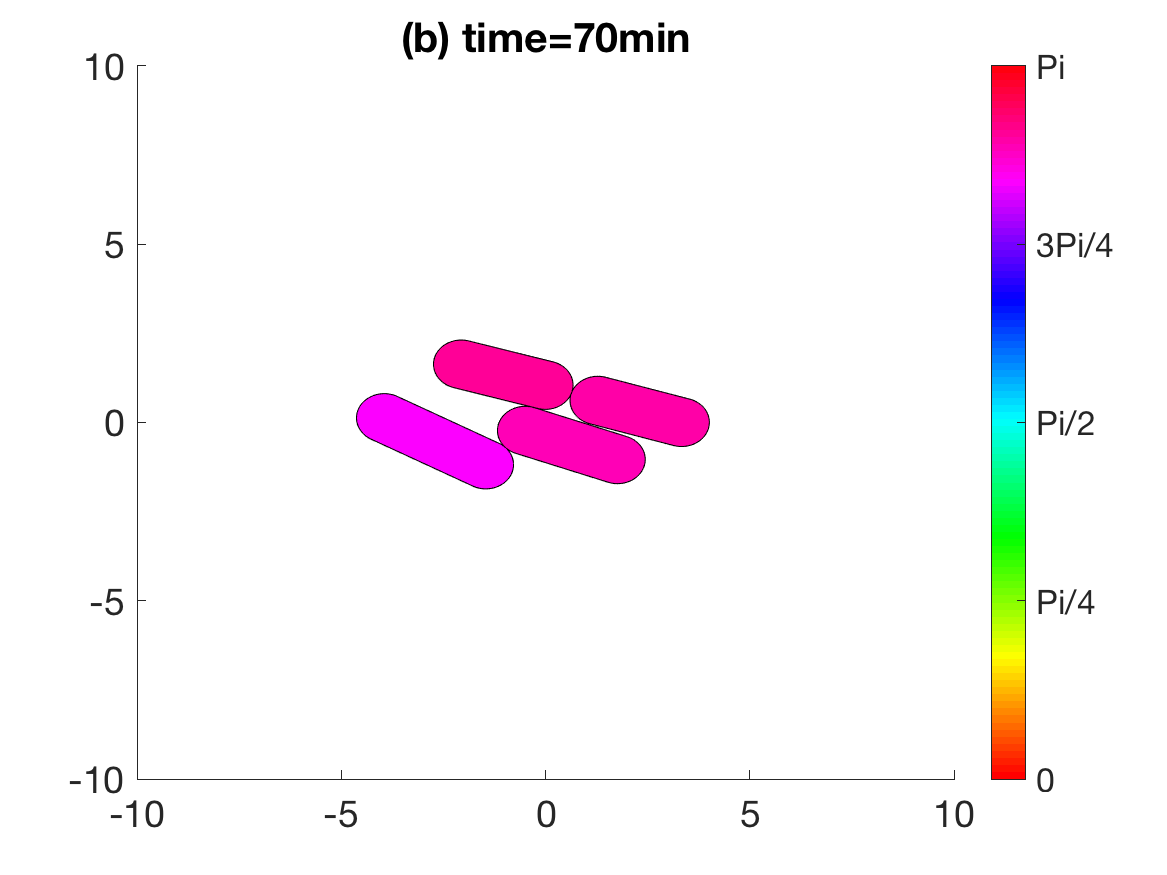}
\caption{Plot of the colony for $\alpha=0.5$ (left) and $\alpha=0.9$ (right) at $t= 70 \mbox{ min}$ (which correspond to the moment where the colony is composed of four cells). The color of the bacteria are given by their angle from the horizontal axis.}
\label{fig:10}
\end{figure}

In Fig.~\ref{fig:10}  we show two colonies right after the second division for $\alpha=0.5$ and $\alpha=0.9$ with $T_\alpha=+\infty$. The parameters used are the ones of Experiment 1 and the results of Table~\ref{Table:3} are averaged over 10 simulations. We observe that  for $\alpha=0.5$ the four bacteria are almost arranged in a line, while for $\alpha=0.9$ some of the bacteria are side by side. Although cells are still not perfectly arranged in a four-cell array structure, introducing an asymmetric mass distribution enables to get closer to the experimental results.

In Table~\ref{Table:3} we show the value of the distance $d_2$ for different values of $(\alpha,T_\alpha)$: $(\alpha,T_\alpha)=(0.5,\infty)$, $(\alpha,T_\alpha)=(0.6,\infty)$, $(\alpha,T_\alpha)=(0.75,\infty)$, $(\alpha,T_\alpha)=(0.9,\infty)$ and $(\alpha,T_\alpha)=(0.9,12)$. We recall that $d_2$ is a quantifier for the organisation of two-cell colonies. When $d_2$ is close to $0,$ the two bacteria are side by side and give rise to the four-cell array organisation. The values in Table~\ref{Table:3} show that  $d_2$ decreases as $\alpha$ increases, showing that the increase of the asymmetry of the mass allows bacteria to slide side by side. In addition, the value of $T_\alpha=T^{\mbox{\scriptsize div}}/2= \mbox{12 min}$ reduces the impact of $\alpha>1$ compared to the case $T_\alpha=+\infty$, while maintaining the sliding of the bacteria.

\begin{table}[h!]
\centering
\begin{tabular}{ | l l l l | } 
\hline
$\alpha, T_\alpha$ & average of $d_2$ & minimum of $d_2$ & maximum of $d_2$ \\ 
\hline
\hline
$\alpha=0.5$, $T_\alpha=\infty$ &   0.999999891665219 &  0.999999459667383  & 0.999999999972967 \\ 
\hline
$\alpha=0.6$, $T_\alpha=\infty$ &  0.999985160851674 &  0.999916495527016 &  0.999999975414758 \\ 
\hline
$\alpha=0.75$, $T_\alpha=\infty$  & 0.989321040266743 &  0.933306183165130 &  0.999990698078135 \\ 
\hline
$\alpha=0.9$, $T_\alpha=\infty$  & 0.957609449429128  & 0.828729443220301  & 0.991407320492294 \\ 
\hline
$\alpha=0.9$, $T_\alpha= T^{\mbox{\scriptsize div}}/2 = 12 \mbox{ min}$  & 0.986765218722925  &  0.912749581037532 &  0.999994661249026 \\ 
\hline
\end{tabular}
\caption{\label{Table:3}Combined influence of asymmetric mass $\alpha$ and relaxation time to symetric mass $T_\alpha$ on the four-cell array quantifier $d_2$}
\end{table}

\begin{figure}[!ht] 
   \centering
\includegraphics[scale=0.35]{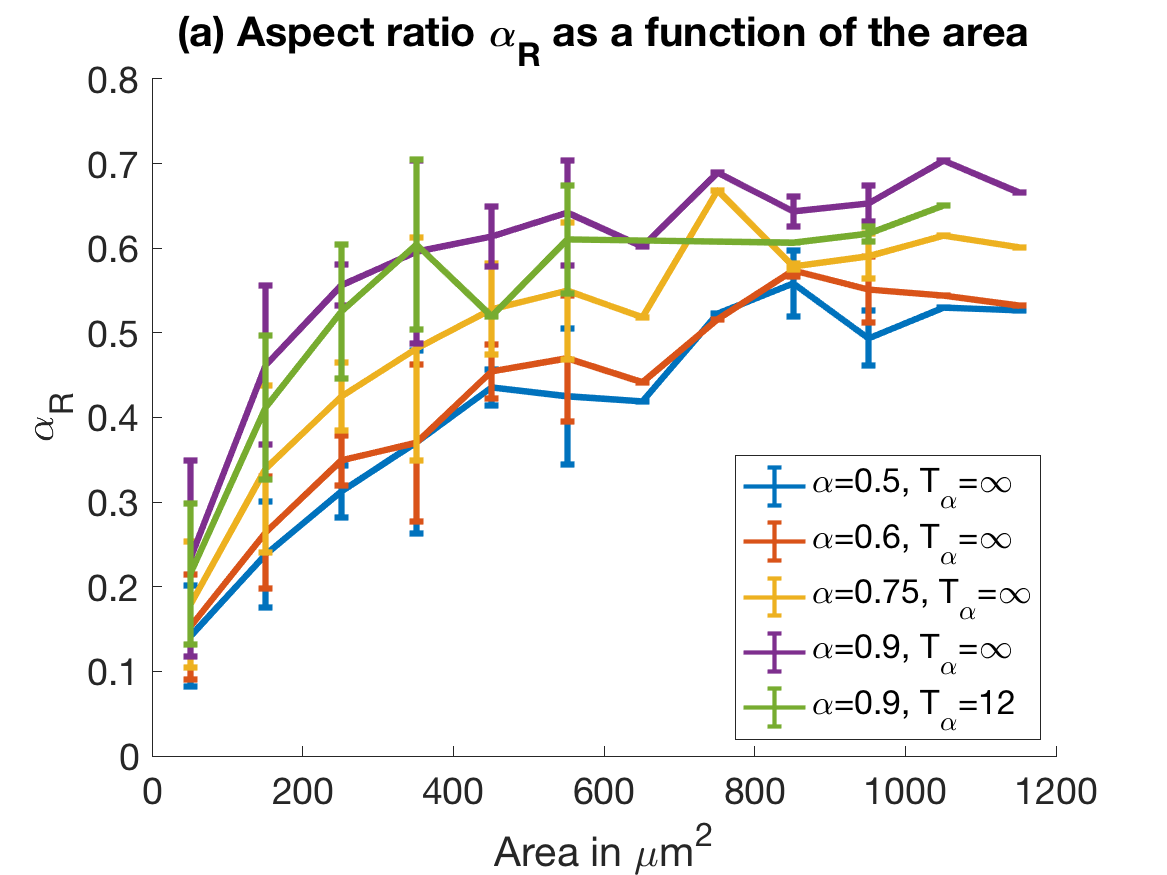}
\includegraphics[scale=0.35]{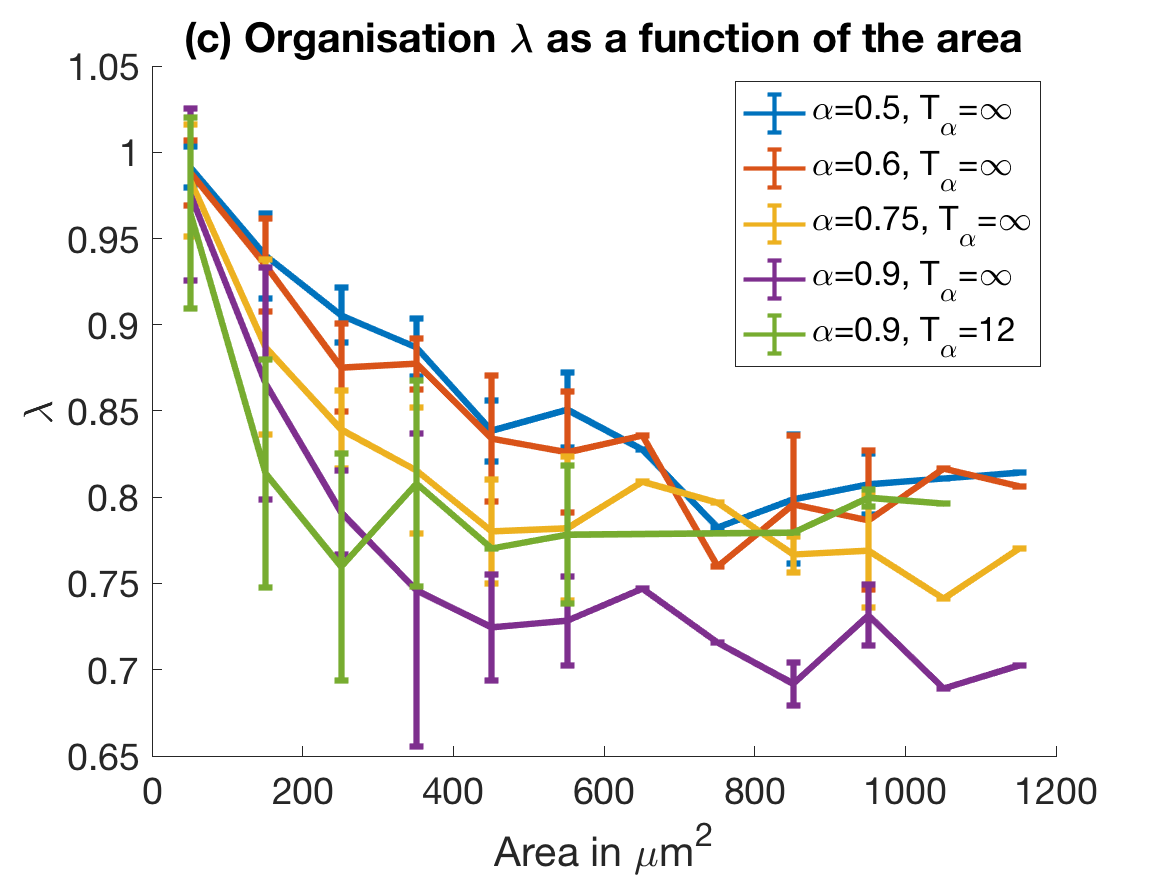}
\includegraphics[scale=0.35]{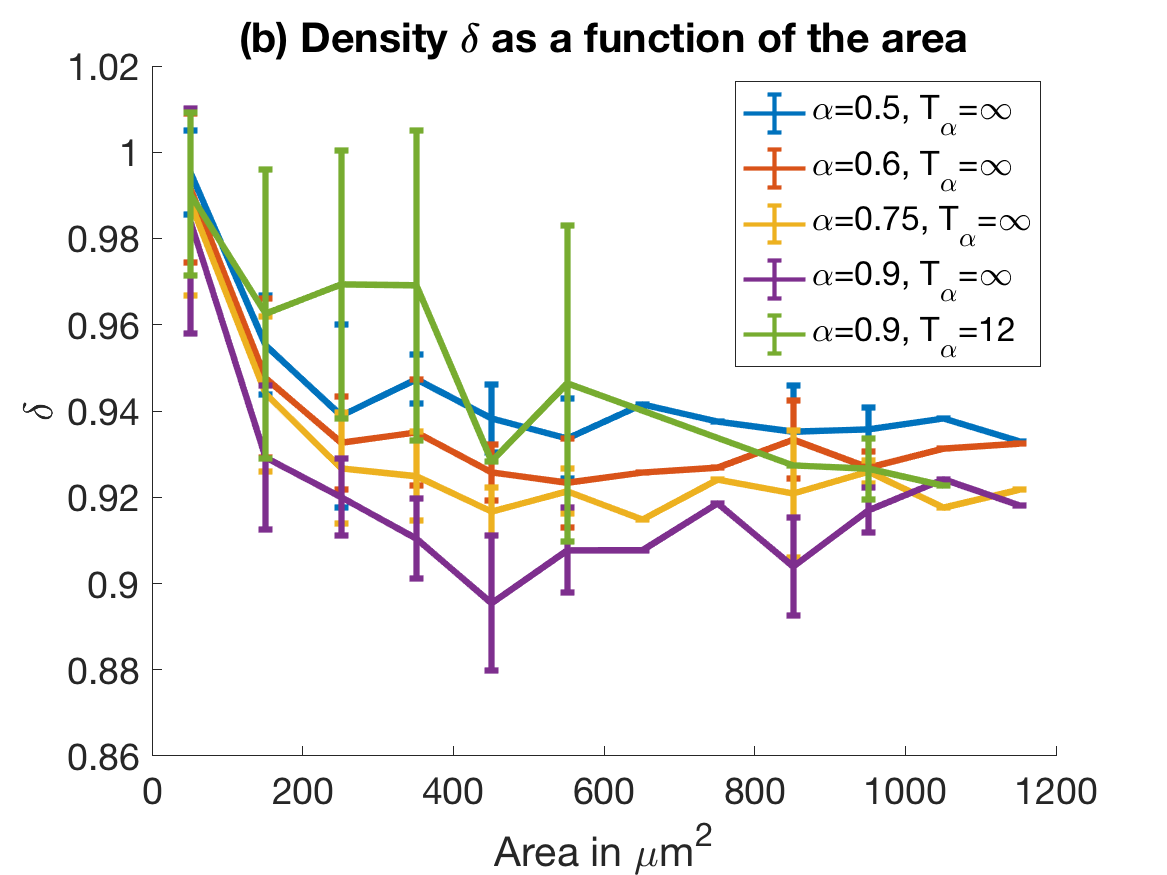}
\includegraphics[scale=0.35]{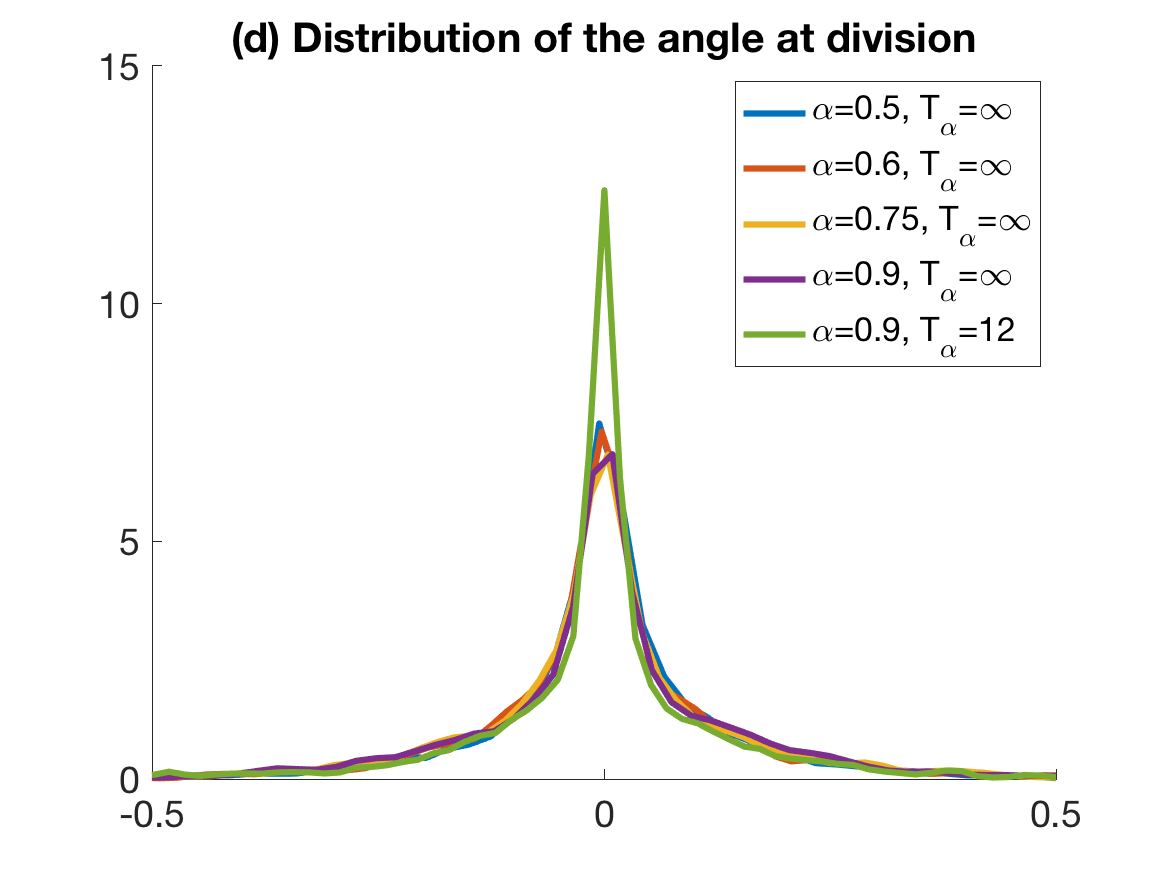}
\caption{Evolution of the aspect ratio $\alpha_R$ (a), the local order quantifier $\lambda$ (b) and the density (c) as functions of the area of the colony, and of the distribution of the angle at division (d) for different values of $\alpha$: $(\alpha,T_\alpha)=(0.5,\infty)$, $(\alpha,T_\alpha)=(0.6,\infty)$, $(\alpha,T_\alpha)=(0.75,\infty)$, $(\alpha,T_\alpha)=(0.9,\infty)$ and $(\alpha,T_\alpha)=(0.9,12)$.
\label{fig:11}}
\end{figure}

The distribution of the mass along the length of the bacteria also influences the shape of the colony and its organisation. Indeed when the mass of a bacterium is located near its old pole, the centre of mass of the cell is shifted to one side and the bacterium is more likely to turn. This observation is illustrated in Fig.~\ref{fig:11} where we present the evolution of the aspect ratio $\alpha_R$, the local order quantifier $\lambda$, the density $\delta$ as functions of the area of the colony and the angle at division distribution for different values of $\alpha$. From Panels~(a) and~(b), it is clear that the increase of $\alpha$ makes the colony more spherical and less organised. However colonies are better organised with asymmetric mass distribution when $T_\alpha<\infty$ (compare the green and purple curves of Fig. \ref{fig:11} (c)). The density is slightly impacted by the decrease of the change of the value of $\alpha$ but the modification is relatively small, given the amplitude of the confidence intervals. Finally, Panel~(d) shows that the increase of $\alpha$ reduces the angle at division, as we already saw for a decrease of $A$: the more the cells are asymmetric, the less they turn during the very early stage of the morphogenesis. On the long range however, their effects appear to be opposite: asymmetric friction continues to inhibit rotation, whereas mass asymmetry seems to favour it.

\subsubsection{The noise at division} \label{Section3.3.3}

\begin{figure}[!ht] 
   \centering
\includegraphics[scale=0.35]{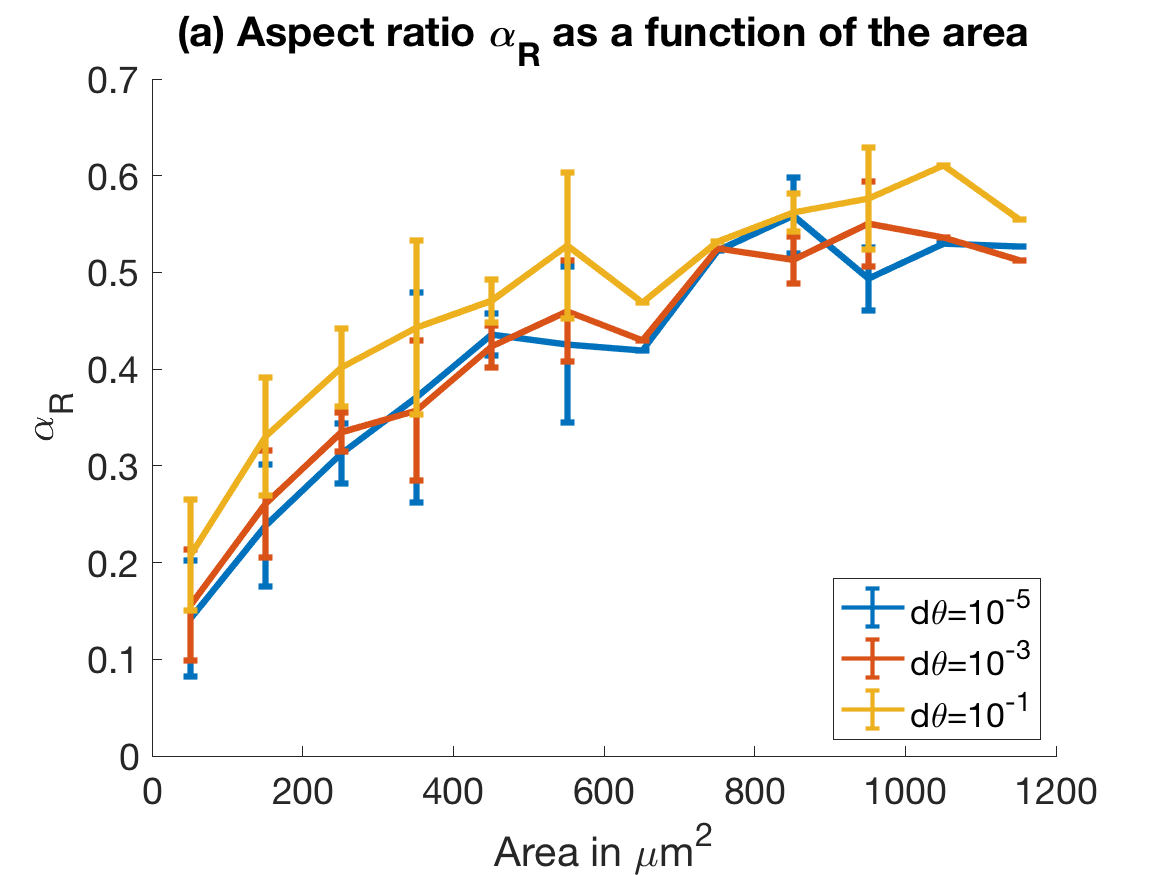}
\includegraphics[scale=0.35]{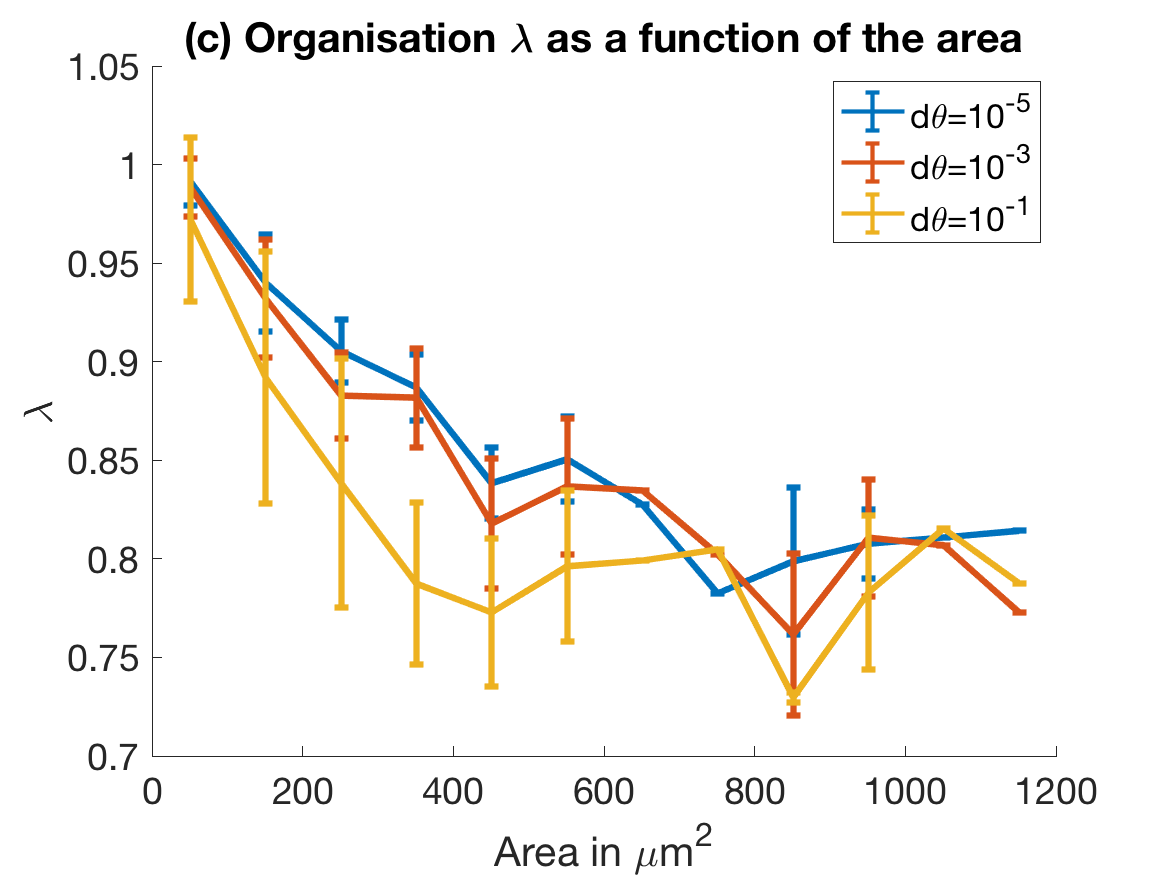}
\includegraphics[scale=0.35]{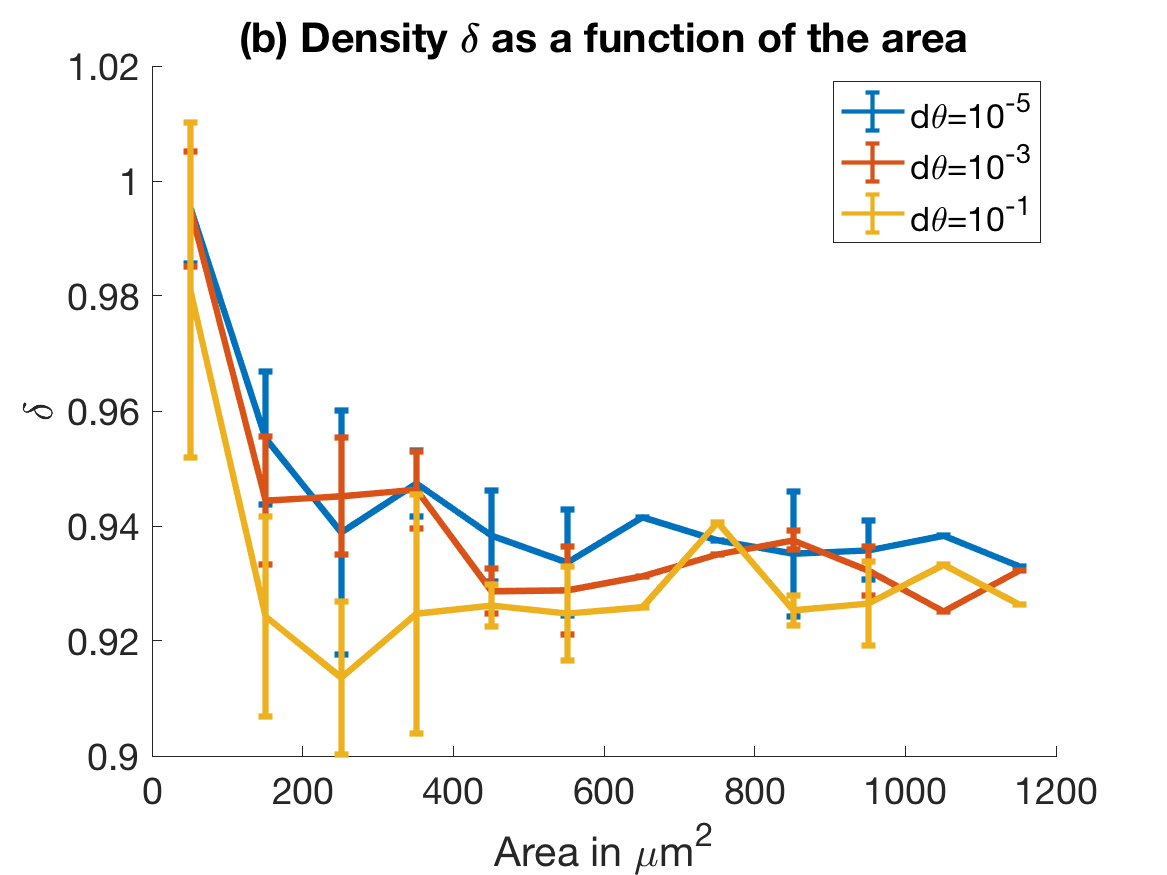}
\includegraphics[scale=0.35]{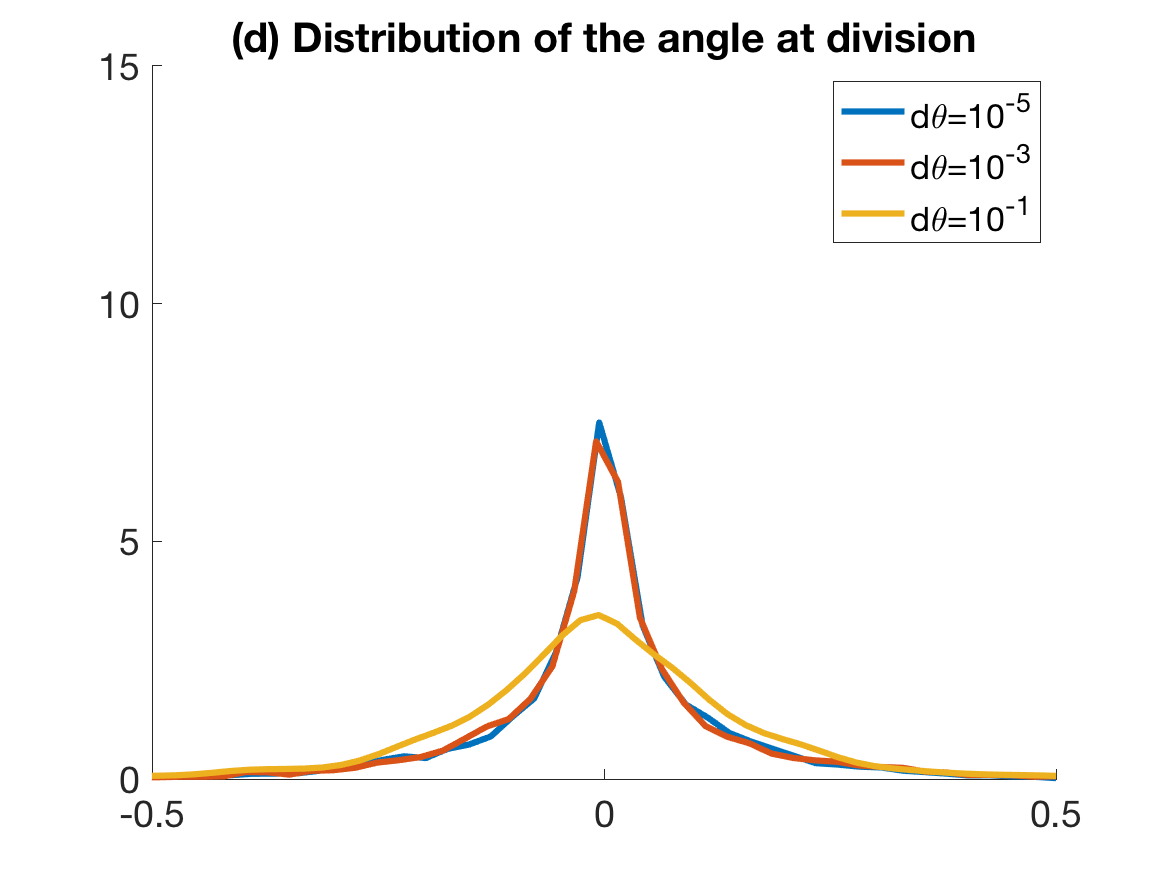}
\caption{Evolution of the aspect ratio $\alpha_R$ (a), the local order quantifier $\lambda$ (b) and the density (c) as functions of the area of the colony, and of the distribution of the angle at division (d) for different values of $\Theta$: $\Theta=10^{-5}$, $\Theta=10^{-3}$, $\Theta=10^{-1}$.}
\label{fig:12}
\end{figure}

Let us now discuss the influence of the noise parameter $\Theta$ on the organisation of the colony. In Fig.~\ref{fig:12} we present the evolution of the four previously-seen quantifiers: the aspect ratio $\alpha_R$, the local order quantifier $\lambda$, the density $\delta$ as functions of the area of the colony as well as the observable angle at division distribution, for different values of $\Theta$. In Table~\ref{Table:4}  the average distance $d_2$ for the different values of $\Theta$ is presented. Fig.~\ref{fig:12} and Table~\ref{Table:4} show that the quantifiers have similar behaviour for $\Theta=10^{-5}$ and $\Theta=10^{-3}$. However, we observe that the increase of $\Theta$ to $10^{-1}$ slightly disorganises the colonies while making them more spherical. It also flattens the angle at division and decreases the value of $d_2$. Therefore, to a certain extent, an important increase of $\Theta$ has an effect similar to the increase of the mass asymmetry at the very early stage (four cell) of the micro-colony, but an opposite - though moderate - effect on the long term. 

\begin{table}[h!]
\centering
\begin{tabular}{ | l l l l | } 
\hline
$\Theta$ & average of $d_2$ & minimum of $d_2$ & maximum of $d_2$ \\  
\hline
\hline
$\Theta =10^{-5}$ &  0.999999891665219 &  0.999999459667383  & 0.999999999972967\\ 
\hline
$\Theta =10^{-3}$ &  0.999087269350617  & 0.995912388894949  & 0.999999729668722\\ 
\hline
$\Theta =10^{-1}$  & 0.953627181718370 &  0.864658486163227  & 0.990629980128104\\ 
\hline
\end{tabular}
\caption{\label{Table:4}Influence of $\Theta$ on the four-cell array quantifier $d_2$}
\end{table}

The meaning of the noise at division can be discussed. It was first introduced to break the symmetry in the division and avoid the growth of one-line colonies.  However, as we model biological systems, it is reasonable to suppose that they are subject to random fluctuations (from the environment), and that cell division is not perfectly symmetric but has a random component. Then comes the question of the amplitude of this noise. A small noise can be easily identified to the division of the bacteria, whereas a large noise is more difficult to justify. It could then be considered as the result of a hidden phenomenon unknown to this date. Throughout the paper we do not restrict the noise at division amplitude.

\section{Comparison of experimental data and numerical simulations} \label{Section4}

In this section, we compare the numerical simulations with experimental data, in order to quantify as much as possible to which extent the model  is suitable to study the two-dimensional evolution of sessile bacteria on a substrate. The comparison is made via the use of the quantifiers presented in Section~\ref{Section3.2} to describe the characteristics of the micro-colonies. The comparisons are done for the 3 sets of experimental data we presented in Section~\ref{Section3.1}. When the parameters are not explicitly mentioned they are defined according to Table~\ref{Table:1}. The study made in Section~\ref{Section3} shows that asymmetric friction results in a more elongated micro-colony while an asymmetric distribution of the mass along the length of the bacteria provides the four-cell array organisation in the early stage of the colony growth. However, the asymmetric mass distribution tends to disorganise the colony and  to make it more spherical. Thus, to fit at best the experimental data, we aim to find a ratio between the influence of both parameters. For each experiment, we compare the experimental data with numerical simulations with four sets of parameters. We restrict the number of simulations to 4 to simplify the comprehension of the paper, given the high number of parameters in the model. The four cases of the numerical simulations are the following:
\begin{enumerate}
\item Symmetric friction $A=1$, uniform mass distribution $\alpha=0.5$ and a small noise at division $\Theta=10^{-5}$. Blue curves in Figs.~\ref{fig:13}, \ref{fig:15}, \ref{fig:17}, Panels~(a) in Figs.~\ref{fig:14_1}, \ref{fig:14_2}, \ref{fig:16_1}, \ref{fig:16_2}, \ref{fig:18_1}, \ref{fig:18_2}.
\item Asymmetric friction $A<1$ and uniform mass distribution $\alpha=0.5$. Red curves in Figs.~\ref{fig:13}, \ref{fig:15}, \ref{fig:17}, Panels~(b) in Figs.~\ref{fig:14_1}, \ref{fig:14_2}, \ref{fig:16_1}, \ref{fig:16_2}, \ref{fig:18_1}, \ref{fig:18_2}.
\item Symmetric friction $A=1$ and asymmetric mass distribution $\alpha>0.5$. Honey yellow curves in Figs.~\ref{fig:13}, \ref{fig:15}, \ref{fig:17}, Panels~(c) in Figs.~\ref{fig:14_1}, \ref{fig:14_2}, \ref{fig:16_1}, \ref{fig:16_2}, \ref{fig:18_1}, \ref{fig:18_2}.
\item Asymmetric friction $A<1$ and asymmetric mass distribution $\alpha>0.5$. Purple curves in Figs.~\ref{fig:13}, \ref{fig:15}, \ref{fig:17}, Panels~(d) in Figs.~\ref{fig:14_1}, \ref{fig:14_2}, \ref{fig:16_1}, \ref{fig:16_2}, \ref{fig:18_1}, \ref{fig:18_2}.
\end{enumerate}
For the first set of parameters we aim to reproduce the general model found in the literature \cite{Boyer2011,Damme2019,Volfson2008,You2018} for the spatial forces, combined with the most up-to-date model for growth and division.
For the three other sets of parameters, the choice of the parameters $A$, $\alpha$ and $\Theta$ is made in order to fit qualitatively as best as possible the experimental data.

\subsection{Dataset 1} \label{Section4.1}

The parameters which have been chosen to fit the experimental data are listed below:
\begin{enumerate}
\item symmetric friction $A=1$ and uniform mass distribution $\alpha=0.5$, angle at division parameter $\Theta=10^{-5}$,
\item asymmetric friction $A=0.4$ and uniform mass distribution $\alpha=0.5$, angle at division parameter $\Theta=10^{-1}$,
\item symmetric friction $A=1$, asymmetric mass distribution  $\alpha=0.9$ with $T_\alpha =12 \mbox{ min}$,  angle at division parameter $\Theta=10^{-5}$,
\item asymmetric friction $A=0.5$ and asymmetric mass distribution  $\alpha=0.9$ with $T_\alpha =12 \mbox{ min}$,  angle at division parameter $\Theta=10^{-1}$.
\end{enumerate}

\begin{figure}[!ht] 
   \centering
\includegraphics[scale=0.35]{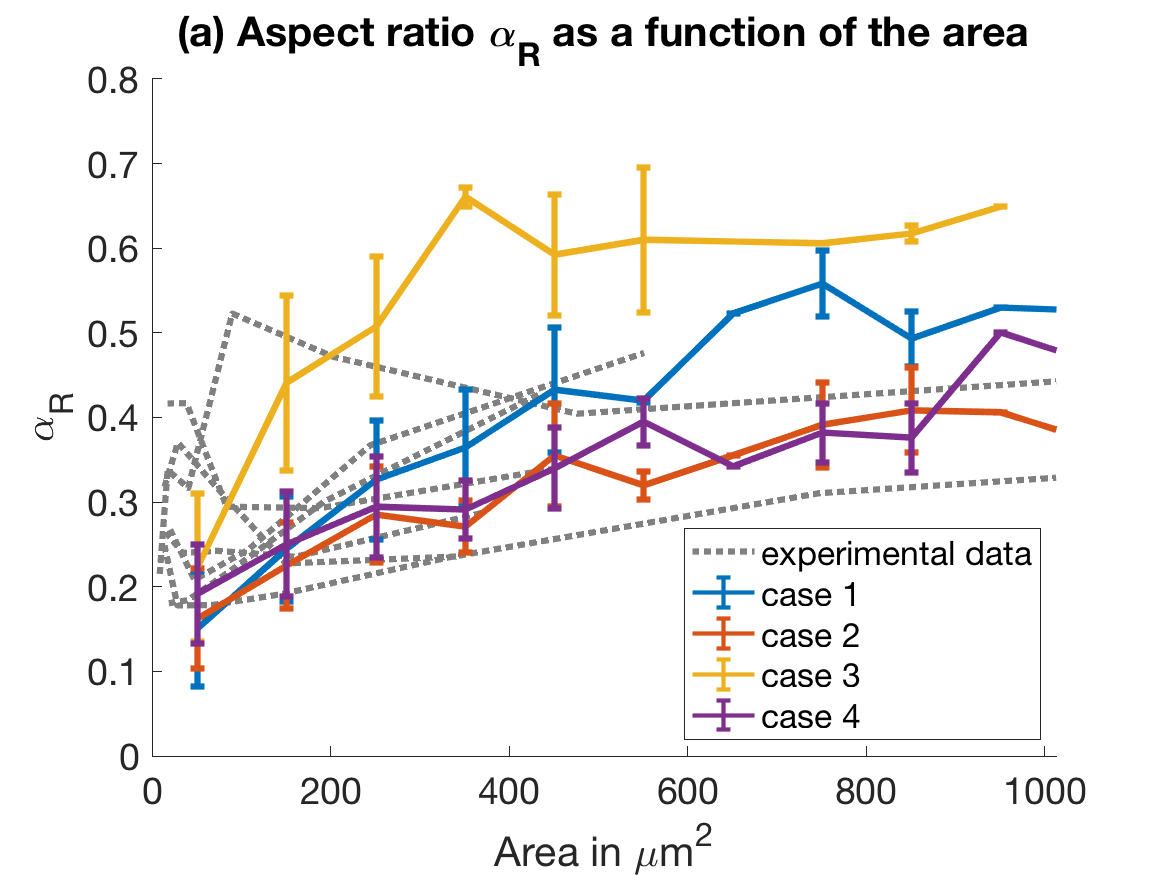}
\includegraphics[scale=0.35]{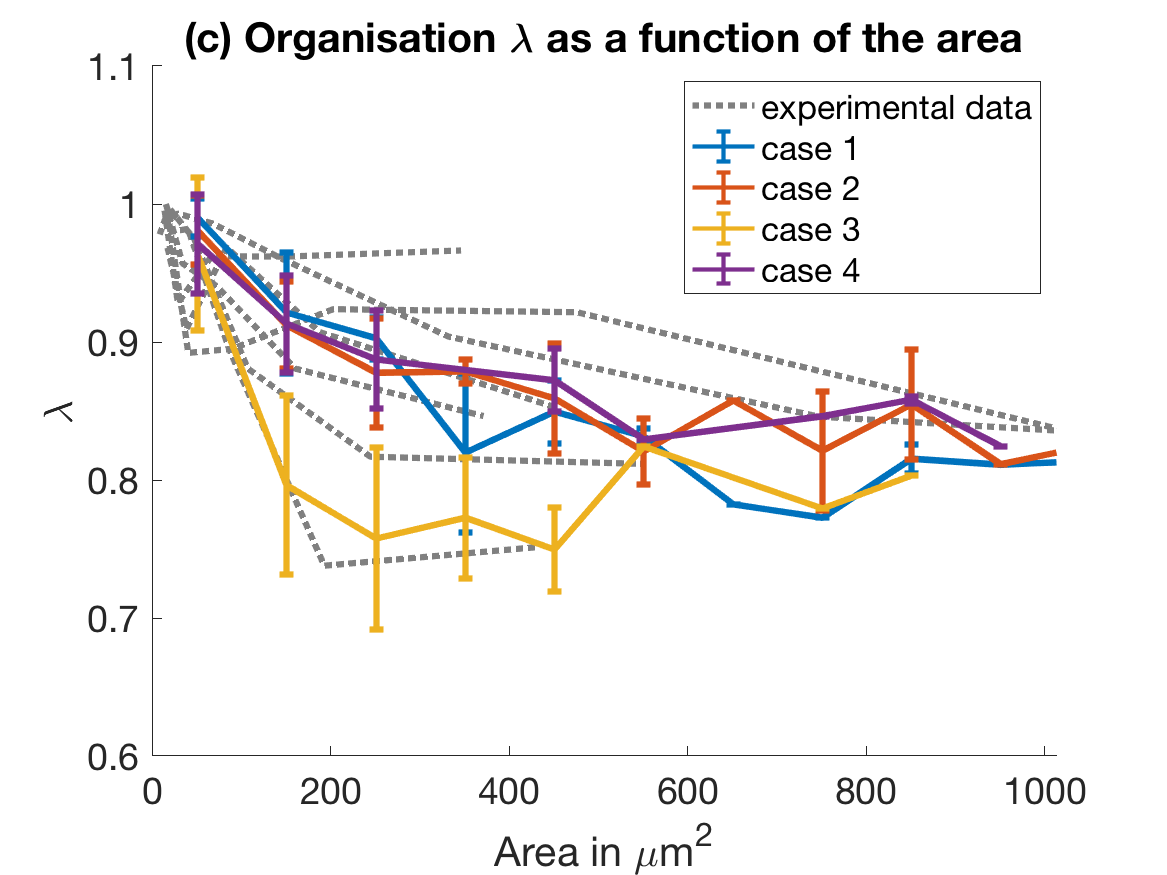}
\includegraphics[scale=0.35]{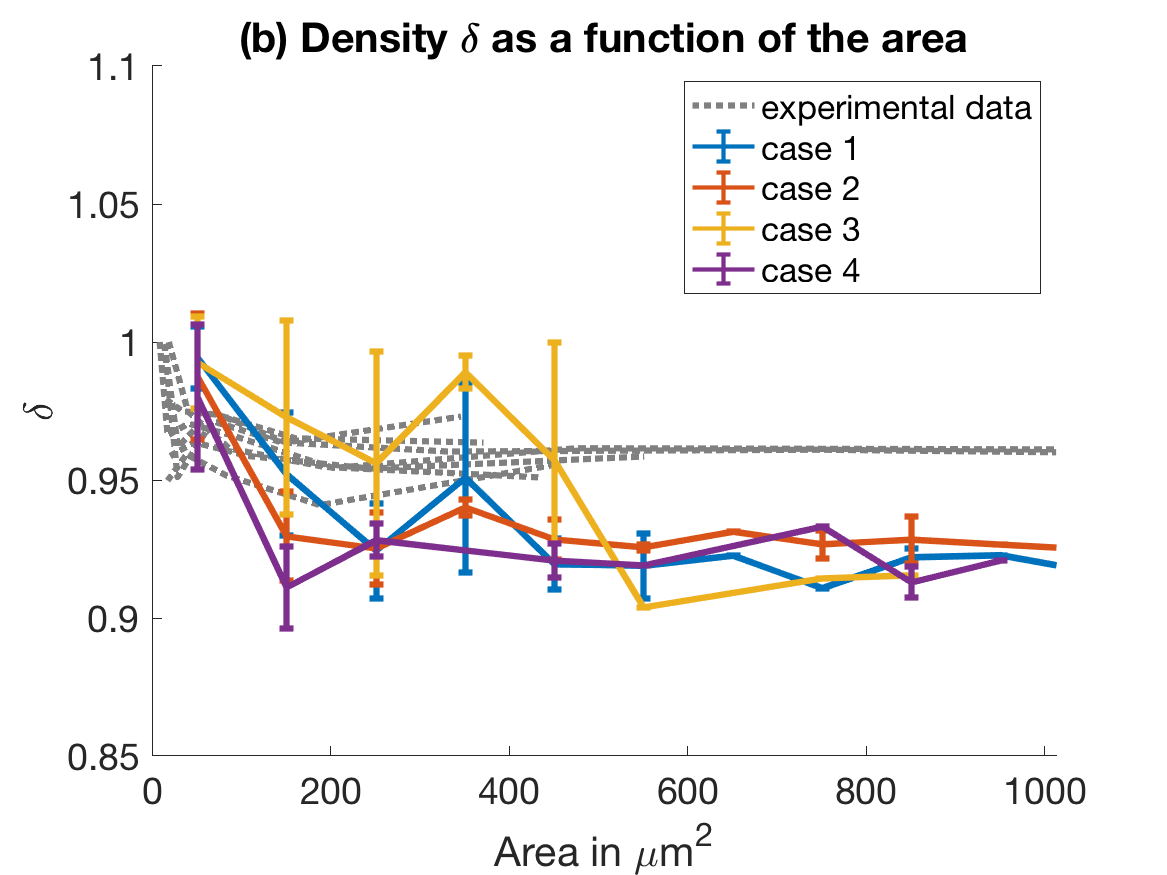}
\includegraphics[scale=0.35]{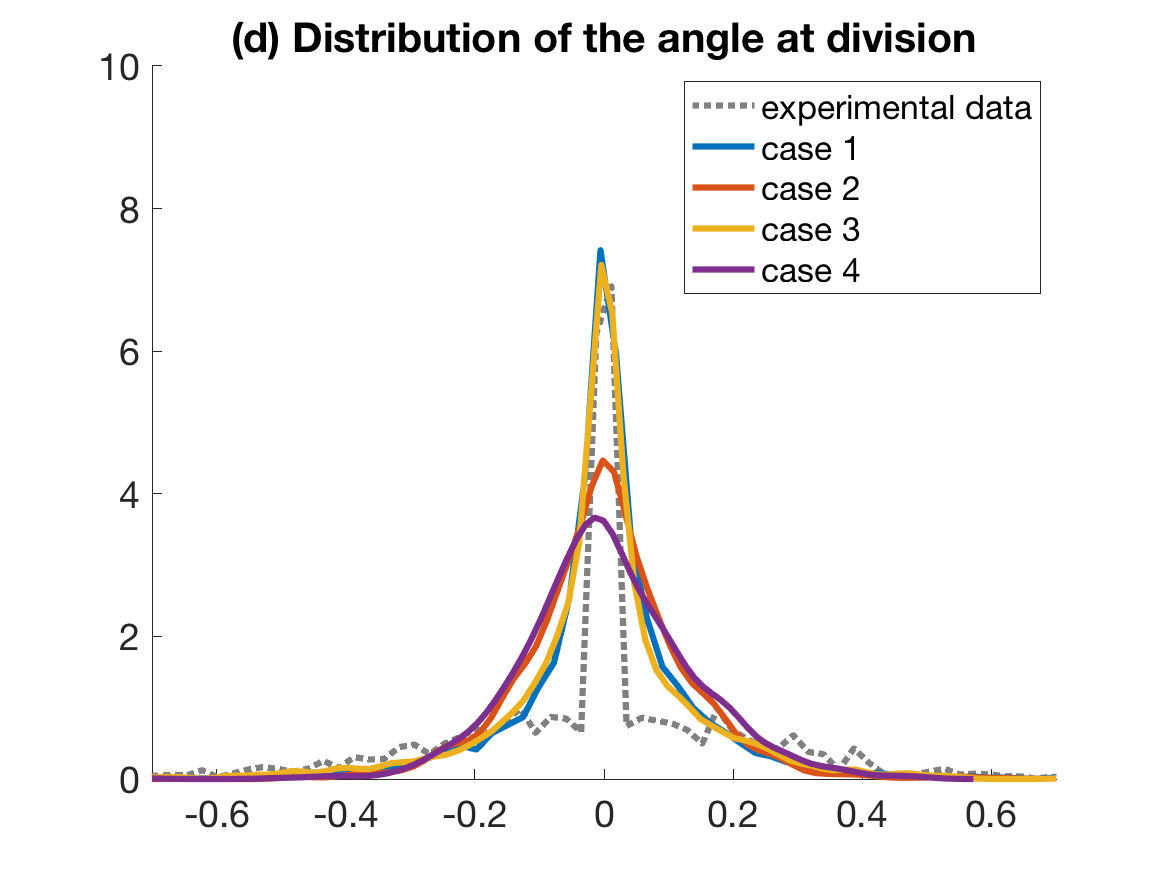}
\caption{Dataset 1: plots of the aspect ratio $\alpha_R$ (a), the local order quantifier $\lambda$ (b) and the density (c) as functions of the area of the colony, and of the distribution of the angle at division (d) for the experimental data (grey dashed curve), and numerical simulations for the case 1 (blue curve), case 2 (red curve), case 3 (yellow curve) and case 4 (purple curve). The plots of the numerical data are averaged over 10 simulations.}
\label{fig:13}
\end{figure}

In Fig.~\ref{fig:13} we show the evolution of the aspect ratio $\alpha_R$, the local organisation parameter $\lambda$, and the density quantifier $\delta$ as functions of the colony area as well as the distribution of the observable angle at division. The grey curves correspond to the evolution of the quantifiers computed on the experimental data. Note that these are not averages over the number of experimental colonies, due to the high variability in the values of the quantifiers. Table~\ref{Table:5} presents the average values of $d_2$ for the experimental data and numerical simulations.

\begin{table}[h!]
\centering
\begin{tabular}{ | l l l l | } 
\hline
Dataset 1 & average of $d_2$ & minimum of $d_2$ & maximum of $d_2$ \\  
\hline
\hline
Experimental data &  0.749486923556648  & 0.491327335418617 &  0.976272403852093\\ 
\hline
Case 1 &  0.999999877816758 &  0.999999459667383 &  0.999999999631776 \\ 
\hline
Case 2  & 0.988597129401271  & 0.948318305860428  & 0.999980901984201 \\ 
\hline
Case 3  & 0.986823375651587 &  0.912749581037532  & 0.999994661249026\\ 
\hline
Case 4  & 0.950405437714561 &  0.844623355182293  & 0.999990867503669\\ 
\hline
\end{tabular}\centering\caption{
\label{Table:5}
{Four-cell array quantifier $d_2$:  comparison of the four parameter choice cases with the experimental dataset~1.}}
\end{table}

\begin{figure}[!ht] 
   \centering
\includegraphics[scale=0.17]{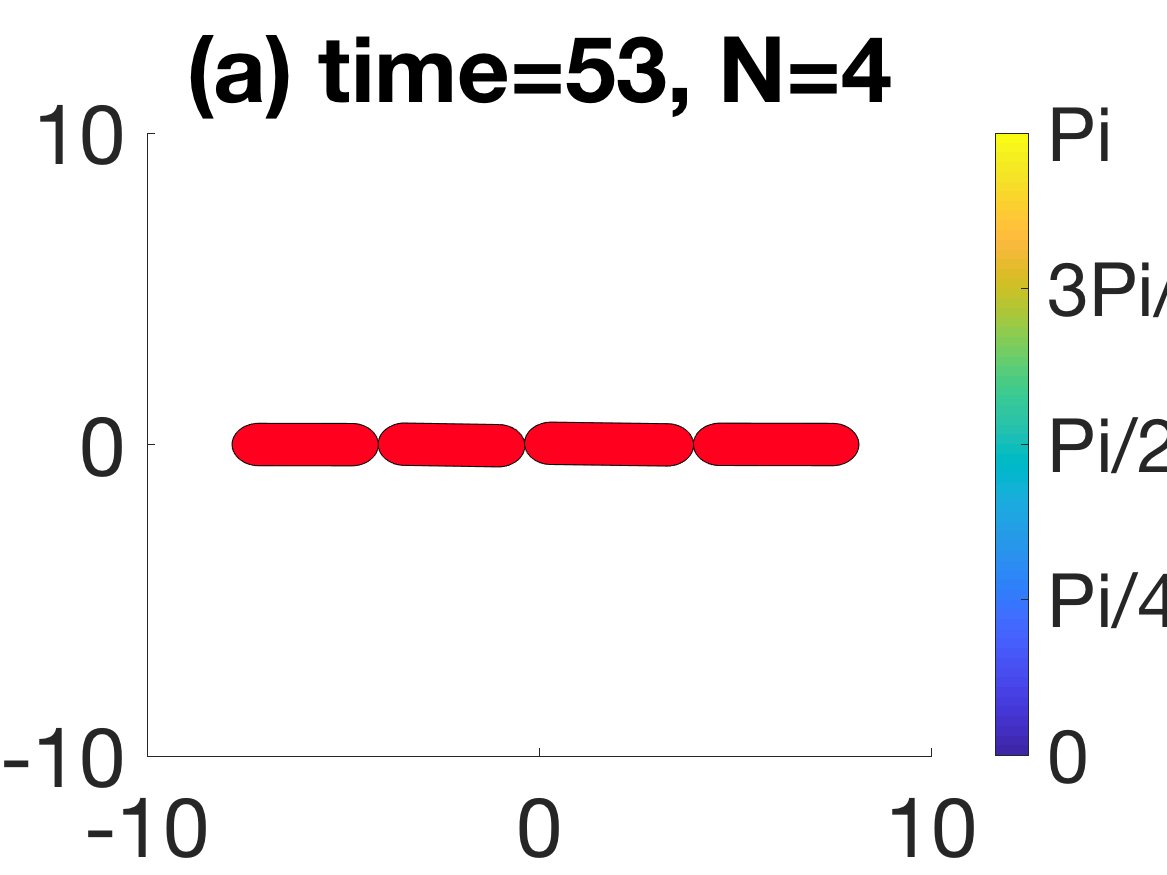}
\includegraphics[scale=0.17]{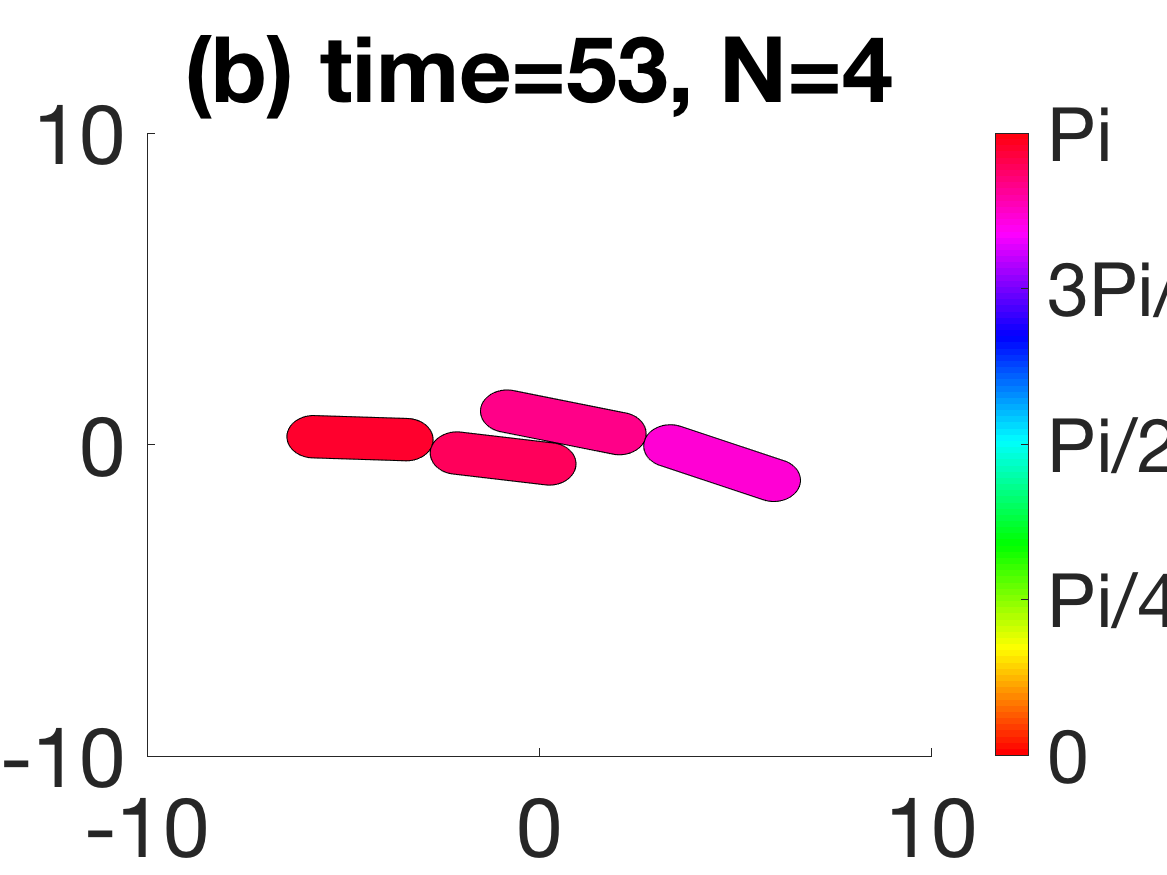}
\includegraphics[scale=0.17]{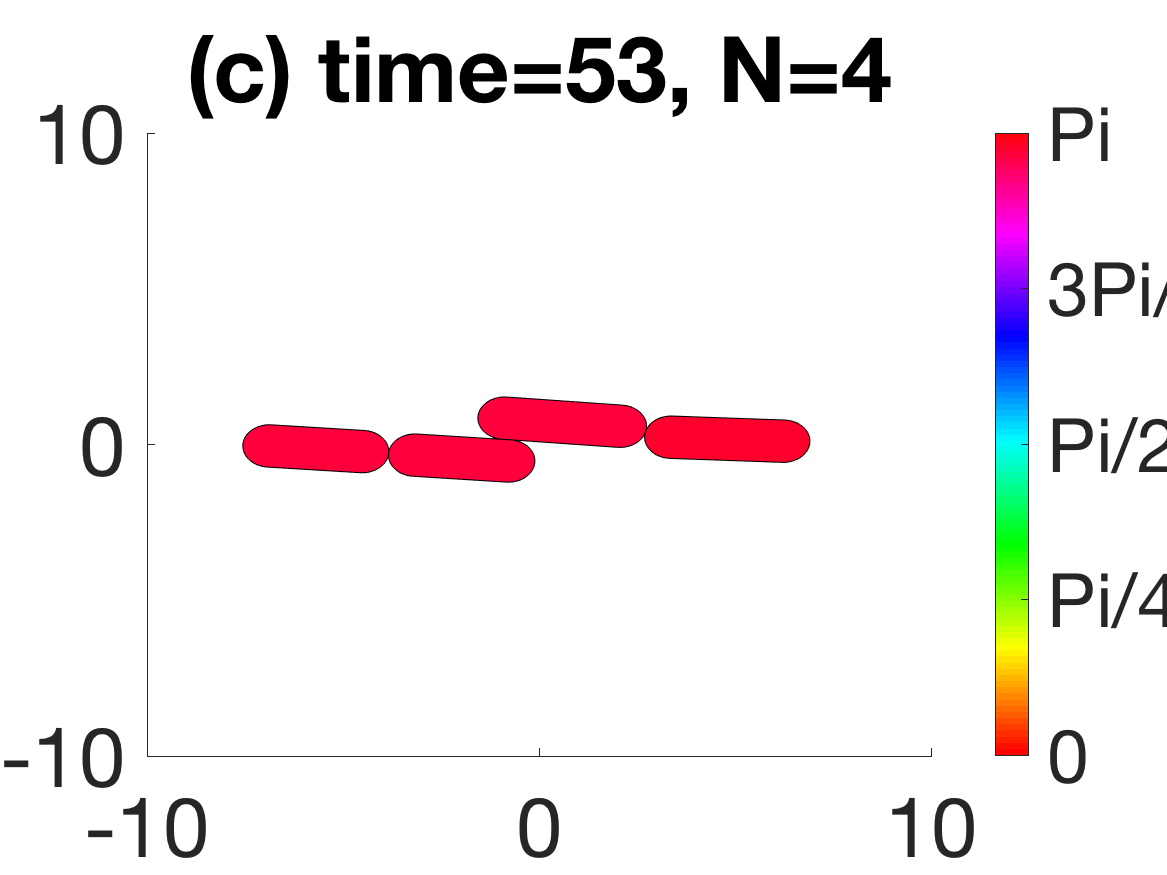}
\includegraphics[scale=0.17]{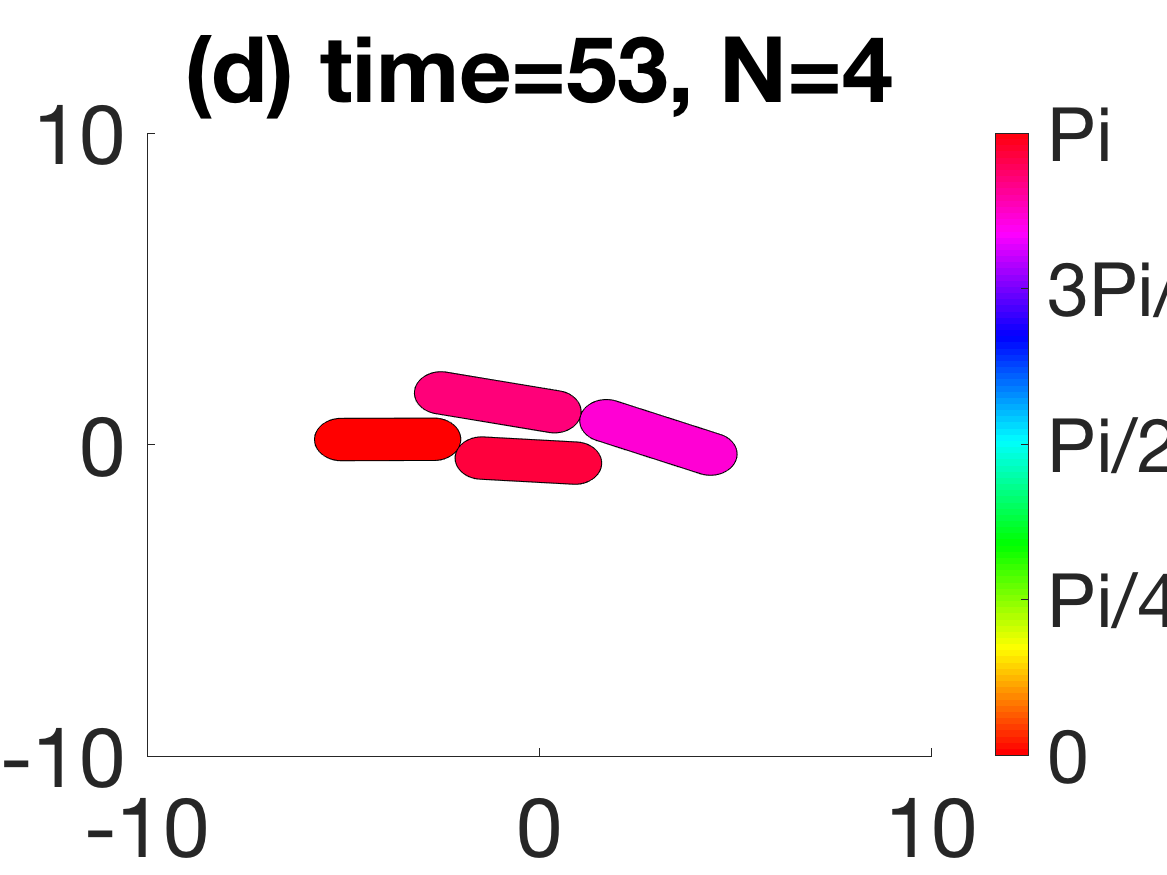}
\caption{Dataset 1: Plot of simulation at time $t=53  \mbox{min}$ for Case 1 (a), Case 2 (b), Case 3 (c) and Case 4 (d). The colors of the bacteria are given by their orientation.  These figure can be compare to Fig.~\ref{fig:7} Panels~(a).}
\label{fig:14_1}
\end{figure}

As stated previously, the variability of the quantifiers for the experimental data in Fig.~\ref{fig:13}, Panel~(a) and~(b) makes the comparison with numerical simulations difficult. The observation of Panel~(a) shows that the best fit for the aspect ratio are the cases 2 and 4.  The case 1 could also be considered even though the large colonies tend to be too spherical, but we can exclude the case~3 on the basis of the aspect ratio. This is also observed for the case 1 for colonies bigger than $600 \mu m^2$. Concerning the local organisation in the colony (Panel~(c) Fig.~\ref{fig:13}), because the range of values taken by the experimental data is wide, we can conclude that the four cases are acceptable choices. For the cases 2 and 4 in particular, the quantifier $\lambda$ follows rather well one of the experimental colony. Panel~(c) shows that the densities of the
 numerical colonies is systematically smaller than the one of the simulated colonies. 
 An exception can be made for the case 3 which is denser than the other cases for areas smaller than $500 \mu m^2$. 
 The angle at division of the experimental colonies 
 (Panel (d)) is composed of a peak centred 
 in zero and has then an almost uniform distribution spread from
  $-0.5$
    to $0.5$. 
 The distributions observed numerically take the form of normal distributions, which makes the comparison with the experimental data difficult.  We distinguish two cases: for the cases 1 and 3, the peak of the experimental angle is reached; for the cases 2 and 4, the distribution, similarly to the experimental one, 
 spreads up to 
 $-0.4$ and $0.4$. 
 Finally concerning the organisation of the colony at early stages of development, Table~\ref{Table:5} shows that the distance $d_2$ is not as small for the numerical data as for the experimental data.  However, the closest values are taken for the case~4, which is also confirmed by visual comparison in Fig.~\ref{fig:14_1}. Therefore, given these observations, we can conclude that the best choice of parameters is first the case 4 and second the case 2, whereas the cases 1 and 3 may be excluded.
 
\begin{figure}[!ht] 
   \centering
\includegraphics[scale=0.35]{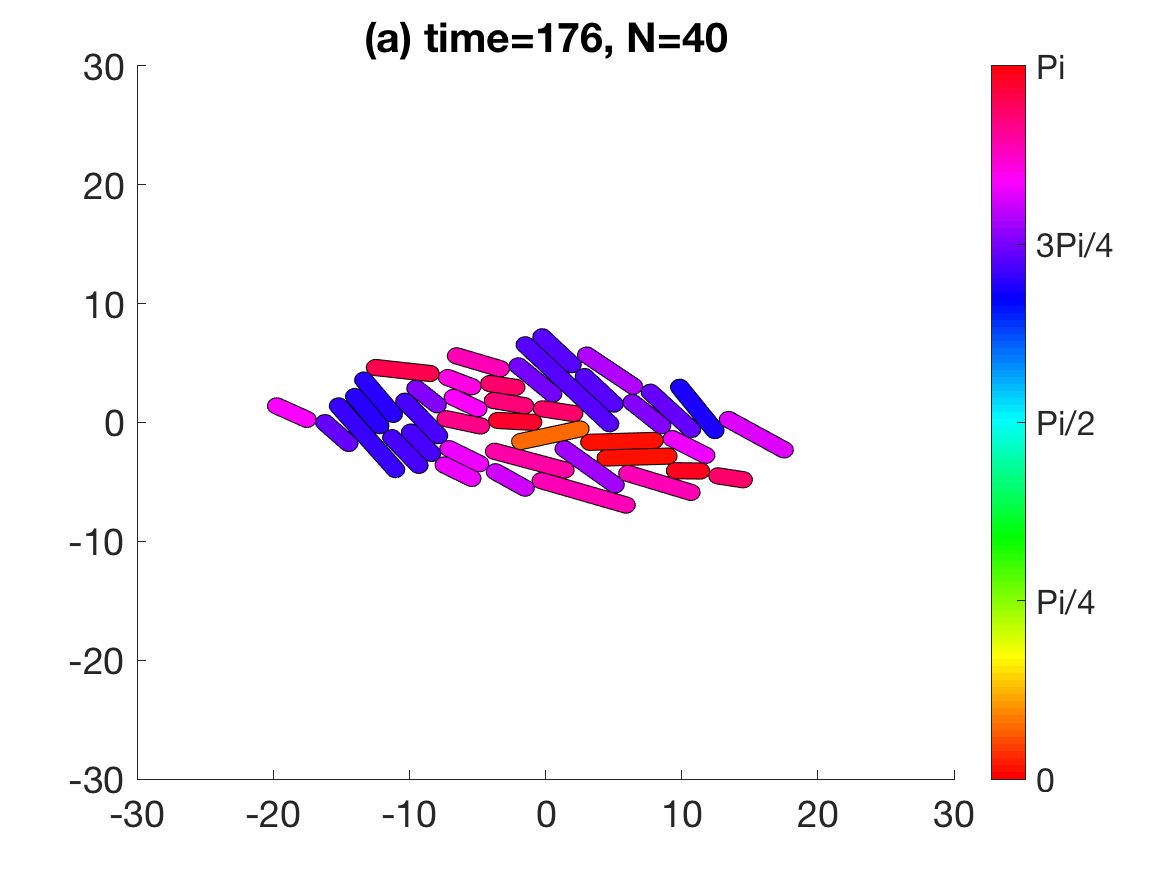}
\includegraphics[scale=0.35]{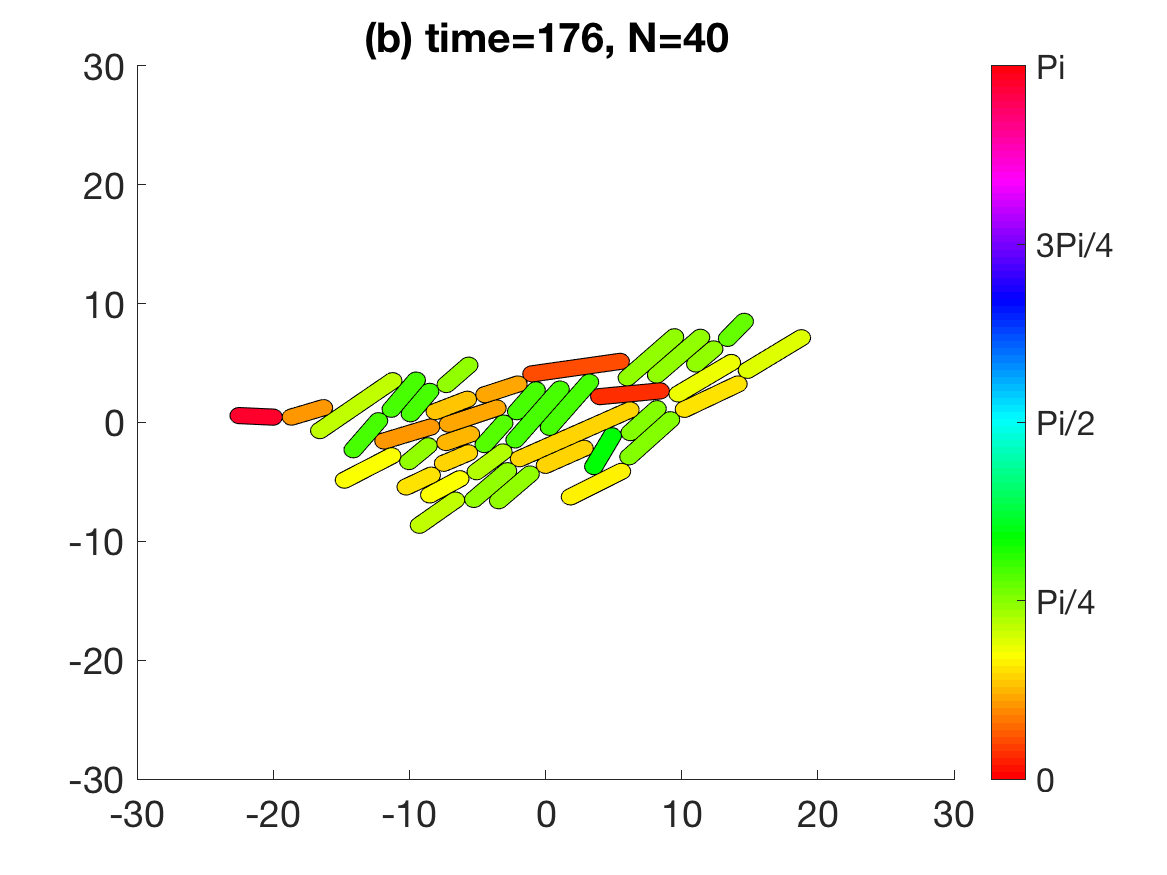}
\includegraphics[scale=0.35]{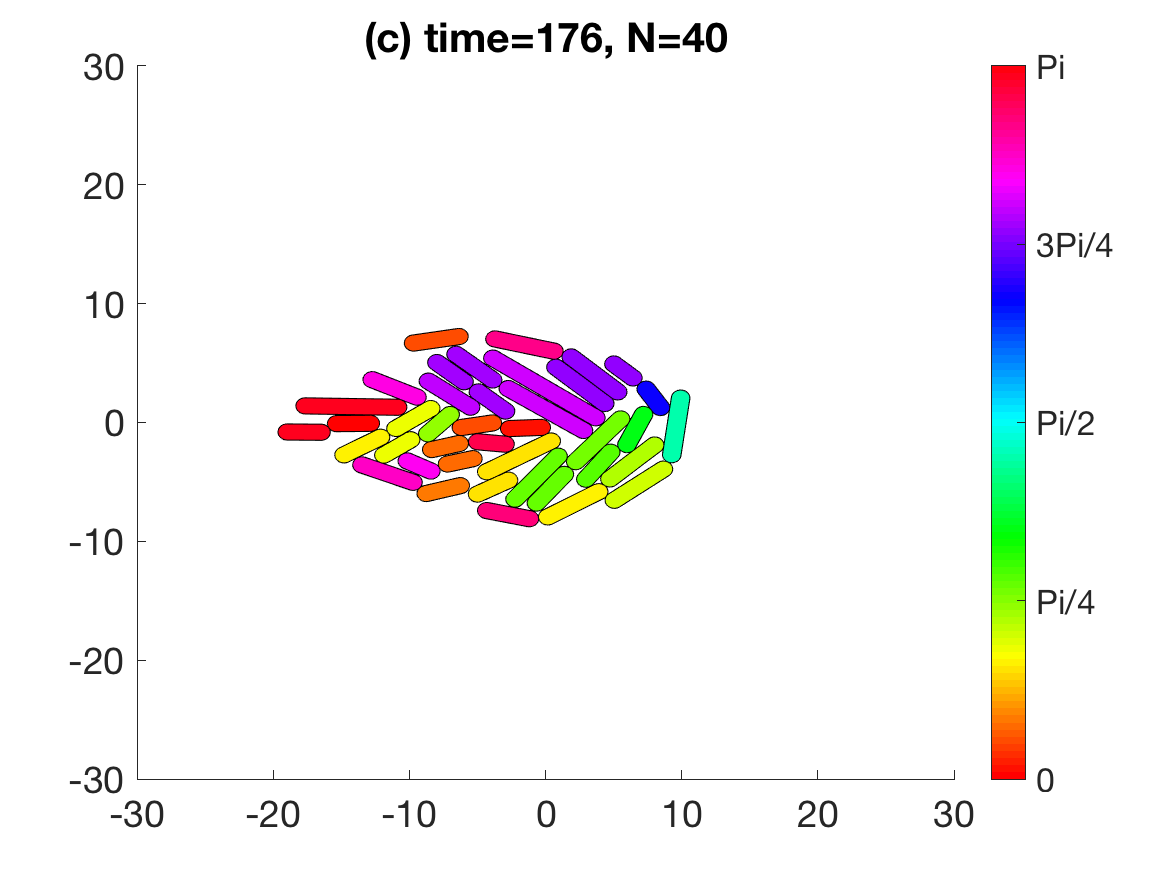}
\includegraphics[scale=0.35]{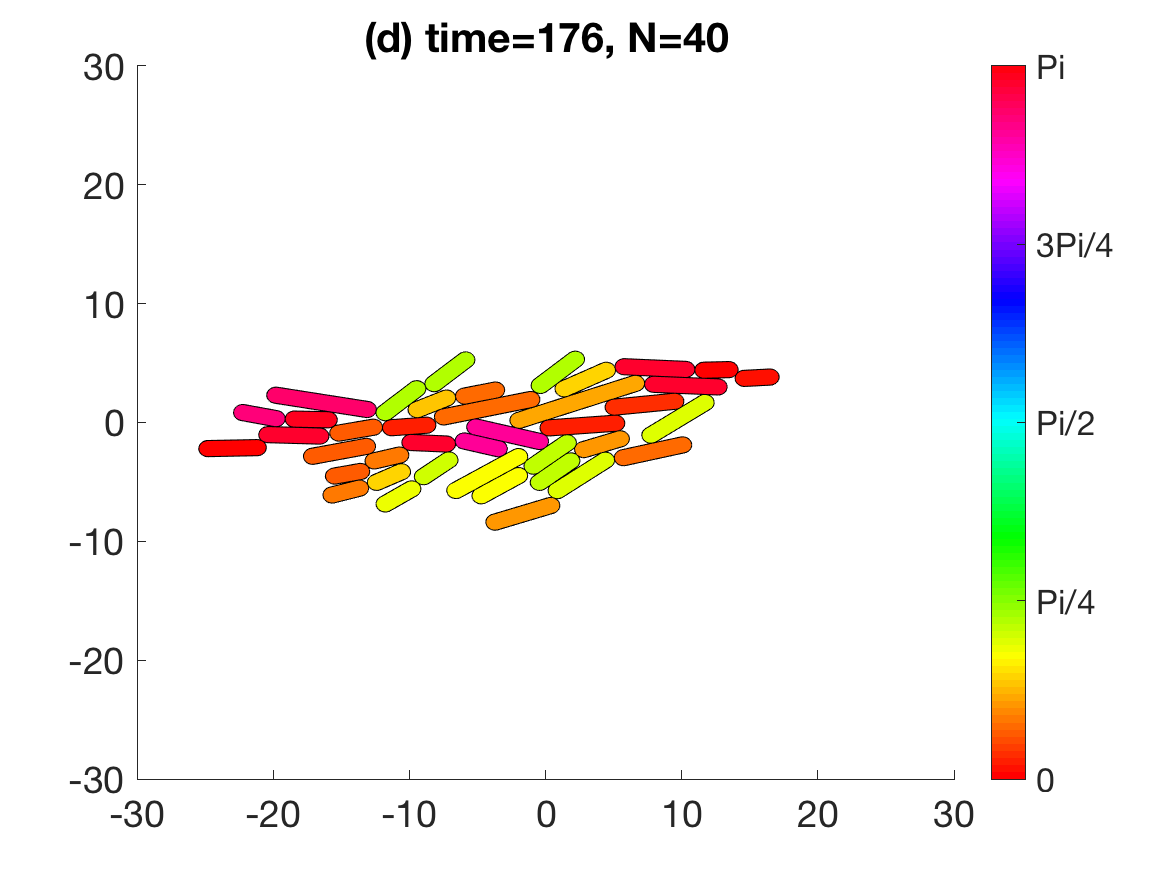}
\caption{Dataset 1: Plot of simulation at time $t=200  \mbox{min}$ for Case 1 (a), Case 2 (b), Case 3 (c) and Case 4 (d). The colors of the bacteria are given by their orientation. These figure can be compare to Fig.~\ref{fig:7} Panels~(a).}
\label{fig:14_2}
\end{figure}

 In Fig.~\ref{fig:14_1} and \ref{fig:14_2} we present plots of the colonies for cases 1, 2, 3 and 4 at after $53$ and $176$ minutes respectively. The first time is the first occurrence where the colony is composed of four cells and the second time has been chosen so that the number of bacteria in the colonies is equal to $40$, which is similar to the number of bacteria present in the plots \ref{fig:7} (2). Fig.~\ref{fig:14_1} shows that the best four-cell array configuration is obtained in Panel~(d) corresponding to the case~4. Besides, by comparing Figs.~\ref{fig:14_2} and \ref{fig:7} (a) we visually observe that the colonies which are the most similar to the experimental colony are the cases 2 and 4. This supports our previous statement.  Therefore for this set of data, our model suggests that the overall anisotropy of the colony could be mainly due to asymmetric friction of the bacteria, and that cell division could be accompanied by an asymmetric mass distribution.
 
\subsection{Dataset 2} \label{Section4.2}

\begin{figure}[!ht] 
   \centering
\includegraphics[scale=0.35]{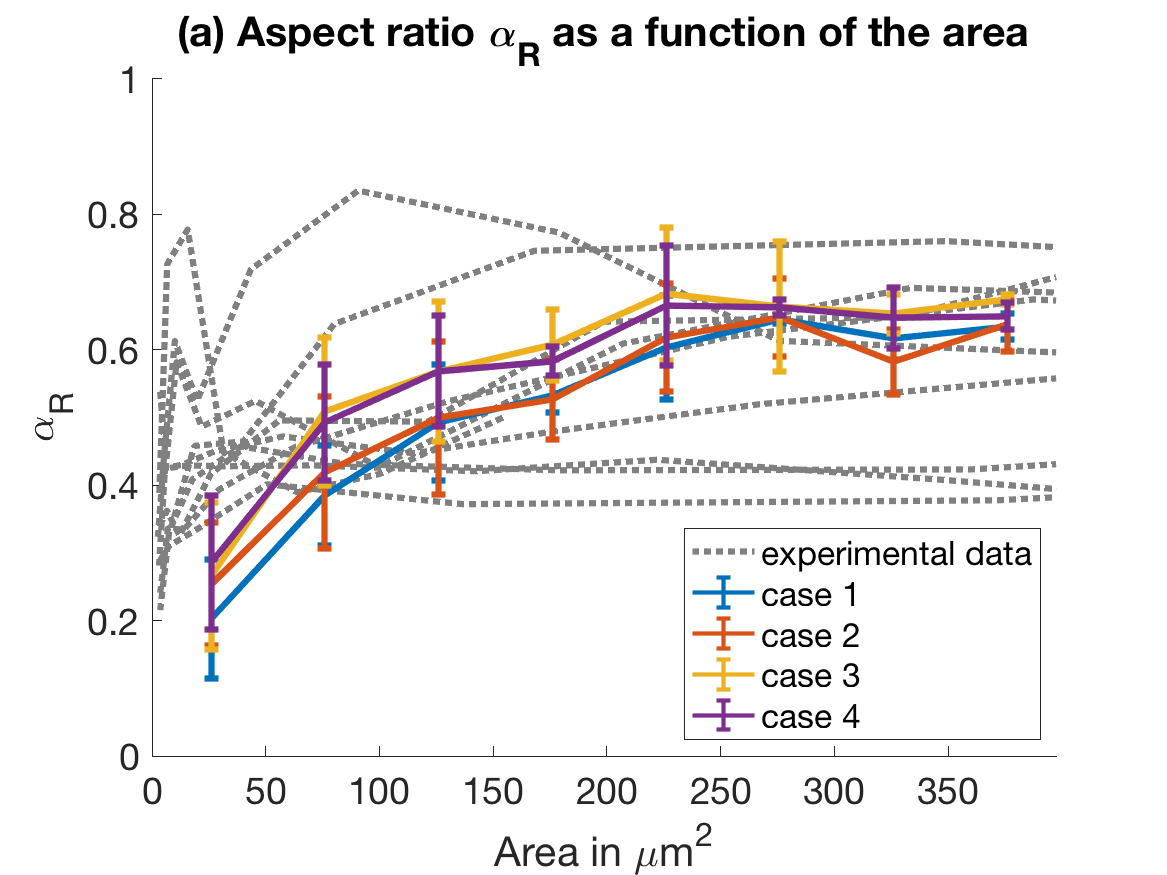}
\includegraphics[scale=0.35]{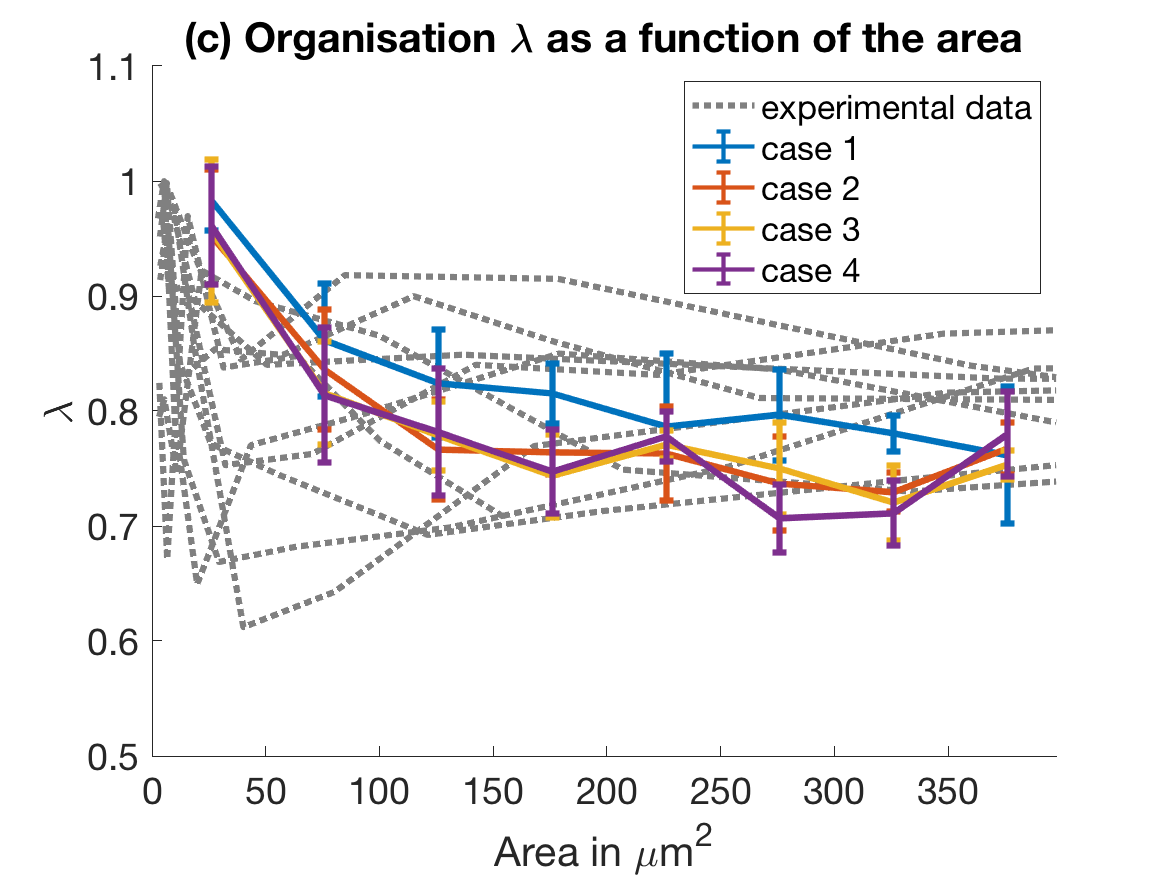}
\includegraphics[scale=0.35]{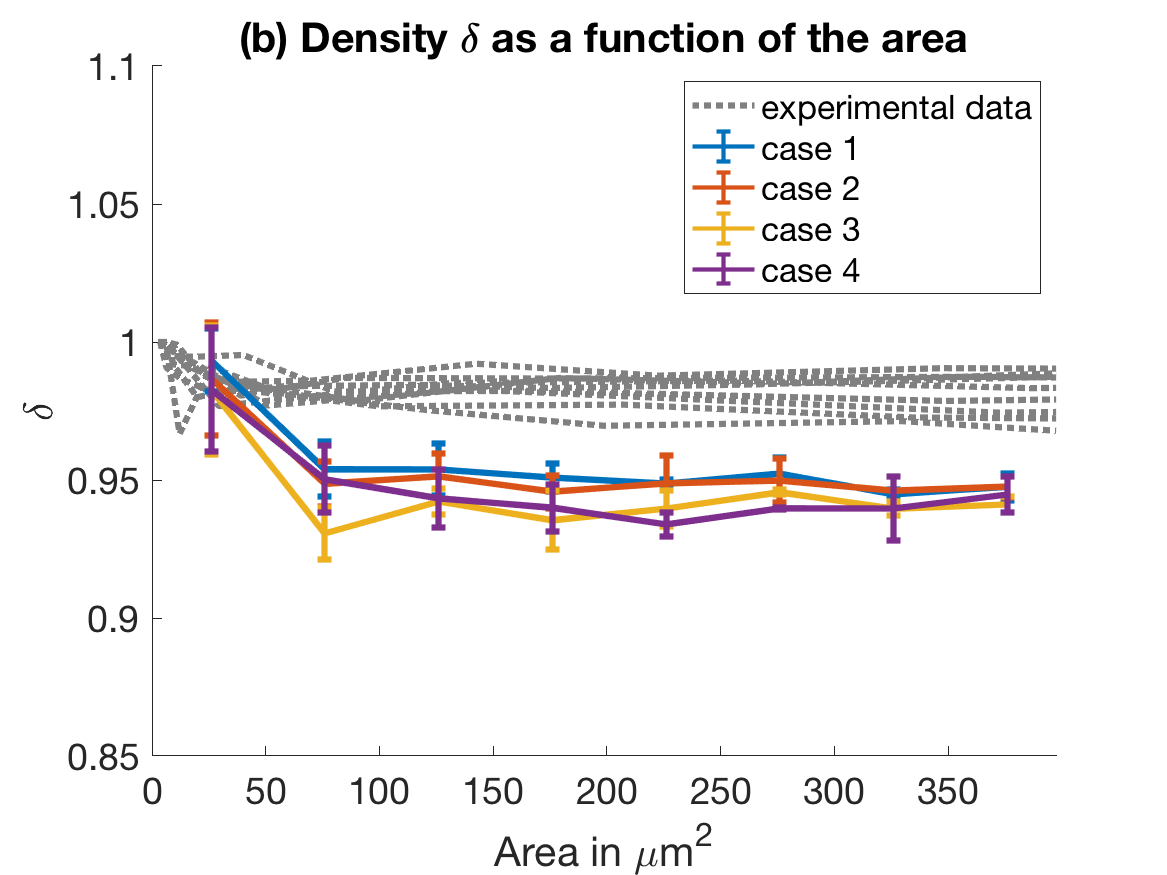}
\includegraphics[scale=0.35]{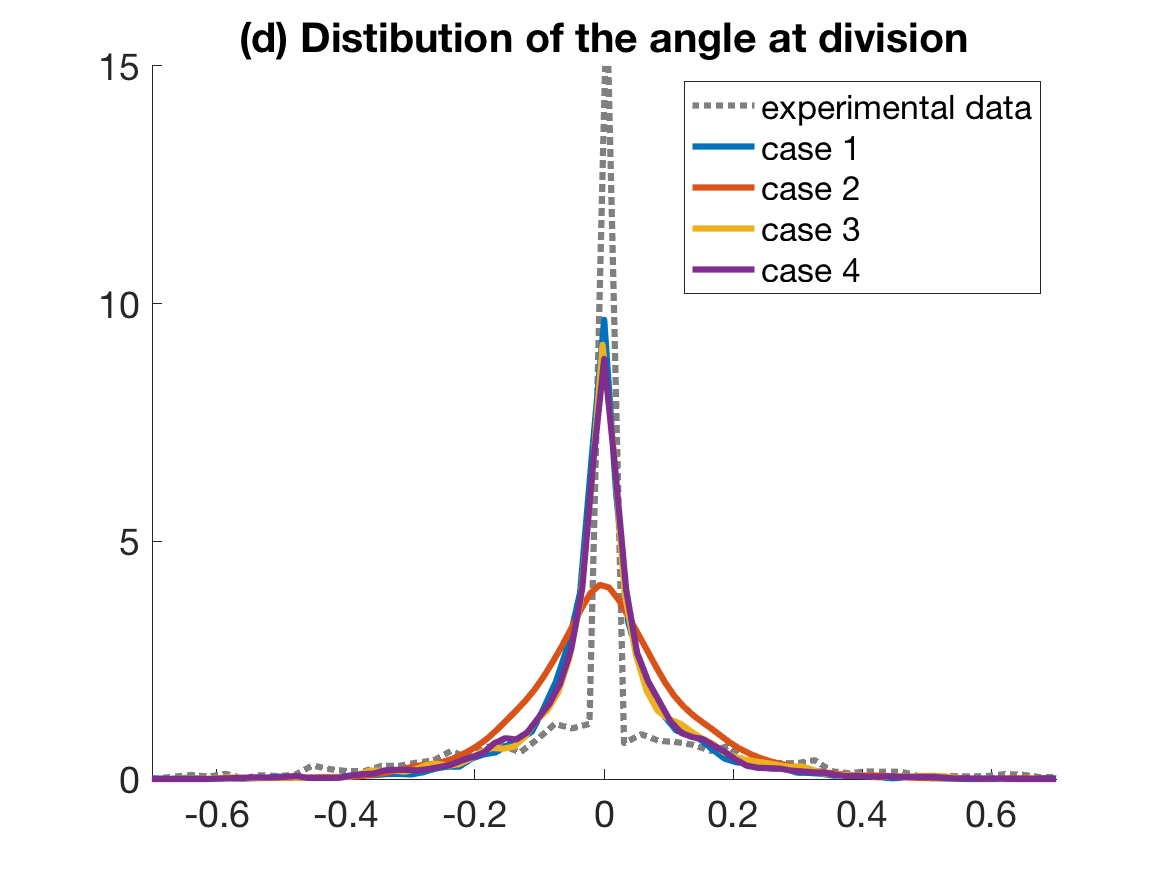}
\caption{Dataset 2: plots of the aspect ratio $\alpha_R$ (a), the local order quantifier $\lambda$ (b) and the density (c) as functions of the area of the colony, and of the distribution of the angle at division (d) for the experimental data (grey dashed curve), and numerical simulations for the case 1 (blue curve), case 2 (red curve), case 3 (yellow curve) and case 4 (purple curve). The plots of the numerical data are average over 10 simulations. 
\label{fig:15}}
\end{figure}

The parameters which have been chosen to fit the experimental data are listed below:
\begin{enumerate}
\item Symmetric friction $A=1$ and uniform mass distribution $\alpha=0.5$, angle at division parameter $\Theta=10^{-5}$,
\item Asymmetric friction $A=0.8$ and uniform mass distribution $\alpha=0.5$, angle at division parameter $\Theta=10^{-1}$,
\item Symmetric friction $A=1$ and asymmetric mass distribution $\alpha=0.9$ with $T_\alpha =19 \mbox{ min}$, angle at division parameter $\Theta= 10^{-5}$,
\item Asymmetric friction $A=0.8$ and asymmetric mass distribution  $\alpha=0.9$ with $T_\alpha =19 \mbox{ min}$,  angle at division parameter $\Theta=10^{-3}$,
\end{enumerate}
In Fig.~\ref{fig:15} we show the evolution of the aspect ratio $\alpha_R$, the local organisation quantifier $\lambda$, the density $\delta$ as functions of the colony area and the distribution of the angle at division. In addition, Table~\ref{Table:6} presents the average values of $d_2$ for the experimental data and numerical simulations. Similarly as for Dataset~1, we observe a large variability in the values of the aspect ratio $\alpha_R$ and the local organisation quantifier $\lambda$ for the experimental data (see the grey curves of Fig.~\ref{fig:15} Panels~(a) and~(b)). Moreover, Panel~(a) shows that the colony of \textit{pseudomonas} (Dataset~2) are less elongated than the one of \textit{E. coli} (Datasets~1 and~3). Indeed on Fig.~\ref{fig:15} the aspect ratio $\alpha_R$ takes values between $0.2$ and $0.8$ while in Fig.~\ref{fig:13} its values are between $0.2$ and $0.5$. The \textit{pseudomonas} colonies are also less organised, with the local order parameter $\lambda$ taking values down to $0.65$ (compared to $0.75$ for Dataset~1). Note that the difference of shape between the \textit{pseudomonas} colonies and the \textit{E. coli} colonies explains why  the value of $A$ considered for this dataset is closer to $1$ than for Dataset~1:  the colonies being less elongated, we do not need to consider a strong asymmetry in the friction. 

\begin{table}[h!]
\centering
\begin{tabular}{ | l l l l | } 
\hline
Dataset 2 & average of $d_2$ & minimum of $d_2$ & maximum of $d_2$ \\ 
\hline
\hline
Experimental data &  0.4288    &     0  &  0.8062  \\ 
\hline
Case 1 &  0.999890738047451 &  0.998927231703967  & 0.999999995072841 \\ 
\hline
Case 2  & 0.908447782653370 &  0.633902849679844  & 0.998085101017730 \\ 
\hline
Case 3  & 0.924943407198252 &  0.652822846580324 &  0.998015220844816 \\ 
\hline
Case 4  & 0.878801675567598 &  0.547594782220091 &  0.952911252992901 \\ 
\hline
\end{tabular}
\caption{\label{Table:6}Four-cell array quantifier $d_2$: comparison of the experimental dataset 2 with the four parameter choice cases.
}
\end{table}

Fig.~\ref{fig:15} (a) shows that, except for colonies of area smaller than $50 \mu m^2$, the numerical colonies aspect ratio for the four cases are between the bounds of the experimental data. The difference at early stage might be due to the averages made. Panel~(c) also shows that the  local organisation of the four cases is acceptable. The density of the experimental colony observed in Panel~(b) of Fig.~\ref{fig:15} takes values higher than  for Dataset~1 (see Fig.~\ref{fig:13} (c)). This might be due to the different bacteria considered in the two datasets. Also, Fig.~\ref{fig:15} (c) shows that the numerical colonies are not as dense as the experimental ones; as discussed earlier, this might also be an artefact of segmentation. Looking at the angle at division $\Theta$ in Panel~(d), we observe that none of the numerical distributions reaches the peak of the experimental one. However their spread is qualitatively alike the experimental one. Finally Table~\ref{Table:6} shows that the values of $d_2$ taken by the experimental colonies is much smaller than for Dataset~1 (see Table~\ref{Table:5}). Likewise, the values taken by the numerical colonies is smaller for Dataset~2, going down to $0.87$ (compared to $0.95$ for Dataset~1). Therefore these differences might be due to the shape of the bacteria. Nevertheless the diminution of $d_2$ observed in the numerical data is not enough to reach the experimental value $0.42$. The smallest values of $d_2$ are taken for the case 4, followed by the case 3. We conclude that the best fit for Dataset~2 is obtained for the case 4 where an asymmetric distribution of mass is considered along with a small asymmetry in the friction. Nonetheless, the case 2 with only asymmetry for the friction can be considered as a good fit.

\begin{figure}[!ht] 
   \centering
\includegraphics[scale=0.17]{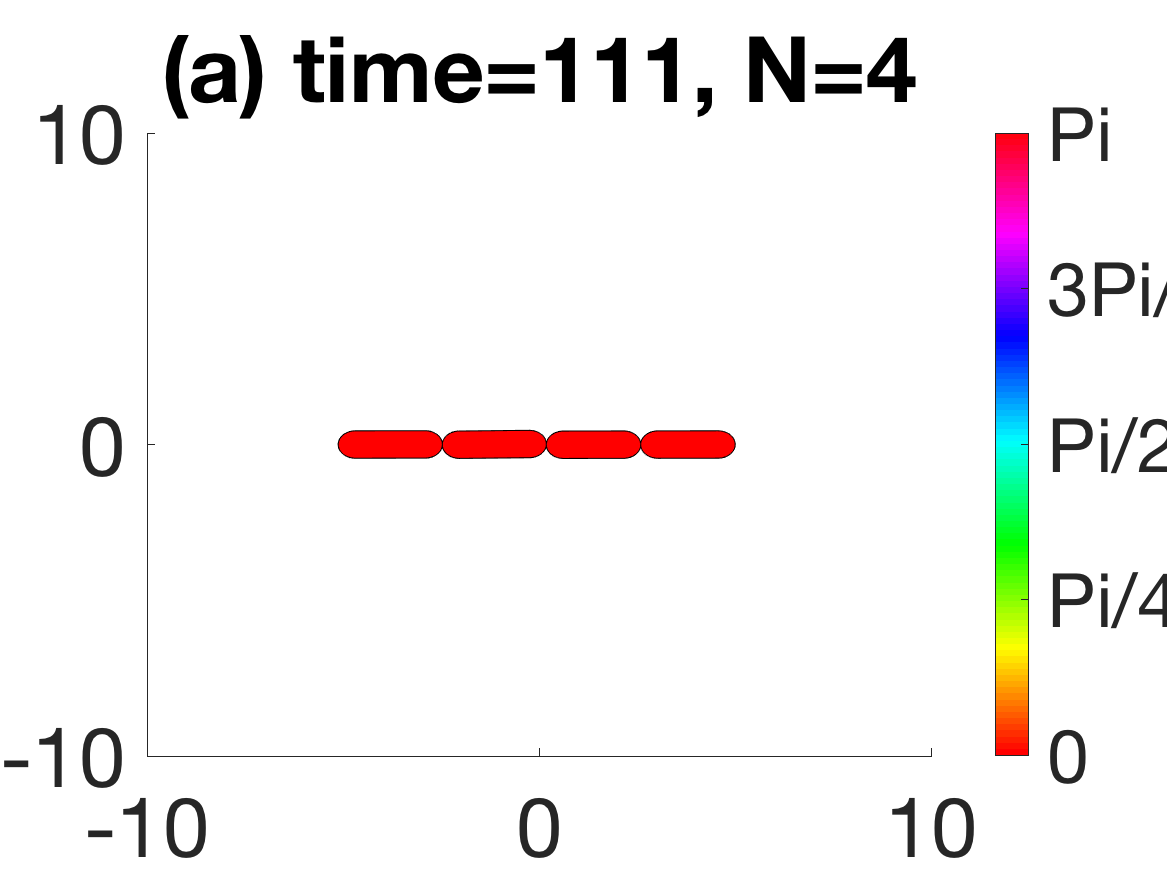}
\includegraphics[scale=0.17]{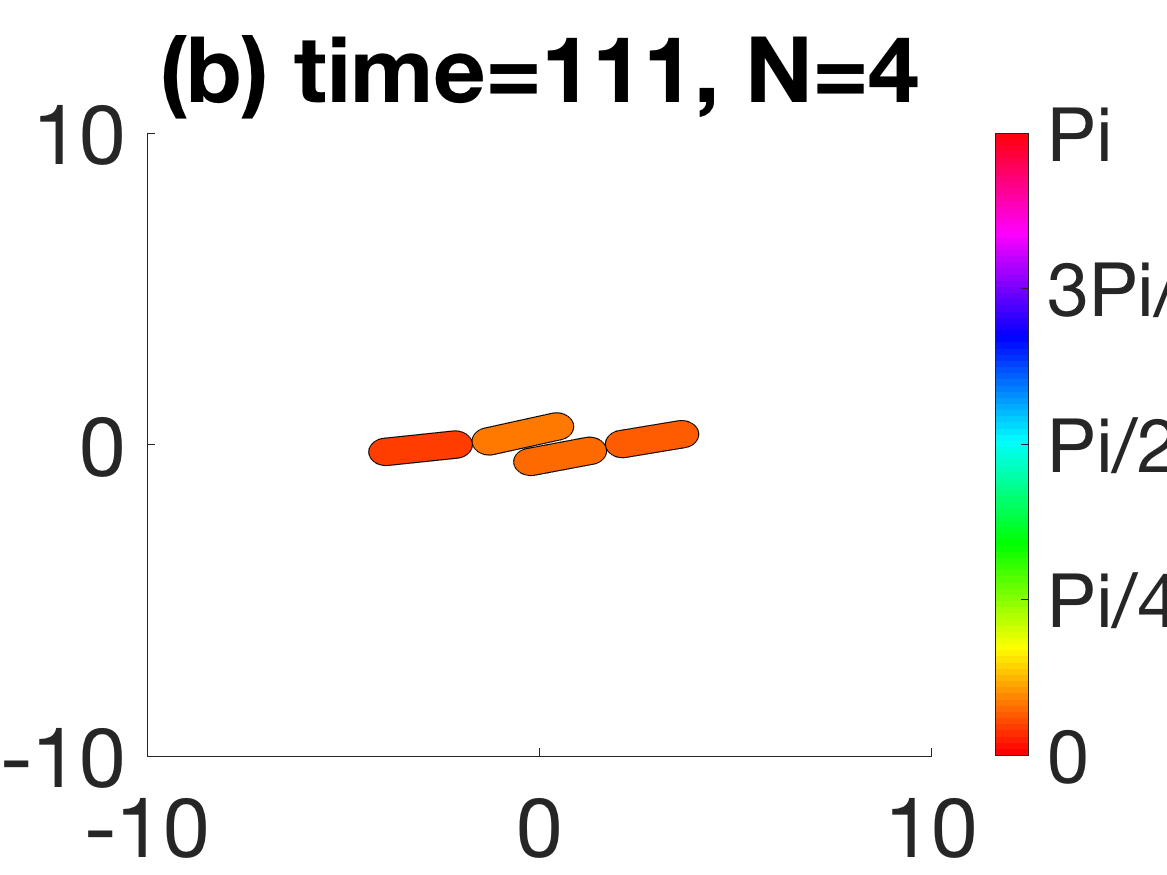}
\includegraphics[scale=0.17]{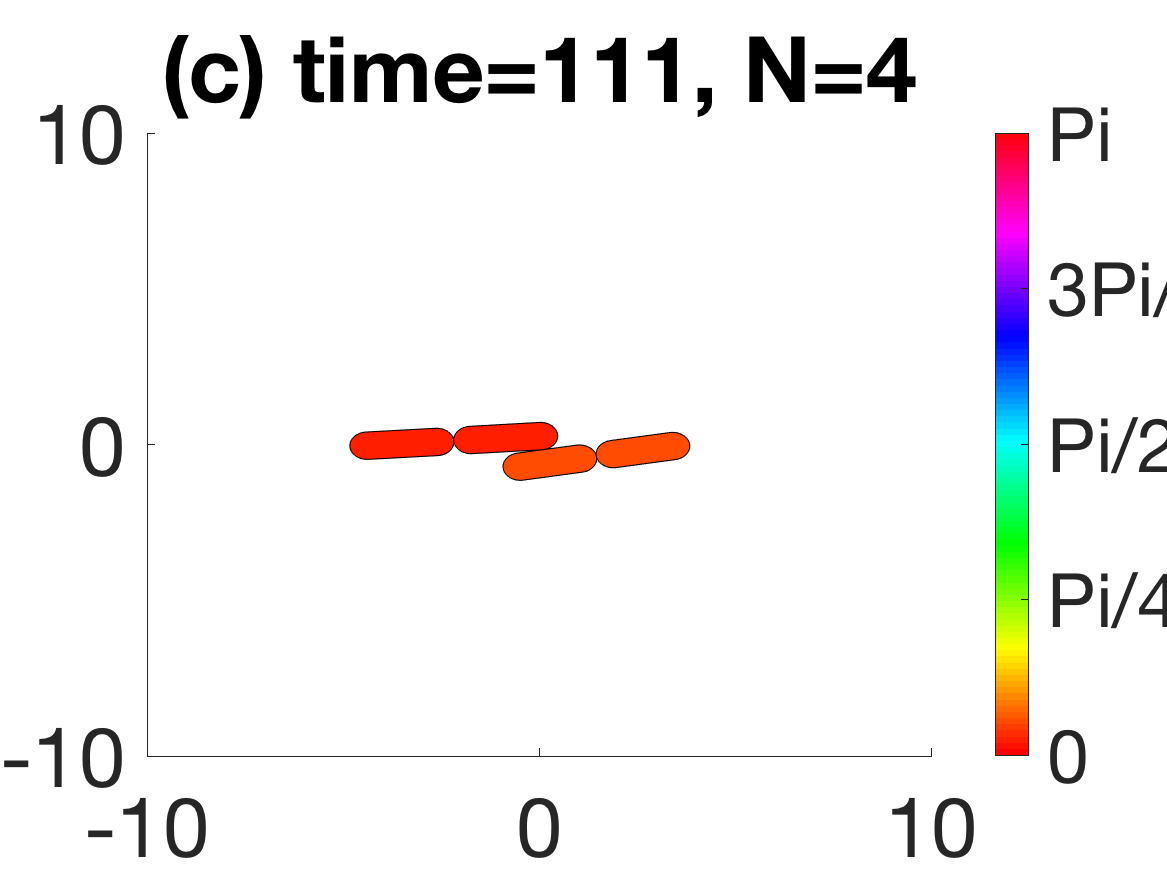}
\includegraphics[scale=0.17]{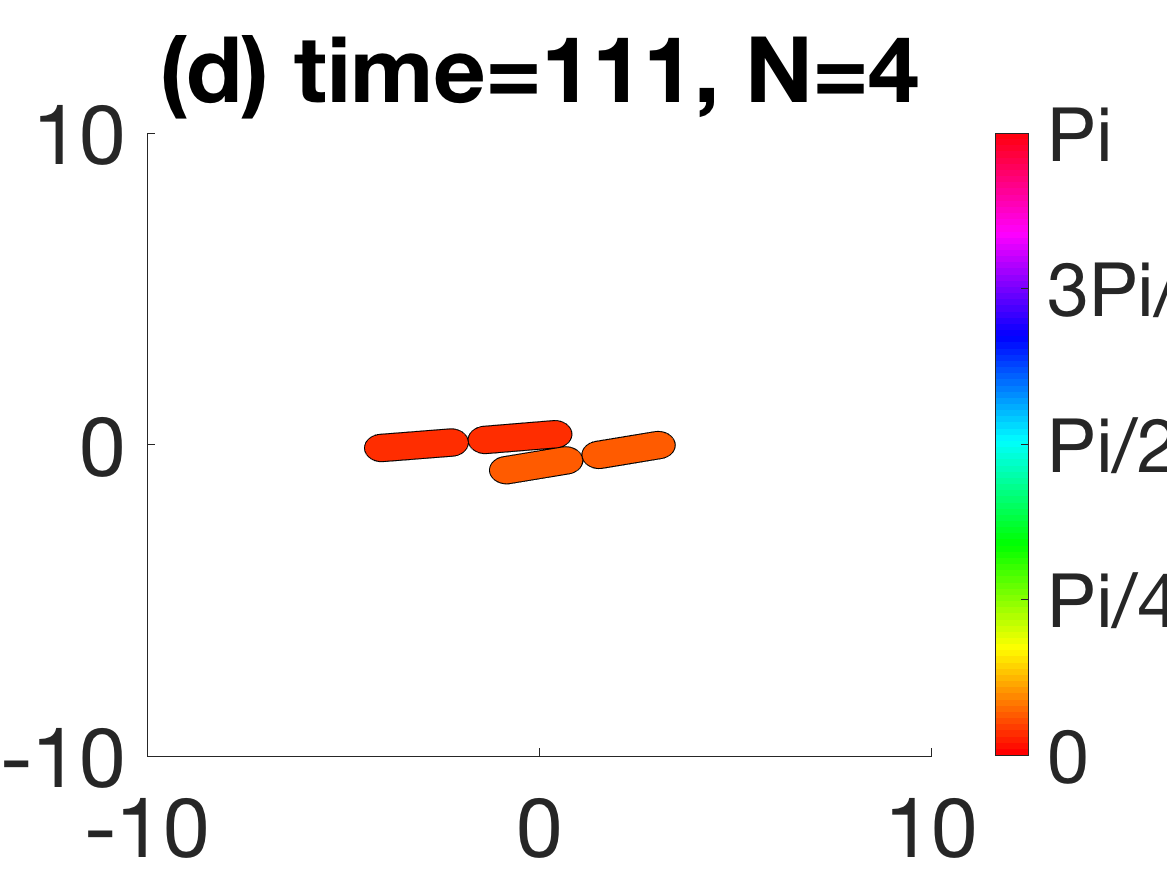}
\caption{Dataset 2: Plot of simulation at time $t=111   \mbox{min}$ for Case 1 (a), Case 2 (b), Case 3 (c) and Case 4 (d). The colors of the bacteria are given by their orientation.  These figures can be compare to Fig.~\ref{fig:7}(1) Panel~(b).}
\label{fig:16_1}
\end{figure}

\begin{figure}[!ht] 
   \centering
\includegraphics[scale=0.35]{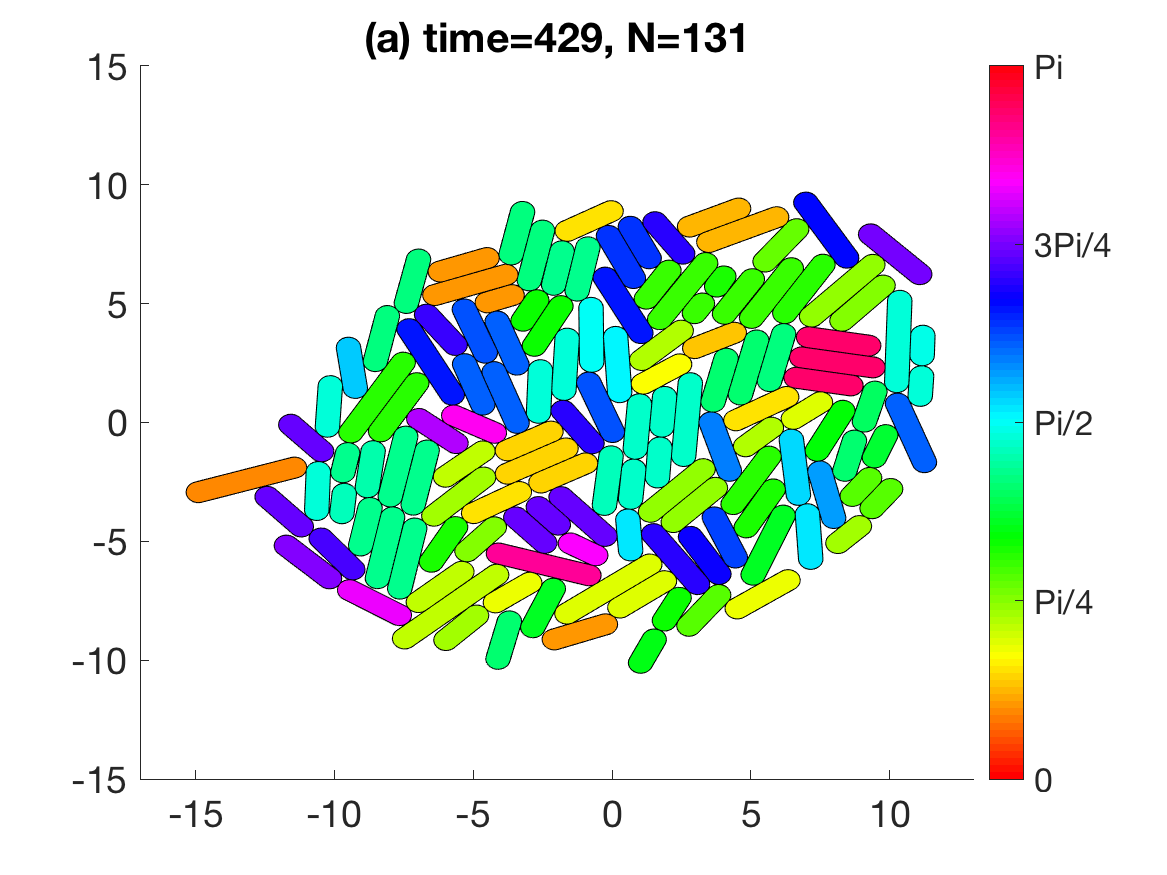}
\includegraphics[scale=0.35]{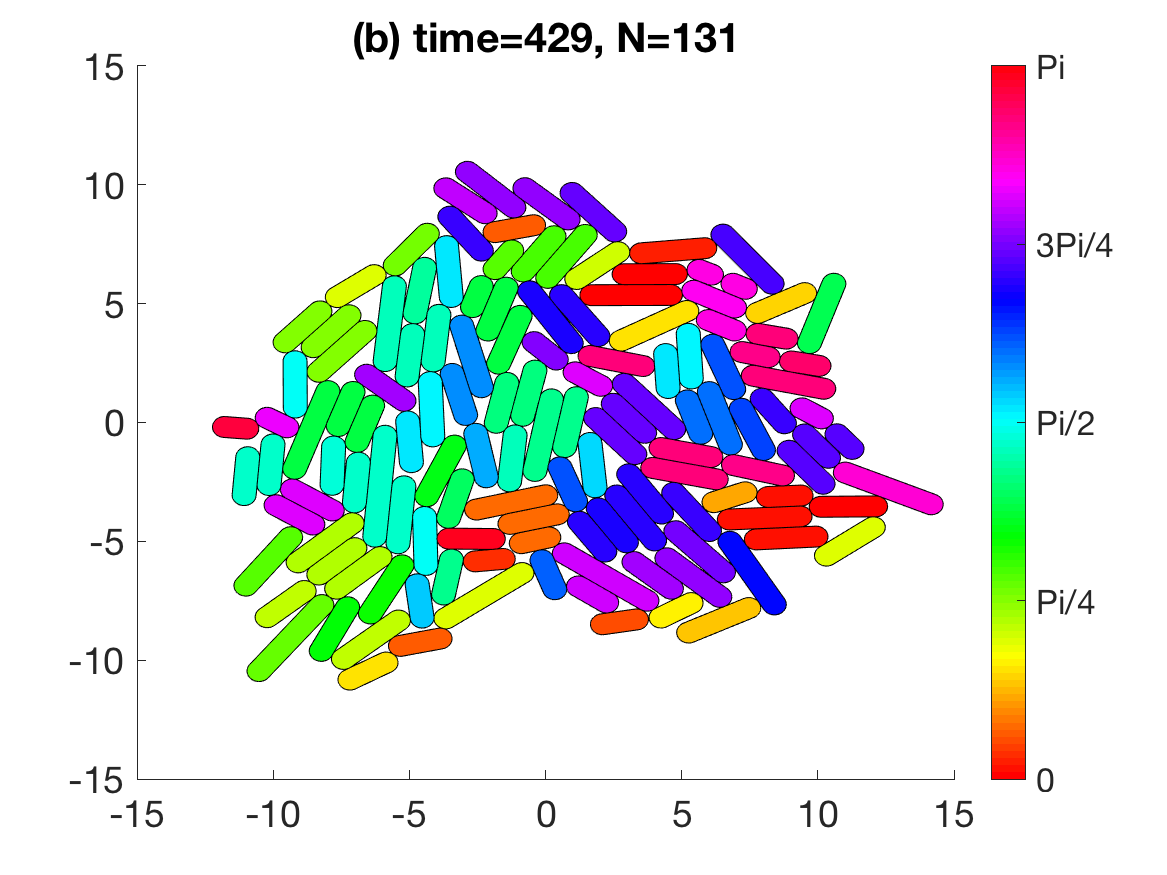}
\includegraphics[scale=0.35]{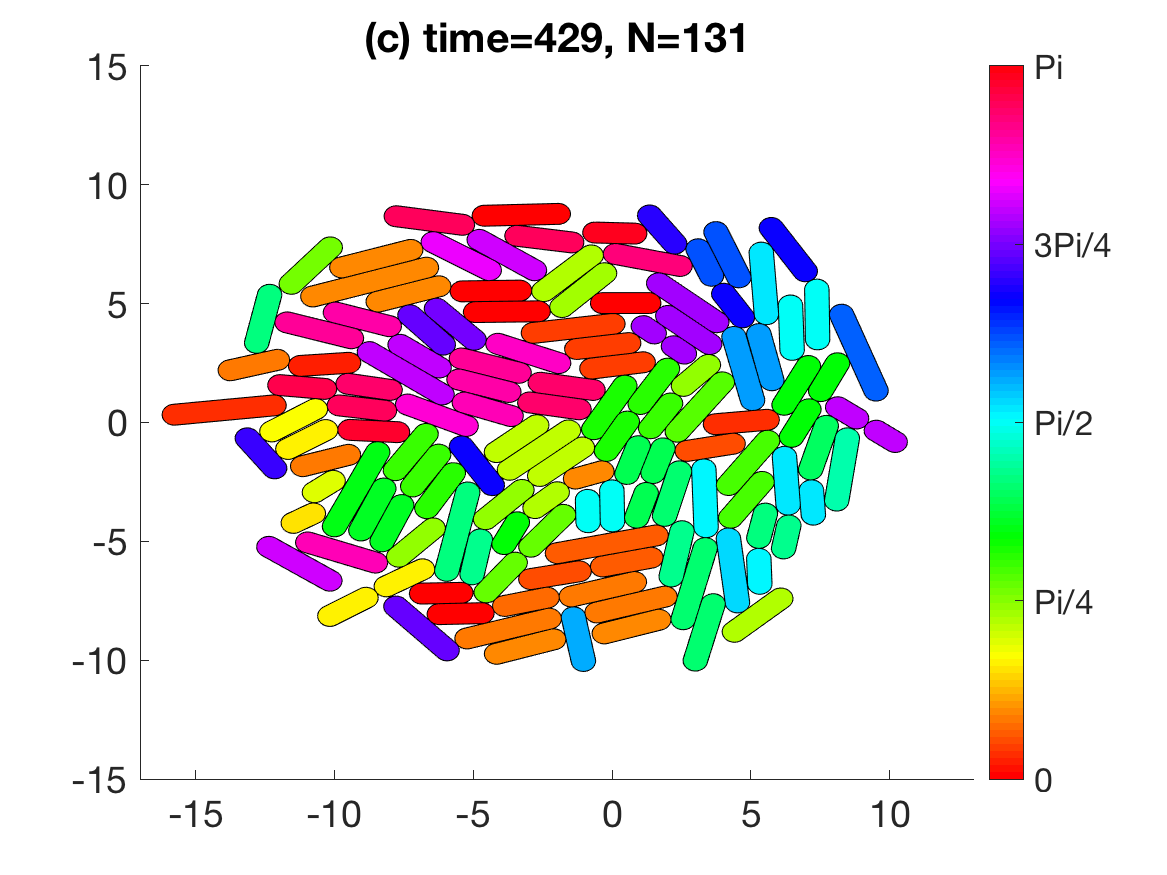}
\includegraphics[scale=0.35]{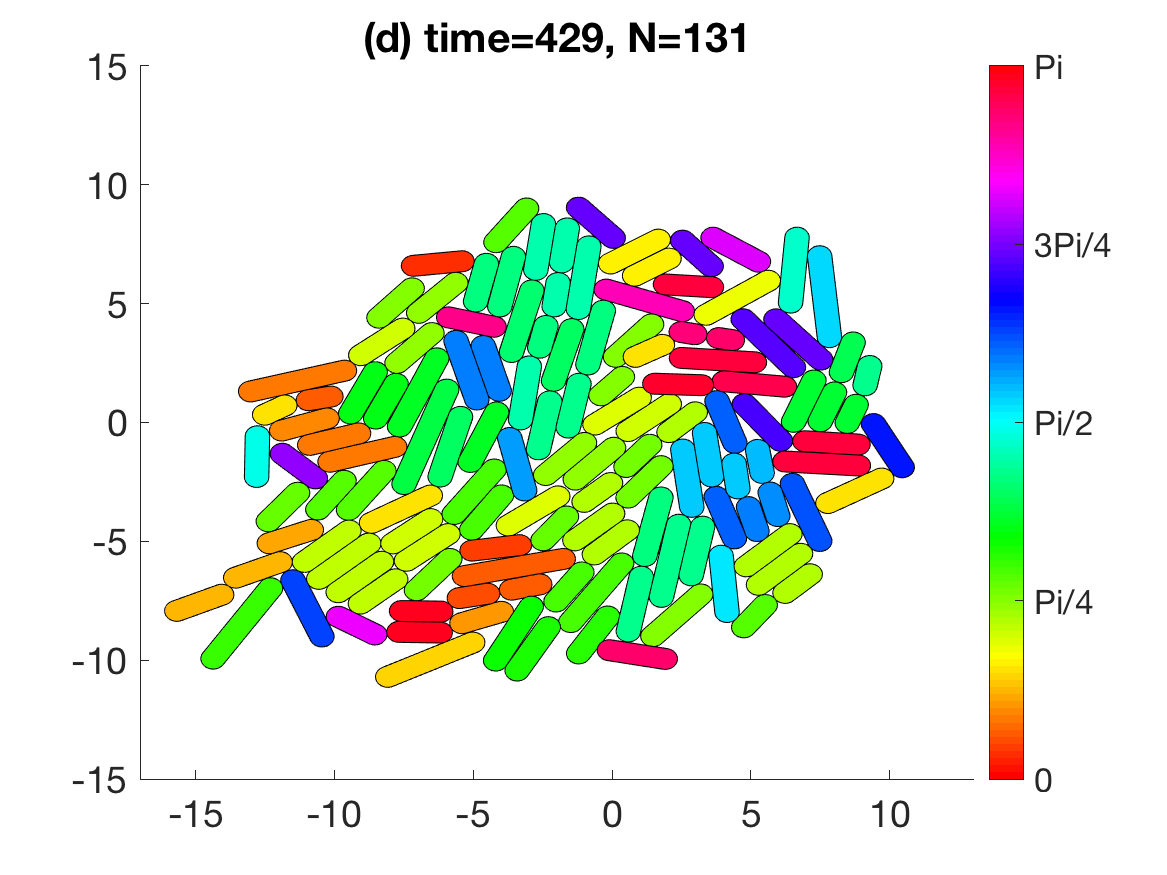}
\caption{Dataset 2: Plot of simulation at time $t=200  \mbox{min}$ for Case 1 (a), Case 2 (b), Case 3 (c) and Case 4 (d). The colors of the bacteria are given by their orientation. These figure can be compared to Fig.~\ref{fig:7}(3) Panel~(b).}
\label{fig:16_2}
\end{figure}

 In Figs.~\ref{fig:16_1} and \ref{fig:16_2} we present plots of the colonies for the cases 1, 2, 3 and 4 at after $111$ and $429$ minutes respectively. Figs.~\ref{fig:16_1} shows the four-cell arrangement of the colony and can be compare to Fig.~\ref{fig:7} (b). We observe that in the numerical colony the best case is for Panel~(d) (case 4) and is not as good as for the experimental colony. This supports the results presented in Table~\ref{Table:6}. In fig.~\ref{fig:16_2} the time has been chosen so that the number of bacteria in the colonies is close to $130$, which is similar to the number of bacteria present in the plots \ref{fig:7} (b). The experimental colony in \ref{fig:7} (b) presents some triangular features. The only colony in Fig.~\ref{fig:16_2} which could have a similar shapes is in Panel~(b). It corresponds to the case 2, which contradicts the observations made with the quantifiers previously. Therefore, these plots show that there might be missing features in the model combining the asymmetric friction and mass distribution without counteracting each other effect.

\subsection{Dataset 3} \label{Section4.3}

\begin{figure}[!ht] 
   \centering
\includegraphics[scale=0.35]{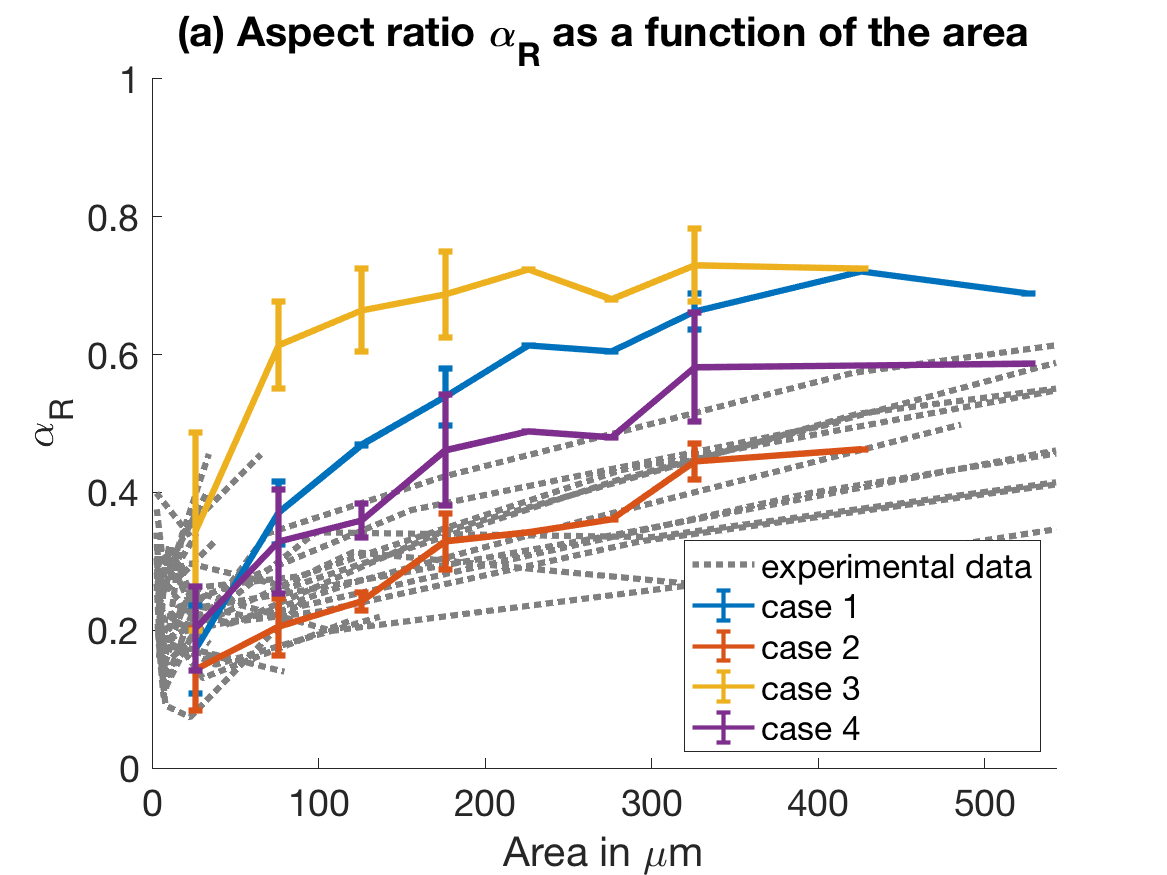}
\includegraphics[scale=0.35]{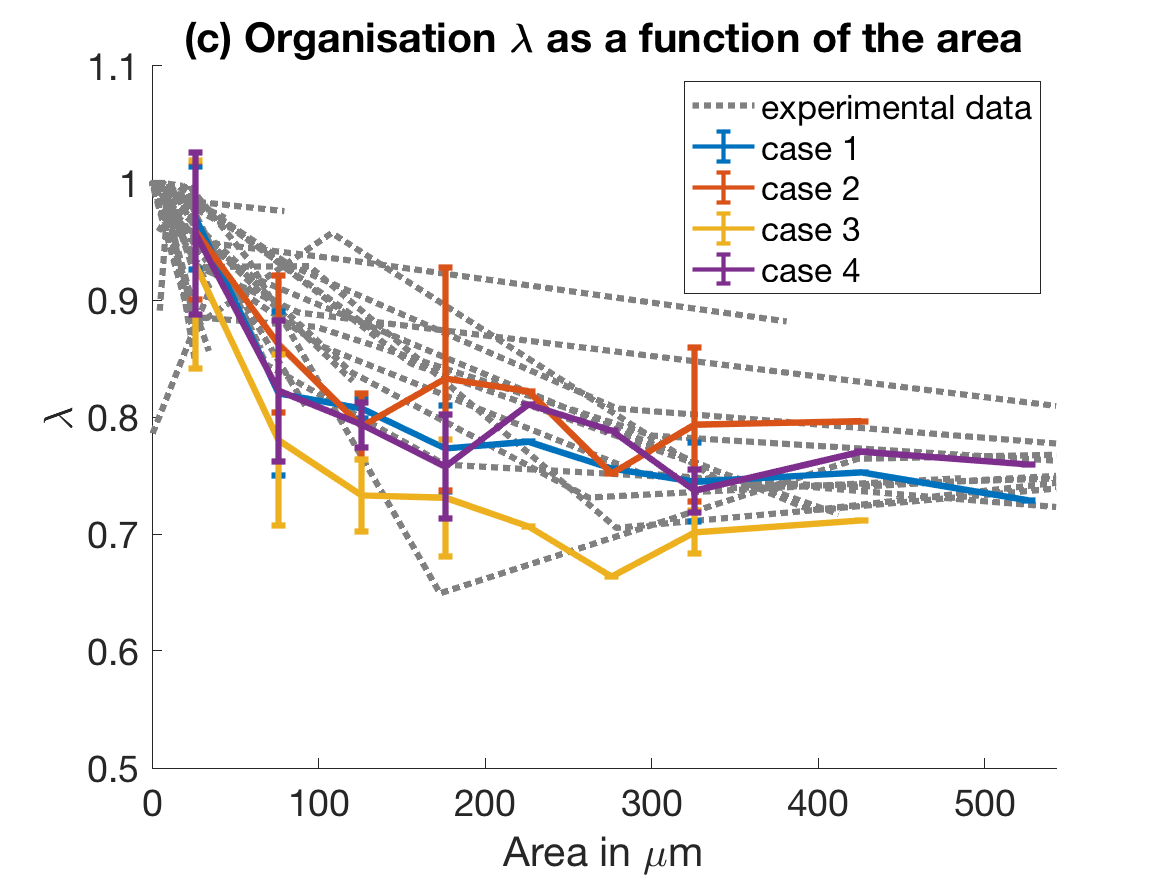}
\includegraphics[scale=0.35]{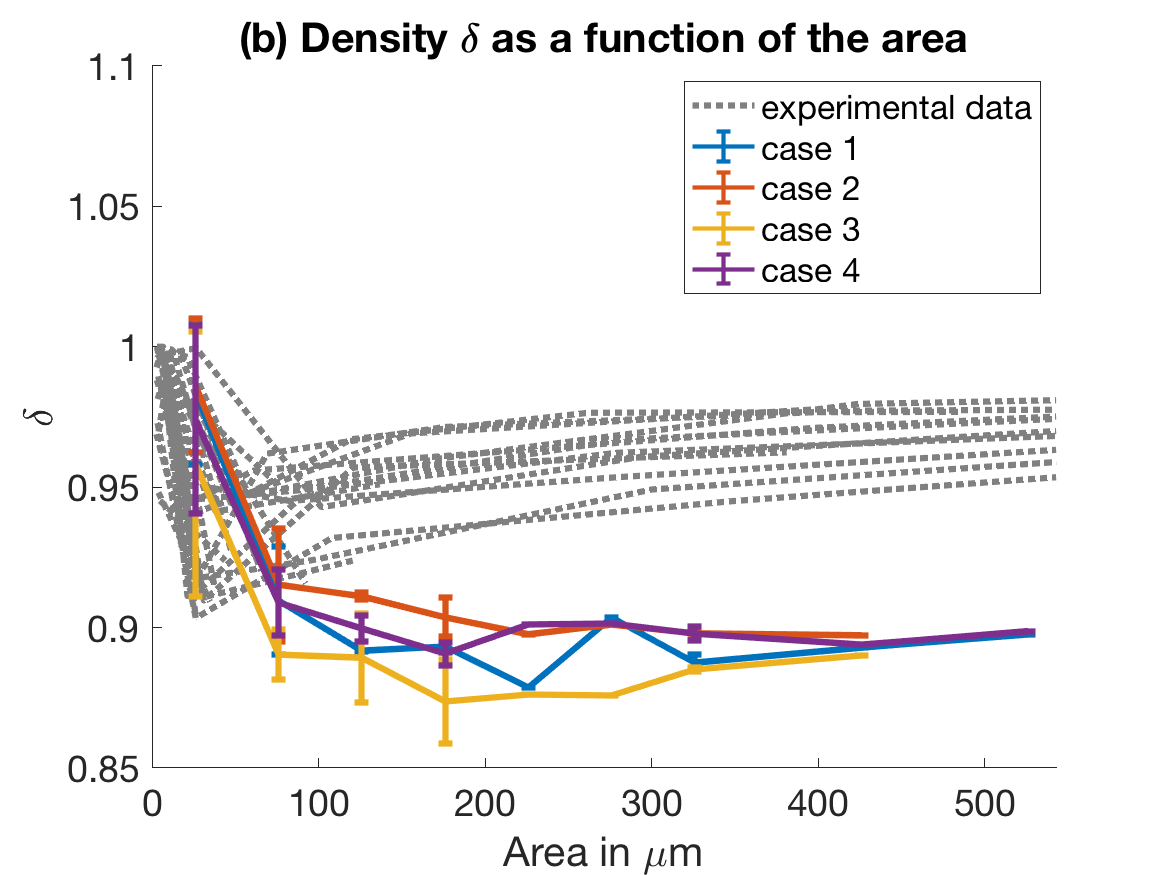}
\caption{Dataset 3: plots of the aspect ratio $\alpha_R$ (a), the local order quantifier $\lambda$ (b) and the density (c) as functions of the area of the colony for the experimental data (grey dashed curve), and numerical simulations for the case 1 (blue curve), case 2 (red curve), case 3 (yellow curve) and case 4 (purple curve). The plots of the numerical data are averaged over 5 simulations.}
\label{fig:17}
\end{figure}

\begin{figure}[!ht] 
   \centering
\includegraphics[scale=0.17]{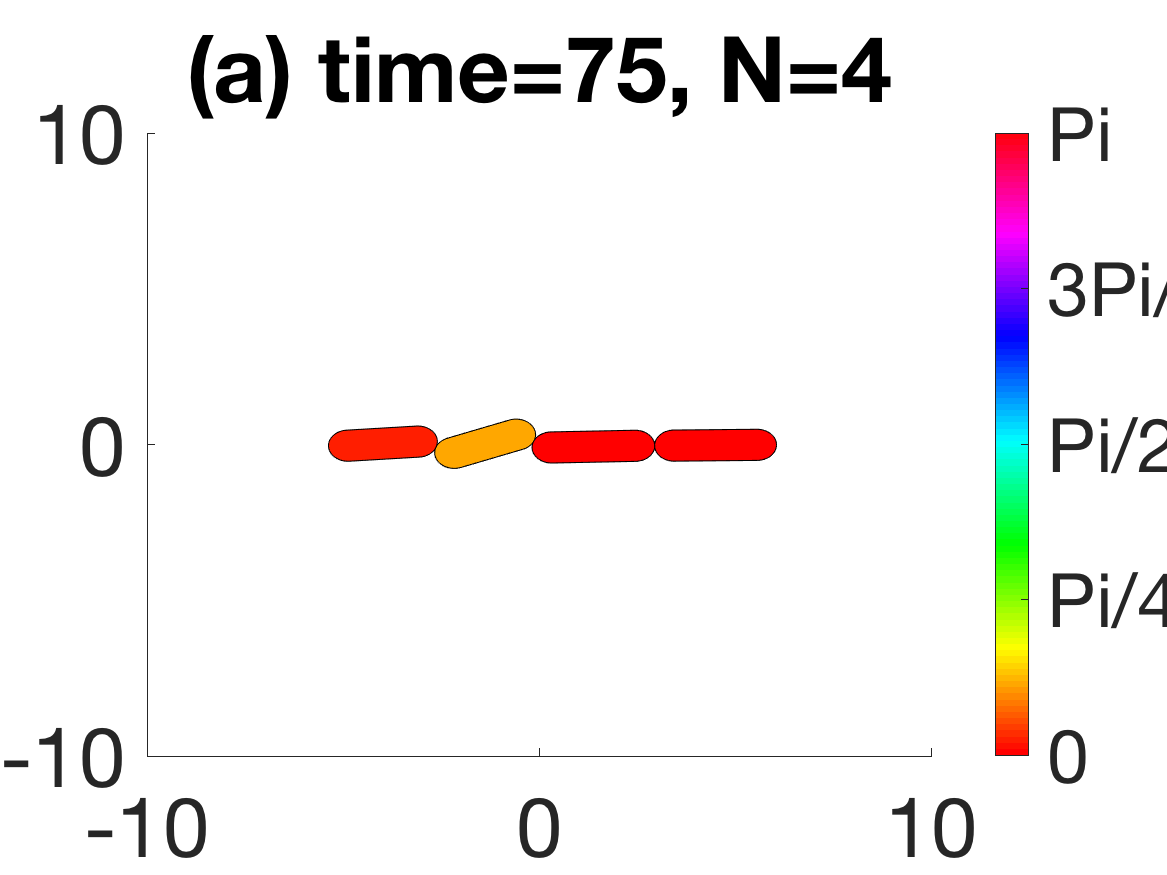}
\includegraphics[scale=0.17]{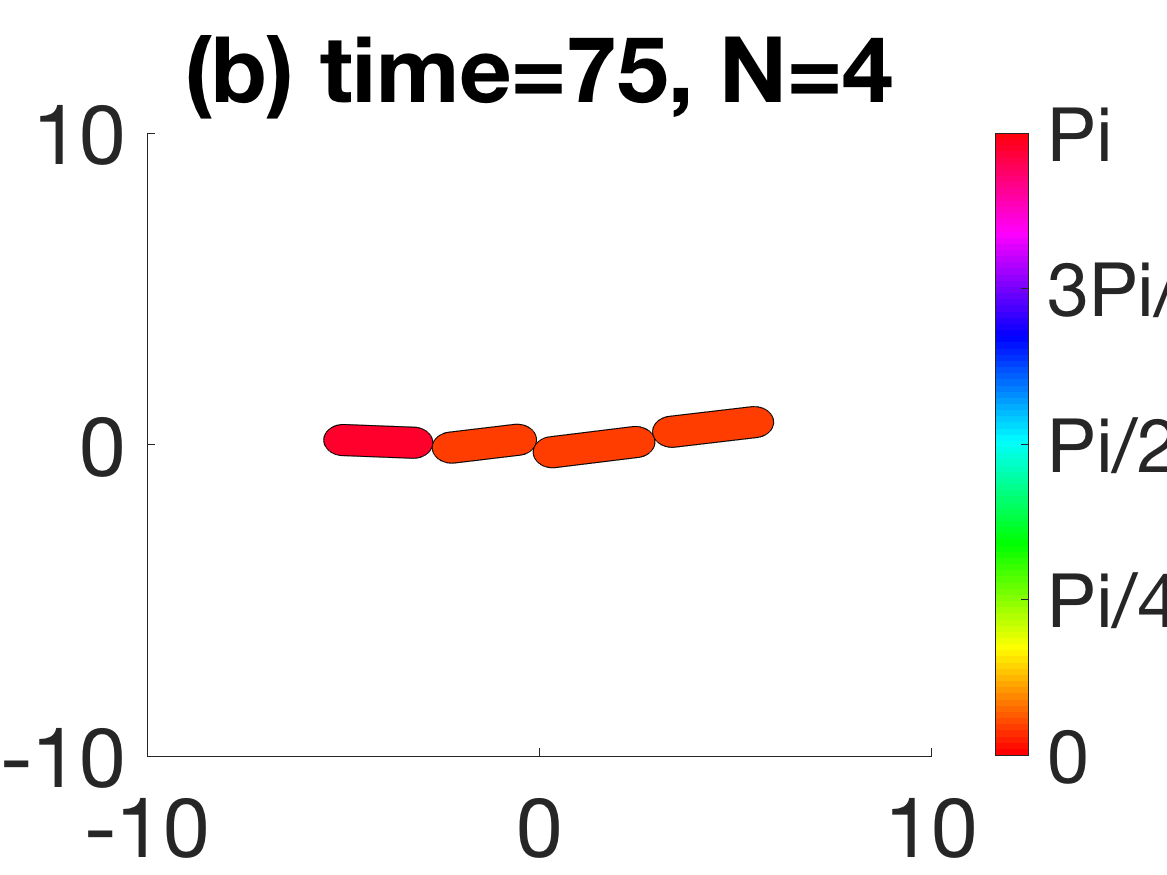}
\includegraphics[scale=0.17]{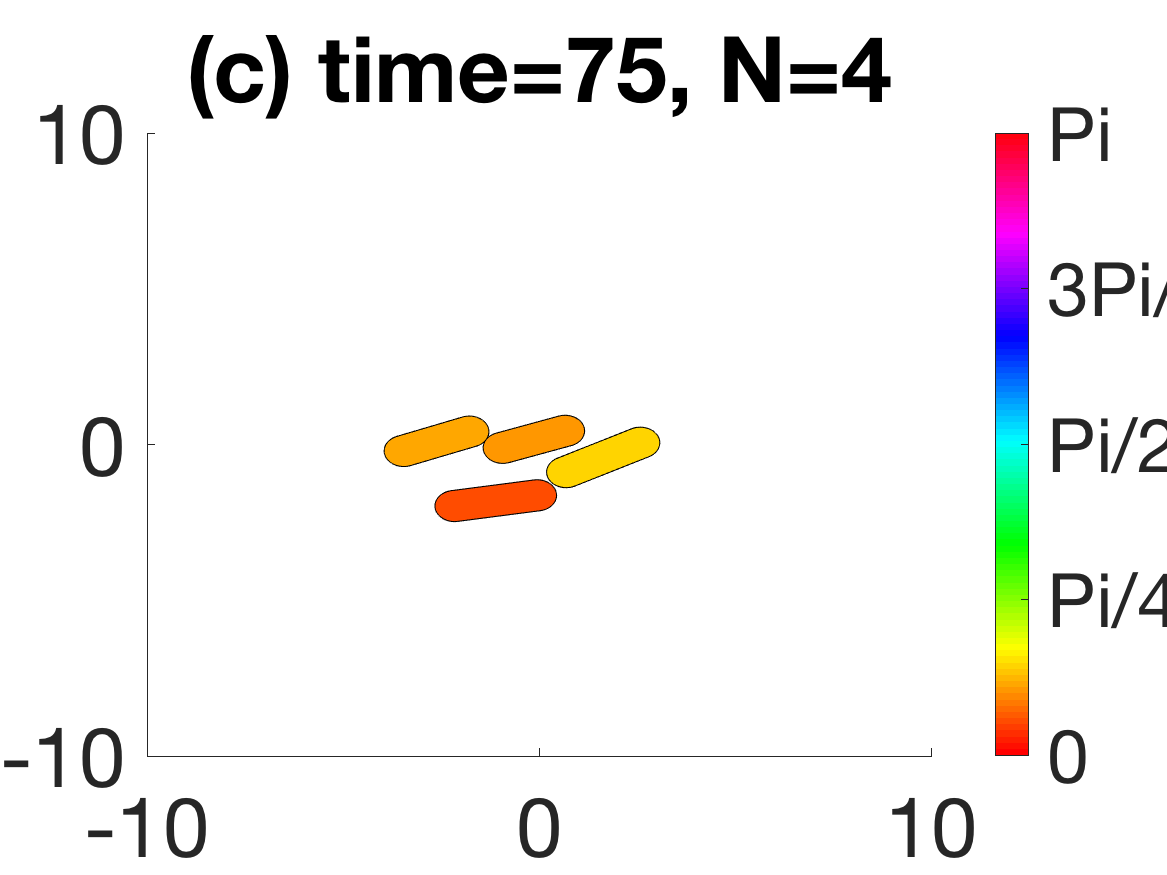}
\includegraphics[scale=0.17]{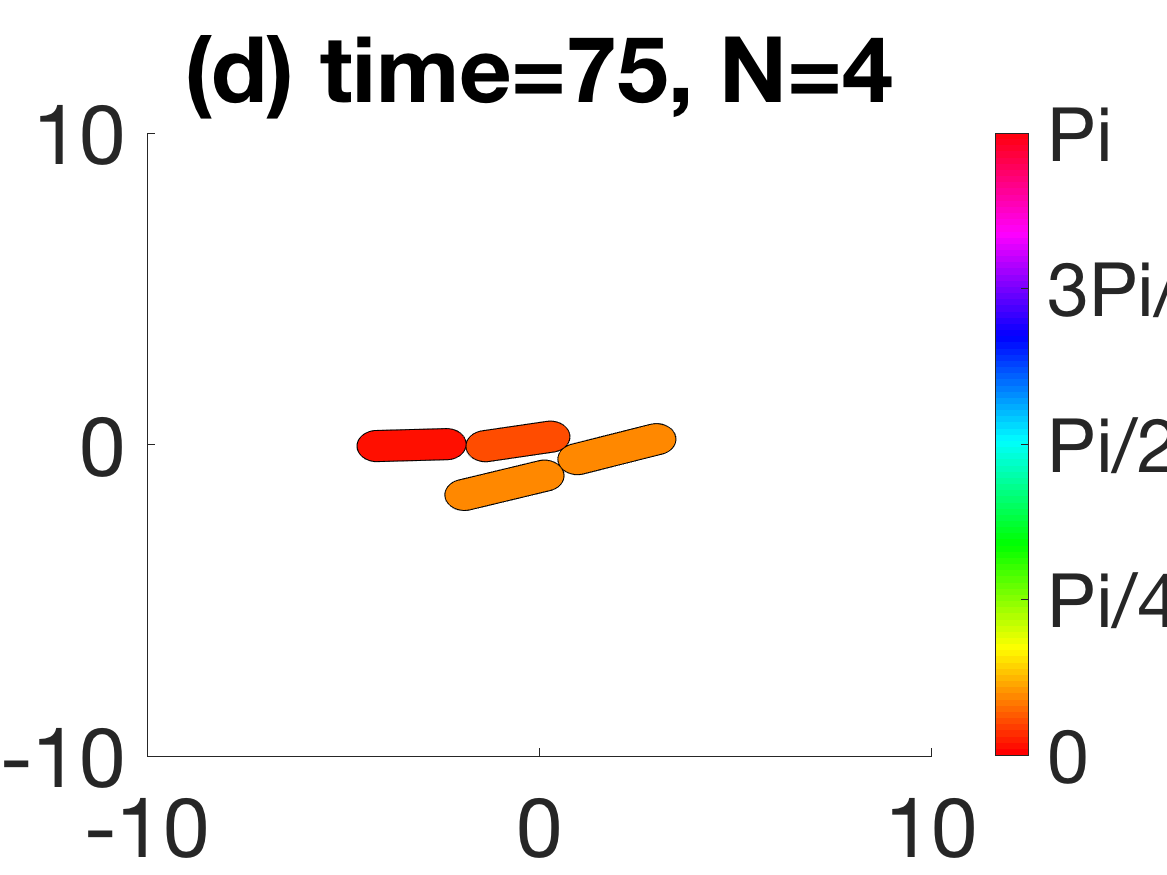}
\caption{Dataset 3: Plot of simulation at time $t=111   \mbox{min}$ for Case 1 (a), Case 2 (b), Case 3 (c) and Case 4 (d). The colors of the bacteria are given by their orientation.  These figure can be compare to Fig.~\ref{fig:7} Panel~(c).}
\label{fig:18_1}
\end{figure}

\begin{figure}[!ht] 
   \centering
\includegraphics[scale=0.35]{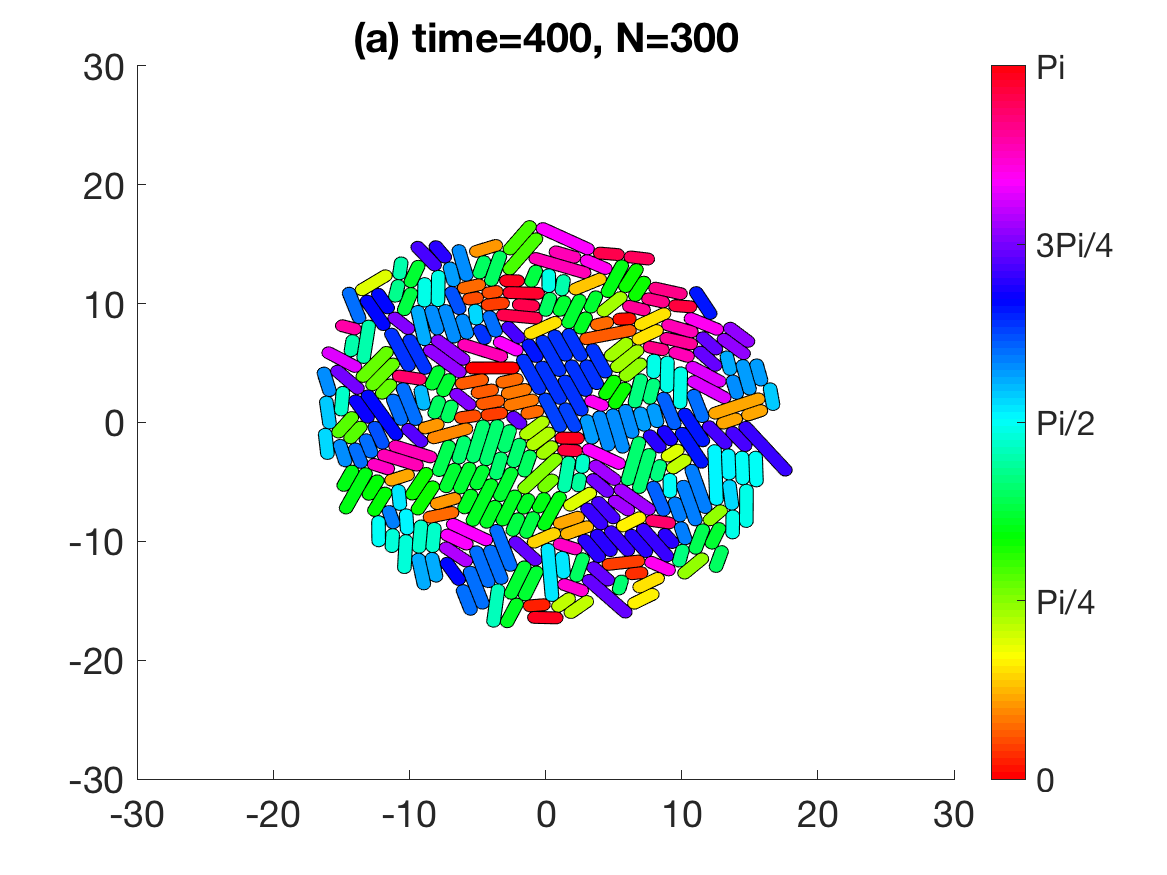}
\includegraphics[scale=0.35]{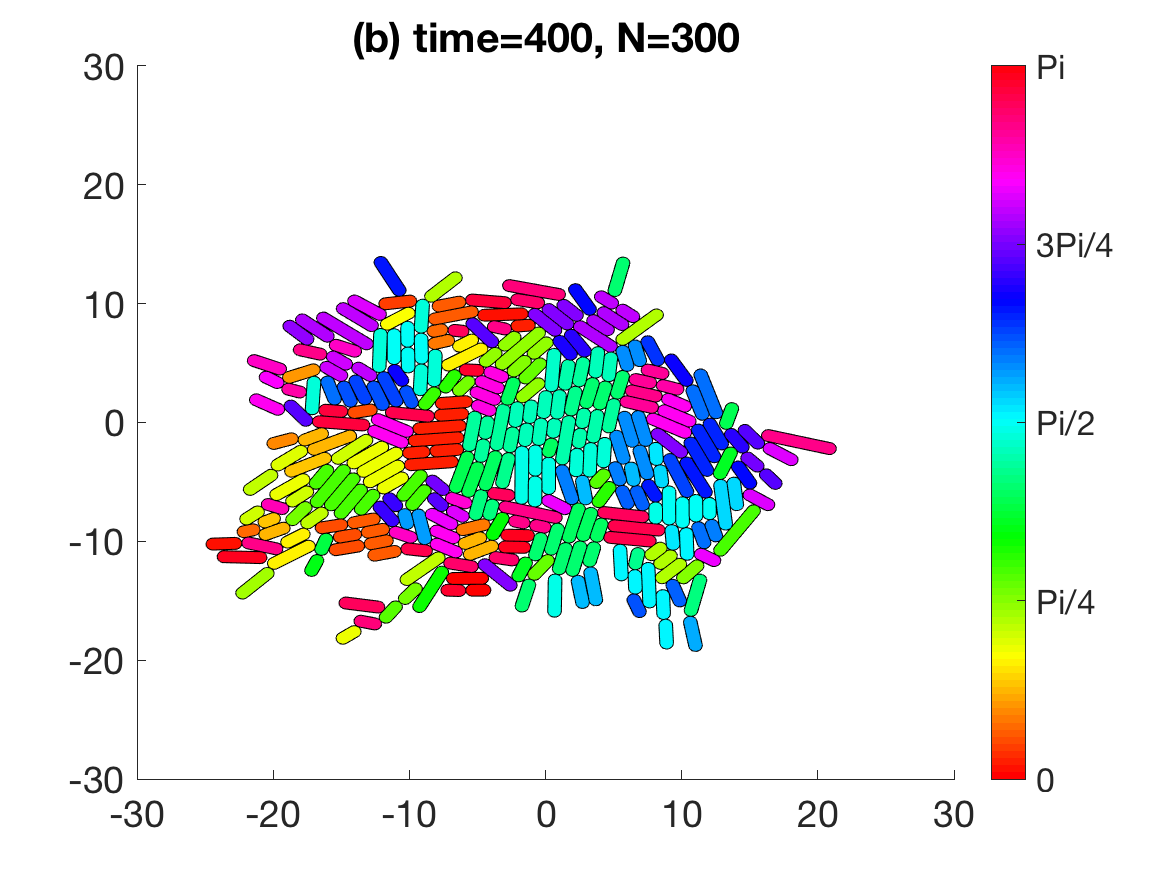}
\includegraphics[scale=0.35]{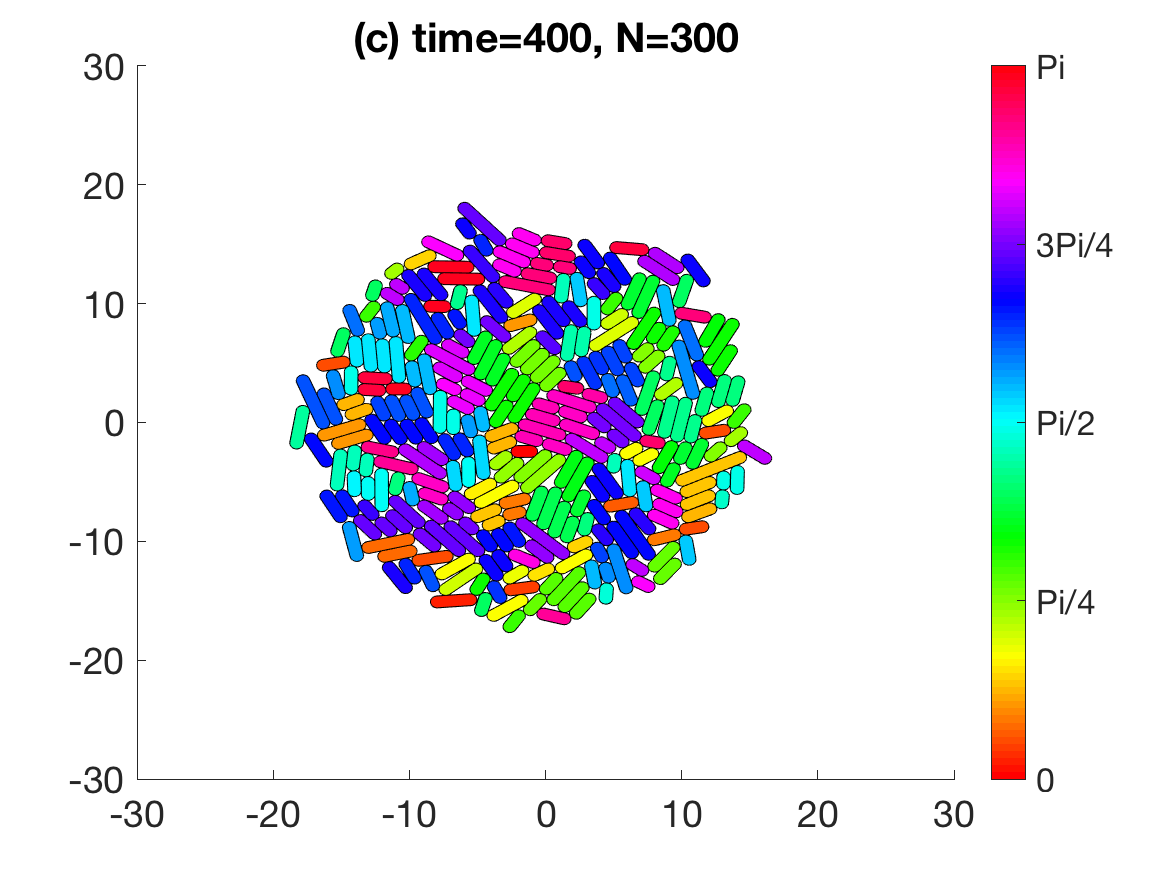}
\includegraphics[scale=0.35]{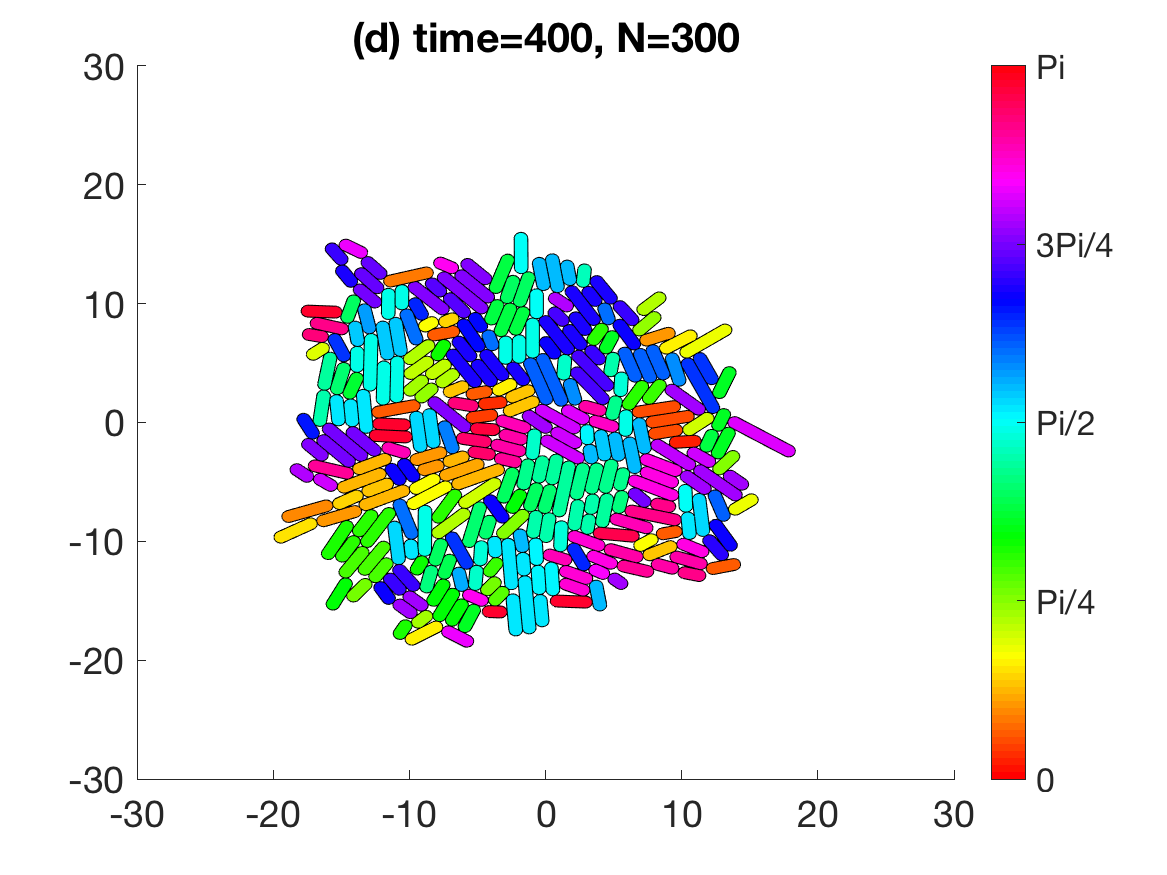}
\caption{Dataset 3: Plot of simulation at time $t=200  \mbox{min}$ for Case 1 (a), Case 2 (b), Case 3 (c) and Case 4 (d). The colors of the bacteria are given by their orientation. These figures can be compared to Panel~(c) of Fig.~\ref{fig:7}.}
\label{fig:18_2}
\end{figure}

The parameters which have been chosen to fit the experimental data are listed below:
\begin{enumerate}
\item Symmetric friction $A=1$ and uniform mass distribution $\alpha=0.5$, angle at division parameter $\Theta=10^{-5}$,
\item Asymmetric friction $A=0.4$ and uniform mass distribution $\alpha=0.5$, angle at division parameter $\Theta=10^{-1}$,
\item Symmetric friction $A=1$ and asymmetric mass distribution $\alpha>0.9$ with $T_\alpha =13 \mbox{ min}$, angle at division parameter $\Theta= 10^{-5}$,
\item Asymmetric friction $A=0.6$ and asymmetric mass distribution  $\alpha>0.9$ with $T_\alpha =13 \mbox{ min}$,  angle at division parameter $\Theta=10^{-3}$,
\end{enumerate}
As for Dataset 1, Dataset 3 corresponds to \textit{E. coli} colonies but in different experimental conditions. We recall that in this dataset, the segmentation of the bacterial colonies presented some issue and therefore some data have been ignored. In addition, the data do not give access to the mother/daughter link and therefore it is not possible to consider the angle at division. Therefore, in Fig.~\ref{fig:17}  we only show the evolution of the aspect ratio $\alpha_R$, the local organisation quantifier $\lambda$,and  the density $\delta$ as functions of the colony area. The grey curves correspond again to the values of the quantifiers computed on the experimental data. Besides, Table~\ref{Table:7} presents the average values of $d_2$ for the experimental data and numerical simulations. 

\begin{table}[h!]
\centering
\begin{tabular}{ | l l l l| } 
\hline
Dataset 3 & average of $d_2$ & minimum of $d_2$ & maximum of $d_2$ \\ 
\hline
\hline
Experimental data &  0.543964112373229  & 0.039986971063898  & 0.999803179785625 \\ 
\hline
Case 1 &  0.999999837839351  & 0.999999441519551 &  0.999999998925162 \\ 
\hline
Case 2  & 0.993090800524502 &  0.984636756967105 &  0.999848938964758 \\ 
\hline
Case 3  & 0.834388133145832 &  0.700129206144369 &  0.913543326468035 \\ 
\hline
Case 4  & 0.937676566731141  &  0.848372616838960  & 0.984599839318571 \\ 
\hline
\end{tabular}
\caption{Four-cell array quantifier $d_2$: comparison of the experimental dataset~3 with the four parameter choice cases.\label{Table:7}
}
\end{table}

Fig.~\ref{fig:17} (a) shows that the  values of the aspect ratio $\alpha_R$ in cases 1 and 3 is largely above the experimental ones, showing that for these cases the colonies are way more spherical than in the experiments. On the contrary,  the case 2 seems to be a reasonably good fit of the experiments in terms of aspect ratio of the colonies, while case 4 is slightly above the real data. However, we consider the case 4 to be also an acceptable set of parameters as the aspect ratio follows one of the experimental curves. The local organisation quantifier presented in Panel~(b) indicates that the four cases can be good fits. Additionally, Panel~(c) shows that the the density of the numerical simulations is smaller than the one of the experimental data, in line with the two previous datasets. Indeed, colonies larger than $100 \mu m^2$ have density below $0.9$ for the numerical data compared to $0.95$ for the experimental one. Finally, Table~\ref{Table:7} shows that values taken by $d_2$ are closest to the experimental values $0.4$ for the case~3 with $0.83$, followed by the case~4 with $0.93$. However the gap between the experimental and the numerical values is important. It indicates that the model is not yet good enough to produce a consistent four-cell array organisation. Considering the four quantifiers we can conclude that the case 4 is the best fit for Dataset~3. Yet, the fit can be improved, in particular for the shape of the micro-colony and the four-cell organisation.

 We note that we reach the same conclusion for the colonies of Dataset~3 and for the colonies of Dataset~1: even though they are not in the same experimental conditions, the main mechanism at play for E. coli colonies seems to be the asymmetric friction of the bacteria. Indeed, for both systems we have seen that asymmetric friction was necessary to obtain the elongated shape of the growing colony, and that this mechanism alone enables to recover reasonably good colony shape characteristics. Our results also suggest that an asymmetric mass distribution during cell division could also be at play in these systems, particularly in the early stages of development of the colony. 

\section{Conclusion} \label{Section5}

This paper presented a model for the development of rod-shaped bacteria colony based on some key components which are the asymmetric friction, the distribution of mass along the length of the bacteria and the noise of the angle at division. We aimed to compare our numerical simulations with experimental data and therefore developed a systematic approach to characterise a colony. Different quantifiers have already been developed in the literature, such as the aspect ratio $\alpha_R$, however we did not find in the literature a clear way to compare rod-shaped bacteria colony with different modelling assumption. The quantifiers we considered characterise the shape, the organisation and the density of the colony.  We first studied the influence of these different modelling assumptions. It showed that an asymmetric friction results in elongated colonies while an asymmetric mass distribution or a large angle at division is necessary to recover the four-cell array organisation of a four-cell colony. Then, we compared the numerical simulations with experimental data. We had access to experimental data for the \textit{E. coli} (7 and 32 colonies) and the \textit{pseudomonas} bacteria (10 colonies). In the case of \textit{E. coli} colonies, the quantifiers show that the fitting of the experimental data is improved in two cases: 
\begin{itemize}
\item an asymmetric friction and a high noise at division,
\item an asymmetric friction and an asymmetric distribution of the mass along the length of the bacteria.
\end{itemize}

These results confirm the importance of taking into account the shape of the bacteria and its effect on friction. However, it does not establish the need of a non-uniform distribution of the mass if  there exists a force creating an important angle at division between two daughter cells. Therefore this hypothesis, to be confirmed, would need to be justified by biological evidence of such a phenomenon. In the case of \textit{pseudomonas} bacteria, the colony are not as elongated as for \textit{E. coli} bacteria, therefore the need of an asymmetric friction is not confirmed. Nevertheless, our study highlights that colonies with an asymmetric distribution of mass are a good fit for the experimental data. This result can be improved with a slight asymmetric friction. Overall our results have showed that asymmetric friction and asymmetric mass distribution are good model assumptions to describe the growth of a rod-shaped micro-colony.

As stated previously, the shape and local organisation quantifiers take a wide range of value, even for colonies coming from the same datasets. This suggests that better quantifiers could be found to improve our study. In the case of the shape, observation of the colonies shows that while being elongated, they are also starred shaped (with two or more tips). A better quantifier could therefore take into consideration the tips of the shape. Considering the local organisation, other quantifiers can be found in the literature, especially looking at liquid crystal studies. Improving the choice of the quantifiers is left for further works.
For a still more comprehensive study, we shall also introduce other quantifiers, {\it e.g.} the orientation of the cells at the boundary of the micro-colony~(see Fig.3 of~\cite{DellArciprete2018}), or the position of the oldest bacteria~(see Supplementary Fig.2 of~\cite{Duvernoy2018}). An important research direction is then to evaluate accurately the relevance of each quantifier and to use them to build adequate distances between calibrated models and data.

Improvement in the model can still be made.  Our model did not succeed in reproducing the evolution of the density and $d_2$ in the colony. One of the features which is commonly added to models to improve the colony density and allows to recover a four-cell array organisation is the attraction between the bacteria. Because it is unclear from biological experiments that attraction between particles does exist, our approach consisted in showing that attraction was not essential to recover features of colony growth such as colony elongation and (at least partially) four-cell array structure. Besides, attraction between non-spherical particles can be implemented in various ways and would, therefore, need to be carefully considered. Note that attraction would also have an impact on the shape of the colony. Along with attraction, the interactions between particles which are usually considered are repulsion and alignment. One can question whether these interactions should be considered. Other models have taken different approaches, such as considering adherence with the substrate \cite{Duvernoy2018}, the extracellular matrix \cite{Ghosh:2015aa}, nutrient consumption \cite{Farrell:2013aa}, bacteria attraction \cite{Duvernoy2018}. A perspective of this work would be to compare a wider range of models from the literature. Additionally, a larger choice of experimental data would ideally be considered.



{\bf Acknowledgments.} We thank Nicolas Desprat and his co-authors~\cite{Duvernoy2018} for sharing their data. We are very grateful to Nicolas Desprat and Lydia Robert for inspiring discussions.
\appendix

\section{The algorithm} \label{app:algo}

We describe the algorithm used to simulate the model described in Section~\ref{Section2}. 
\begin{enumerate}
    \item \textit{Initialisation}: N=1
           \begin{enumerate}
           \item $t_0=0$, $k=0$, $dt=10^{-2}$,
           \item $X^0_1=(0,0)$, $\theta^0_1=0$, $l^0_1=l^{\mbox{\scriptsize ini}}$ and $l_1^{\mbox{\scriptsize b}}=l^0_1$ (the size at birth of the bacterium),
           \item Draw the increment at division $\epsilon_1$ and the growth rate $g_1$ according to the law of the at-division increment and growth rate respectively (see Appendix \ref{app:division}),
           \item Compute  $A_i$, $K_i^0$ and $\alpha_i$ for all $i\in[1,N]$.
           \end{enumerate}
    \item \textit{Time loop} : while $t_k \leq T_{\mbox{\scriptsize max}}$  
        \begin{enumerate}
           \item Compute  the force ${F^o_{i,j}}^k$ by checking the interaction between the bacteria $i$ and $j$ for $(i,j)\in[1,N]^2$,
           \item Update $dt$: $dt=dt/2^{h}$ with $h\geq0$ such that $\| \frac{dX}{dt} \| \leq 0.1 d_0$ and $ \| \frac{d\theta}{dt} \| \leq 0.1 \pi $, 
           \item Update $X_i$, $\theta_i$ and $l_i$ for all $i\in[1,N]$ according to 
           \begin{align*}
           X^{k+1}_i & = X^k_i + dt~{K_i^k}^{-1} \frac{1}{l^{k}_i}\sum_{j=1}^N {F^o_{i,j}}^k, \\
           \theta^{k+1}_i & = \theta^k_i+\frac{dt}{\zeta_{\perp} I_i} \sum_{j=1}^{N}  \Big (({X^{o,j}_{i}}^k-X^k_i) \land  {F^o_{i,j}}^k \Big)\cdot z. \\
           l^{k+1}_i & = l^{k+1}_i + dt~g_i~ l^{k}_i.
           \end{align*}
           \item \textit{Division}: for all $i \in [1,N]$, if $l_i-l_i^{\mbox{\scriptsize b}} \geq \epsilon_i$ the bacterium divides into two daughter cells $i_1$ and $i_2$ which are initialised as follows: for $j \in \{i_1,i_2\}$
           \begin{enumerate}
                \item Define $A_j$, $K_j^{k+1}$, $\alpha_j^{k+1}$ and $\theta_j$
                \item $l_i^{k+1}=(l_i^{k+1}-d_0)/2$, \\ $\theta^{k+1}_j=\theta^{k+1}_i +d\theta_j$ with $d\theta_j$ drawn according to a uniform law $\mathcal{U}(-\Theta,\Theta)$,\\ $X^{k+1}_j = X^{k+1}_i +\big (\frac{3}{4}(1+2\alpha^{k+1}_i)~(l_i^{k+1}-d_0) \pm (l_i^{k+1}+d_0)/4  \big) p_j^{k+1}$ with $p_j^{k+1} = (\cos \theta^{k+1}_j,\sin \theta^{k+1}_j)$
                \item Draw the increment at division $\epsilon_j$ and  the growth rate $g_j$ according to the law of the at-division increment and growth rate respectively (see Appendix \ref{app:division}),
           \end{enumerate}
           \item Update $K_i^{k+1}$, $\alpha_i^{k+1}$, $t^{k+1}=t^k+dt$, $k=k+1$ for all $i\in[1,N]$.
           \end{enumerate}
\end{enumerate}

\section{Estimating the distribution of at-division increments} \label{app:division}

As explained in the main text, we use the incremental model for the cell division cycle, {\it i.e.}, the increment of size triggers the division, as proposed in~\cite{Amir:2014aa,Taheri-Araghi:2015aa} and now widely accepted in the biological community. Denoting $\beta(a)da$ the instantaneous probability  of a cell of increment of size $a$ to divide in the increment interval $[a, a+da]$, this means that to simulate the instant of division of a cell being born at time $t$, growth rate $g$ and length $l$, we first pick up a random variable $\epsilon_d$ according to the probability distribution $f_\beta(a)$ defined by
$$\epsilon_d \sim f_\beta(a):=\beta(a)e^{-\int\limits_0^a \beta(s)ds},$$
which is independent of both $g$ and $l.$ This provides the increment at division of the cell. From this value, we easily deduce its time $t_d$ and length $l_d$ of division, defined by
$$l_d=le^{g(t-t_d)}=l+\epsilon_d \qquad \implies\qquad t_d=t+\f{1}{g}\log(l+\epsilon_d).$$
When simulating a given cell, its instant and length of division is thus fully determined by the law $f_\beta$, its growth rate $g,$and  its size at birth $l$. The question is now to estimate the law $f_\beta$ from the experimental data, which, due to the fact that they are given by the dynamics of a full population of cells {\it till a certain time} and not {\it till a certain generation}, present a bias, lineages of fast-dividing cells being over-represented compared to lineages of slowly-dividing cells, see{\it e.g.}~\cite{hoffmann2016nonparametric}. We propose here two methods to estimate this law: one in the simpler case where we assume that all cells grow exponentially with the same rate $g,$ one in the general case. As a preliminary step, let us recall the equation satisfied in large time by the cell distribution.

\paragraph{Asymptotic cell distribution.}

Considering that the growth-and-division processes are space-independent, let us denote $n_k(t,a,l,g)$ the expectation of the empirical measure of cells at time $t$ of increment $a$, length $l$ and growth rate $g$. We take $k=1$ for the case where only one daughter cell is kept at each division, as in microfluidic devices~\cite{Stewart2005}, and $k=2$ for the case where the two daughter cells remain in the micro-colony, as is our case. We have the following equation, see {\it e.g.}~\cite{DHKR}
\begin{equation}\left\{\begin{array}{l}
  \label{eq:div:timeeq}  
    \f{\p}{\p t} n_k (t,a,l,g) + \f{\p}{\p a} \left(g l n_k(t,a,l,g)\right) 
    + \f{\p}{\p x} \left(g l n_k(t,a,l,g)\right) 
    +\beta(a)gl n_k(t,a,l,g)=0,\\ \nonumber \\
    n_k(t,0,l,g)=4k\rho(g)\int\limits_0^\infty\int\limits_{\cal E} g' l \beta(a)n_k(t,0,2l,g')dadg', \\ \nonumber \\
    n_k(0,a,l,g)=n_k^{in}(a,l,g), \qquad \iiint n_k^{in} (a,l,g)dadldg=1.
    \end{array}\right.
\end{equation}
We have assumed here that at birth, a newborn cell has a probability $\rho(g)$ to get the growth rate $g,$ independently of its mother growth rate.

We know by previous studies on similar equations~\cite{gabriel2019steady,BP,olivier2017does} that, under fairly general assumptions, there exists a unique eigencouple $(\lambda_k,N_k)$, with $\lambda_1=0$ and $\lambda_2>0$, such that $n_ke^{-\lambda_kt}$ converges exponentially fast towards $cN_k,$ with some normalisation constant $c>0,$ and $N_k\geq 0$ solution of the following equation:
\begin{equation}\left\{\begin{array}{l}
  \label{eq:div:eigeneq}  
    \lambda_k N_k(a,l,g) + \f{\p}{\p a} \left(g l N_k(a,l,g)\right) 
    + \f{\p}{\p x} \left(g l N_k(a,l,g)\right) 
    +\beta(a)gl N_k(a,l,g)=0,\\ \nonumber \\
    N_k(0,l,g)=4k\rho(g)\int\limits_0^\infty\int\limits_{\cal E} g' l \beta(a)N_k(0,2l,g')dadg', \qquad 
    \iiint N_k(a,l,g)=1.
    \end{array}\right.
\end{equation}
Let us denote $f_k(a):=\f{\iint \beta(a)g l N_k(a,l,g)dldg}{\iiint \beta(a)gl{ N}_k (a,l,g)dadldg}:$ it represents the distribution of dividing cells, observed either along a genealogical line for $k=1$ or for the whole population till a given time for $k=2.$  This is made obvious in the case $k=1:$ integrating the equation in $l$ and $g$, and denoting ${\cal N}_k(a):=\iint  N_k(a,l,g)gl dl dg$ the marginal probability of the increment, we obtain
$$\f{\p}{\p a} {\cal N}_1(a) + \beta(a) {\cal N}_1(a)=0, {\cal N}_1(0)=\int\limits_0^\infty \beta(a) {\cal N}_1 (a) da,$$
so that we have ${\cal N}_1(a)={\cal N}_1(0) e^{-\int\limits_0^a \beta(s)ds},$ thus $f_k(a)=C\beta(a) {\cal N}_1(a)$ with $C>0$ a normalisation constant: all this leads us to re-obtain  the already-known equality $f_1(a)=\beta(a)e^{-\int\limits_0^a \beta(s)ds}.$

The difficulty comes from the fact that to simulate the stochastic branching tree described above, we want to estimate $f_1(a)$, whereas we have experimental (noisy) measurements for $f_2 (a)$, ${\cal N}_2(a)$, and more generally to the all-cell distribution $N_2(a,l,g)$ or yet to the at-division distribution $\f{\beta(a)gl N_2(a,l,g)}{\iiint \beta(a)glN_2(a,l,g)dadldg}.$

\paragraph{Simpler case: all cells grow with the same exponential rate}
In the case where all cells grow exponentially with the same growth rate $g,$ {\it i.e.} we have $\rho (\bar g)=\delta_{\bar g=g},$ the above equations simplify, we have $\lambda_2=g$ and a quick computation shows that $N_1(a,l,g)=ClN_2(a,l,g)$ with $C>0$ a normalisation constant. In this case, as seen above, we may define $f_1$ by
$$f_1(a):=\f{\iint \beta(a) g l^2N_2(a,l,g)dldg}{\iiint \beta(a) g l^2 N_2(a,l,g)dadldg}=\f{\iint  l {\  \tt f^d_2}(a,l,g)dldg}{\iiint l {\tt{f^d_2}}(a,l,g)dadldg},$$
where $\tt f_k^d(a,l,g)$ denotes the at-division distribution of cells. Assuming that we have a sample $(a_i,l_i)_{1\leq i \leq n}$ of increments and lengths at division of cells taken at random in a whole population issued from one single cell and living during a time interval $[0, T]$, we make the assumption (justified asymptotically in~\cite{hoffmann2016nonparametric}) that this sample is the realization of $n$ random variables $(A_i,L_i)_{1\leq i\leq n}$, independent, identically distributed, of law the marginal $\int {\tt f_2^d}(a,l,g)dg,$ so that we have the empirical distribution $ f_1^n$ defined by
$$ f_1^n (a)=\f{1}{\sum\limits_{i=1}^n l_i}  \sum\limits_{i=1}^n l_i \delta_{a=a_i}.$$
We thus propose an estimate $\widehat{f^n_1}$ of $f_1$ by a kernel density estimation approach: let $K \in {\cal C}_0^\infty ({\mathbb R})$ be a smooth fast decaying function with $\int K(a)da=1$ and $\int a^m K(a)da=0$ for $1\leq m\leq m_0$, $m_0\in \mathbb{N}$, we denote $K_h(a)=\f{1}{h}K(\f{a}{h})$ so that $(K_h)_{h\in(0,1]}$ is a mollifier sequence, we define $\widehat{f_1^n}$ by
$$\widehat {f_1^n} (a):=K_h * f_1^n (a)=\f{1}{\sum\limits_{i=1}^n l_i} \sum\limits_{i=1}^n l_i K_h(a-a_i),$$
and we choose $h$ by a data-driven bandwidth selection method, such as Goldenschluger and Lepski's or the recent Penalized Comparison to Overfitting (PCO) method~\cite{DHRR,lacour2017estimator}.
\paragraph{General case: distributed growth rates}
In the general case, we have no simple relation between $N_1$ and $N_2,$ so that we need to first estimate $\beta(a)$ and then compute the distribution $f_1(a)=\beta(a)e^{-\int\limits_0^a \beta(s)ds}.$ A possibility among others is to write
$$\beta(a)=\f{\beta(a)\iint glN_2(a,l,g)dldg}{\iint glN_2(a,l,g)dldg}=\f{f_2(a) \iiint \beta(a)glN_2(a,l,g)dadldg}{\iint gl N_2(a,l,g)dldg},$$
that is, we have identified the numerator with the at-division distribution of increments $f_2(a)$ up to the constant $\iint glN_2(a,l,g)dldg=\lambda$, that we can measure with the time evolution of the total length for instance, and the denominator with increment-dependent average of $gl$ taken over the distribution $N_2$ of all cells at all times. Finally, assuming two samples: a first sample at division, denoted $(a_i^d,l_i^d,g_i^d)_{1\leq i\leq n_d}$, and a second sample taken in the distribution at any time, denoted $(a_i,l_i,g_i)$, we propose the following estimator for $\beta:$
$$\widehat{\beta_{n,n_d}}(a):=\widehat\lambda\f{ \f{1}{n_d} \sum\limits_{i=1}^{n_d} K_{h_d} (a-a_i^d) }{\max\big(\f{1}{\sqrt{n}},\f{1}{n}\sum\limits_{j=1}^{n} g_j l_j K_h (a-a_i) \big)}.$$
To estimate $\lambda,$ we can either follow the total size of the population and fit it as being its exponential growth rate - other said, its Malthus parameter - or use again the properties of the equation, multiply it by $l$ to obtain
$$\lambda=\f{\iiint gl N_2(a,l,g)dadldg}{\iiint lN_2(a,l,g)dadldg},$$
so that we propose the following estimator
$$\widehat \lambda=\f{\sum\limits_{i=1}^n g_il_i}{\sum\limits_{i=1}^n l_i}.$$

\bibliographystyle{plain}

\end{document}